\documentclass[aps,twocolumn]{revtex4-1}
\usepackage{amsmath,amssymb}
\usepackage{slashed}
\usepackage{graphicx}
\usepackage{bm}
\usepackage[T1]{fontenc}
\usepackage[utf8]{inputenc}
\usepackage{gauss} 
\usepackage[top=1.0in,bottom=1.0in,left=1.0in,right=1.0in]{geometry}
\usepackage[colorlinks,linkcolor=blue,citecolor=blue]{hyperref}
\usepackage{subfig}
\newcommand{\be}{\begin{equation}}
\newcommand{\ee}{\end{equation}}
\newcommand{\bea}{\begin{eqnarray}}
\newcommand{\eea}{\end{eqnarray}}
\newcommand{\ba}{\begin{eqnarray}}
\newcommand{\ea}{\end{eqnarray}}

\begin{document}

\title{Hadronic structure on the light-front  V.\\
 Diquarks, Nucleons  and multiquark Fock components}

\author{Edward Shuryak}
\email{edward.shuryak@stonybrook.edu}
\affiliation{Center for Nuclear Theory, Department of Physics and Astronomy, Stony Brook University, Stony Brook, New York 11794--3800, USA}

\author{Ismail Zahed}
\email{ismail.zahed@stonybrook.edu}
\affiliation{Center for Nuclear Theory, Department of Physics and Astronomy, Stony Brook University, Stony Brook, New York 11794--3800, USA}

\begin{abstract}
This  work is a continuation in our  series of papers,  that addresses quark models of hadronic structure on the light front, motivated
by the QCD vacuum structure and lattice results. In this paper we focus on the importance of diquark correlations, which we describe by a quasi-local
four-fermion effective 't Hooft interaction  induced by instantons. The same interaction is also shown to generate extra quark-antiquark pair of  the ``sea". Its higher order
iteration can be included via ``pion-mediation": both taken together yield a quantitative
description of the observed flavor asymmetry of  antiquarks sea. Finally we discuss the final
step needed to bridge the gap between hadronic spectroscopy and parton observables,
 by forward DGLAP evolution towards the chiral upper scale of $\sim 1\, {\rm GeV}^2$.
 \end{abstract}
\maketitle

\section{Introduction}
Since this paper is  the fifth in our  series \cite{Shuryak:2021fsu,Shuryak:2021hng,Shuryak:2021mlh,Shuryak:2022thi}, it does not need an extended Introduction, other than for issues
not considered in the previous papers. So, we start directly by  outlining its content.

The two introductory subsections \ref{sec_dq_intro} and \ref{sec_5q_intro}, are devoted to diquark correlations in baryons
and multiquark hadrons, respectively. Diquark ``spectrosopy" has a rather long history which includes empirical facts, and dynamical calculations (NJL model, instantons, lattice). Furthermore, by identifying certain four-quark effective interactions,  one naturally can proceed to the evaluation of  their role not only in the $2\leftrightarrow 2$ channels, but also
in the $1\rightarrow 3$ channel, and in the  coupling of the 3q and 5q sectors in baryons. 

The introductory subsection \ref{sec_5q_intro}  outlines our general strategy to ``bridge" hadronic spectroscopy and partonic observables. The $1\rightarrow 3$ processes are the first step towards the creation of the ``hadronic sea" of quarks and antiquarks, which complements perturbative DGLAP evolution,  as it is clear from the flavor asymmetry of antiquarks. 

We start our studies of diquark correlations from the nonrelativistic setting in section \ref{sec_C_and tH}, where
we compare the effects of the perturbative Coulomb and instanton-induced 't Hooft interactions using some simple
variational approaches. Its main conclusion is that  diquark correlations are strong, and that the 't Hooft interaction
is dominant. In section \ref{sec_dq_LFWF}, the diquark problem is treated on the light front, by a Hamiltonian
similar to the meson one, using a quasilocal $qq$ interaction. In section~\ref{SIMPLE} we present a simplified analysis of baryons in the CM frame, using Coulomb and $^\prime$t Hooft interactions only.

The next section deals with baryons on the light front, it starts with \ref{sec_heavy_mass} where
we derive the LFWFs deformation by a heavy quark mass. Note that in our previous analysis in~\cite{Shuryak:2022thi}, 
we only considered flavor symmetric baryons $qqq,sss,ccc,bbb$. Here instead, we consider heavy-light baryons such as
$\Lambda_Q=Qud$ with a single diquark in \ref{sec_Lambda_c}, before addressing the diquark pairing in the nucleon in section  \ref{sec_N}. We further elucidate the observable consequences of this pairing by calculating the formfactors for the isobar Delta and the Nucleon in section
\ref{sec_ff}.

\begin{figure}[t!]
\begin{center}
\includegraphics[width=8cm]{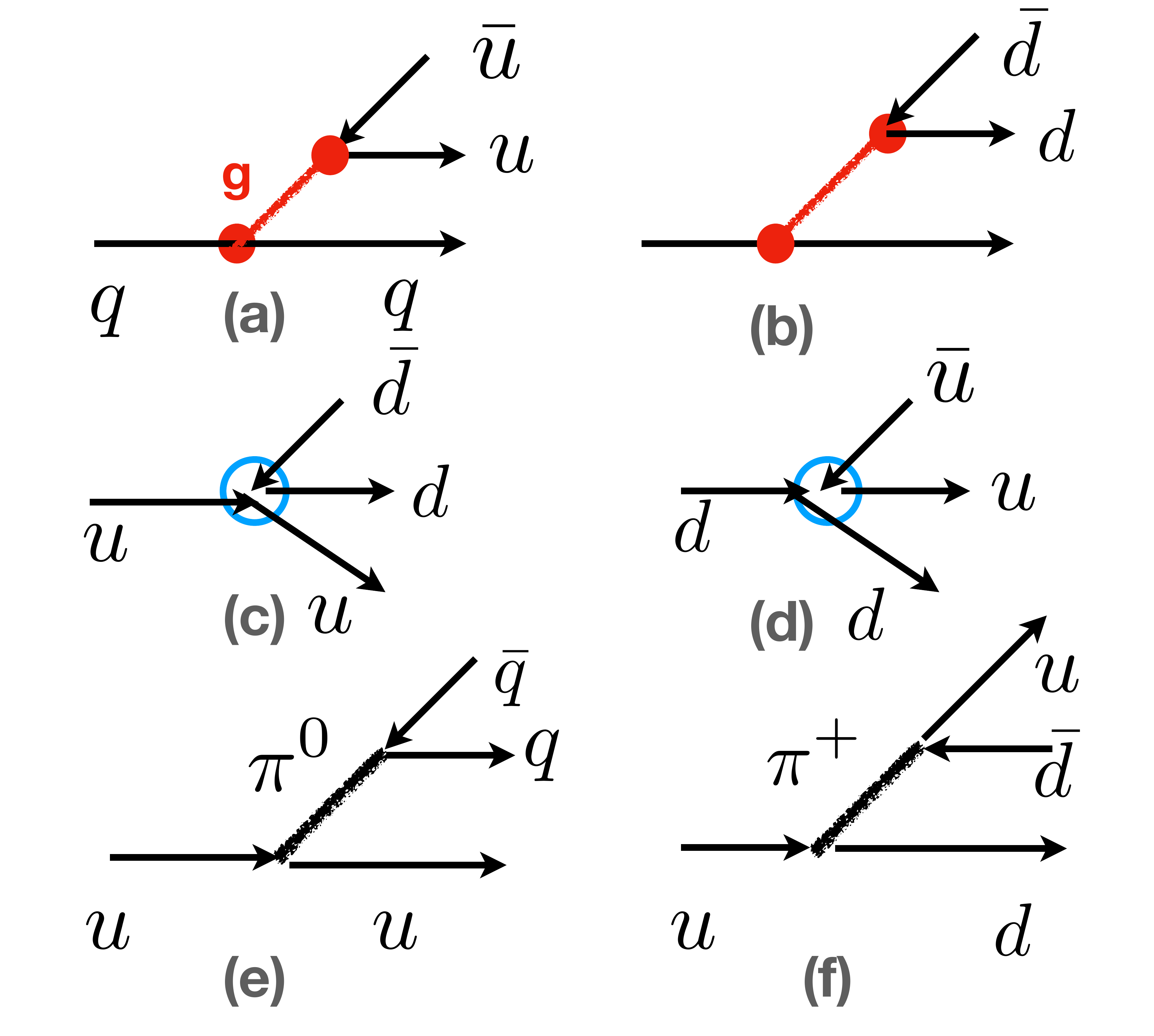}
\caption{Upper raw (a,b): gluon-mediated quark pair production; Middle raw (c,d): instanton-induced 't~Hooft four-fermion
interaction; Lower raw (e,f): pion-mediated  quark pair production, or iterated 't~Hooft Lagrangian in $s$ and $t$ channels.
}
\label{fig_sea}
\end{center}
\end{figure}

In section~\ref{sec_matching_point} we show how to $^\prime$ bridge the gap$^\prime$ between the spectroscopic analysis
and the partonic observables, using the chiral processes discussed in section \ref{sec_chiral_sea}. More specifically, in~\ref{sec_matching_point} we motivate the selection of the scale where the chiral theory and perturbative DGLAP evolution match. In
\ref{sec_flavor_assymetry} we detail the  empirical information on the
flavor asymmetry of the antiquark sea of the nucleon. We present two mechanisms for this effect,
one using the first order in 't Hooft Lagrangian ~\ref{sec_tHooft_sea}, and the other 
a pion-mediated processe  \ref{sec_pion_sea},  each of which is illustrated
in~Fig.\ref{fig_sea}.  The matching of the chiral and perturbative evolutions are discussed in section \ref{sec_DGLAP}.

The last section \ref{sec_summary} summarizes the main results of our series of papers. A number of more
technical issues are discussed in the Appendices.

\subsection{Diquark correlations} \label{sec_dq_intro}
Diquark correlations of light quarks in nucleons and hadronic reactions, 
have been extensively discussed in the  literature in the past  decades, see e.g. \cite{Carroll:1968mlb,Jaffe:2003sg},
and more recently in the review~\cite{Barabanov:2020jvn}. Here, we will not cover their phenomenological contributions
to various hadronic reactions, but rather  address some theoretical considerations about 
their dynamical origins based on semiclassical instantons. We will also cover  recent lattice advances in
diquark studies. 

In two-color QCD with $N_c=2$, diquarks are baryons. In the chiral limit,  QCD with two colors and flavors, admit 
Pauli-Gursey symmetry, an extended SU(4) symmetry that mixes massless baryons and mesons. 
In three-color QCD with $N_c=3$, diquarks play an important role in the light and heavy light baryons. 

The simplest way to understand diquark correlations in hadrons, is in single-heavy baryons where the heavy
spectator quark compensates for color, without altering the light diquark spin-flavor correlations.  A good example are
$Qud$ baryons,  with  $\Sigma_Q$ composed of a  light quark with a flavor symmetric assignment $I=1, J^P=1^+$,
and $\Lambda_Q$ composed of a light quark pair with a flavor asymmetric assignment $I=0, J^P=0^+$ state, the
so called {\it bad} and {\it good} diquark states. Note that the latter has no spin, and
thus no spin-dependent interaction with the heavy quark $Q$, while the former does. However, assuming that the
standard spin-spin interactions are of the form $(\vec \sigma_1 \vec \sigma_2)$, this spin interaction can be eliminated 
as follows
\bea  \label{eqn_binding}
&& M( 1^+ud)-M( 0^+ud)\\
&&\approx \big((2M(\Sigma^*_Q)+M(\Sigma_Q)/3\big)-M(\Lambda_Q)  \approx 0.21\, {\rm GeV}  \nonumber
\eea 
 with the numerical value thus obtained from experimental masses of $cud,bud$ baryons yields the 
 binding of two types of light quark diquarks.  We note tha the 
 mass difference between heavy-light baryon and meson  iof $m(Qud)-m(Qu)\approx 329 \, MeV$,
 is close to a constituent quark mass, but does not seem to include any extra contribution to the kinetic energy
 of the extra quark. Apparently, it is cancelled by some attraction.

  With antisymmetric color and spin wave function, scalar diquarks must also be antisymmetric in flavor:
 so those can only be $ud, us, sd$ pairs. Those are called ``good" diquarks in the literature, in contrast to  the
 ``bad" ones made of same flavor $dd,uu,...bb$ and, by Fermi statistics, with a symmetric spin $S=1$ wave functions.

The role of the light fermionic zero modes induced by instantons, at the origin of the  't Hooft effective Lagrangian for chiral symmetry breaking, 
and their importance for pions and other aspects of chiral symmetry breaking are well known, for a review see e.g. \cite{Schafer:1996wv}.
Diquark correlations induced by the   't Hooft  interaction  were found in studies of the nucleons in instanton ensembles
in \cite{Schafer:1993ra}. In particular, the "good" diquark mass was found to be $m(0^+)\approx 420\, MeV$,
while the  ``bad" vector  diquark mass was found to be $m(1^+)\approx 940\, MeV$, with a difference as large as $500\, MeV$.

Although  only a  Fiertz transformation is needed from a meson to   a diquark channel, this phenomenon has been 
originally considered
only by few~\cite{Betman:1985jj,Praschifka:1989fd,Thorsson:1989fw}, before the 
 realization that diquarks  would turn to  Cooper pairs in dense quark matter, as pointed out in 
\cite{Rapp:1997zu,Alford:1997zt}. (For subsequent review on ``color superconductivity" see \cite{Schafer:2000et}.)

Theoretically, it was important to note that in $SU(2)$ color theory, the  scalar diquarks
are massless partners of the Goldstone mesons~ \cite{Rapp:1997zu}. By continuation to $SU(3)$ color, 
one then expects ``good" scalar diquarks to be deeply bound as well.  
The ratio of the color factors,  between the pseudoscalar meson (pions or $\eta'$) channels, and 
the scalar diquark channel, is the same for the perturbative one-gluon exchange and the instanton-induced 't Hooft vertex. 
\be \frac{G_{qq}}{G_{\bar q q} }={1 \over N_c-1}  \ee
Note that it
  is $1$ for $SU(2)$ color, supporting Pauli-Gursey symmetry between diquarks (baryons in this theory)  and mesons. 
  It is $1/2$ for the $SU(3)$ color case of interest, and zero in the $N_c\rightarrow \infty$ limit.

 Calculation of (pseudoscalar and vector) meson and (scalar and vector) diquark DAs using Bethe-Salpeter equation with NJL kernels 
 were originally carried in~\cite{Thorsson:1989fw},  and more recently in~\cite{2103.03960}. 

Lattice studies of light diquarks have also a long history, with the ealy analyses in \cite{1012.2353} to
recent studies in \cite{2201.03332}. Diquarks are either studied inside dynamical baryons, or by tagging a Wilson line
to a  $qq$ pair as a heavy-light  $Qqq$-type baryon. We already noted  the mass difference between 
the  vector and scalar $ud$ diquarks, with the lattice estimate putting it at $m(1^+)-m(0^+)\approx 200\, MeV$.
 The lattice studies show that in $Qud$ baryon, the  light quarks are
 correlated together, but in a region of about $r_0\approx 0.6\, fm$ size, which is twice larger than 
 suggested in earlier papers.

To complete our  introduction to diquarks, we briefly note the  issue of heavy diquarks, e.g.
made of two charmed quarks $cc$. This issue reappeared after the recent discovery of  the tetraquark 
$T_{cc\bar u \bar d}^{++}$ by the  LHCb collaboration. If the only force is Coulomb, the $QQ$
coupling is  half of that in $\bar Q Q$. Now,  since for a $1/r$ potential the binding scales as the $square$ of the coupling,
we readily get $  B( Q Q)={1 \over 4} B(\bar Q Q) $.
Yet we do know that charm quarks are not heavy enough to ignore the confining forces in charmonium, and
so this relation is not expected to hold.  The static potentials between heavy quarks
were discussed in detail in our previous paper~\cite{Shuryak:2022thi}.

Karliner and Rosner \cite{1408.5877,1707.07666} {\em conjectured} a different relation
\be  B( Q Q)={1 \over 2} B(\bar Q Q)
\ee
which turned out to be phenomenologically successful.
(While it resembles what we called in our previous paper ``Ansatz A" for the quark-quark static interaction, it is  
not the same, a half for $potentials$ is not half for $bindings$. 
For charmonium binding in their analysis  $B(\bar Q Q)\approx -258\, MeV$, so  $B( Q Q)\approx -129\, MeV$,
which led them to a predict a mass  of $M(T_{cc\bar u \bar d}^{++})=3882 \, MeV $ just $7\, MeV$ above  the subsequent experimentally measured value.)  

Currently we have not performed any calculations for tetraquarks. We had done some
preliminary studies of heavy-heavy-light $QQq$ baryons with some model wave functions,
and concluded that for two charm quarks $QQ=cc$ their separation into 
quasi-two-body (heavy diquark plus light ``atmosphere") is not really justified.   This is in 
qualitative agreement  with the relatively small binding of a $cc$ diquark in the Karliner-Rosner
conjecture. So, in this work, we will focus on the light-light ``good diquarks", known to be
more strongly bound.

 \subsection{Bridging the gap between hadronic spectroscopy and partonic physics} \label{sec_5q_intro}
 In this subsection we outline our plan for bridging this gap. 
 
 Our {\em starting point} is the well known  traditional quark model used in hadronic spectroscopy. 
 The main phenomenon included in this model is the lhenomenon of chiral symmetry breaking, with
an effective mass for the  ``constituent quarks". For light quarks it is $m_q \sim 1/3 \, GeV$. This mass is much smaller 
than the induced mass on gluons, so 
 hadronic spectroscopy is traditionally described as bound states of 
 these constituent quarks, with gluonic states or excitations
described as ``exotica".  The traditional states are
 two-quark mesons and three-quark baryons, but of course there are also
 $tetraquarks$ $q^3 \bar q$ and $pentaquarks$
 $q^4 \bar q$ states, recently discovered with heavy quark content.

The {\em first ark of the bridge} (described in detail in these series of works) is to transfer such quark models from
the CM frame to the light front. For some simplest cases -- like heavy quarkonia -- it amounts to a transition
from spherical to cylindrical coordinates, with subsequent transformation of longitudinal momenta into Bjorken-Feynman variable $x$.  But in general, it is easier to start with light-front Hamiltonians $H_{LF}$ and perform its quantization.
One of the benefit is that no nonrelativistic approximation is needed,  therefore heavy and light quarks are treated in the same way. 

The {\em second ark of the bridge} is built via {\em chiral dynamics} ,  which seeds the quark sea by producing extra
quark-antiquark pair.  In section \ref{sec_chiral_sea} we discuss how it can be done, in the first order in 't Hooft
effective action as well as via intermediate pions.
 
 We will then argue that as the {\em third ark of the bridge} one should use  the well known DGLAP evolution of
 the PDFs (perhaps modified),  down to the scale  at which there are  no gluons. There
 the $q\bar q$ sea
 should  be reduced to only the  part generated by chiral dynamics (step two).  
 The antiquark flavor asymmetry $\bar d -\bar u$ is the tool allowing us to tell gluon and chiral
contributions, as  it cannot be generated by ``flavor blind" gluons. 
 
\section{Dynamical binding of diquarks}
\subsection{Nonrelativistic studies of the role of  Coulomb and 't Hooft attractions} \label{sec_C_and tH}
In the previous papers of this series we have shown how two basic nonperturbative phenomena can be included
in the light front formulation:  \\
(i) chiral symmetry breaking represented by  ``constituent" quark masses, \\
(ii) confinement represented by classical relativistic string. \\
By adding the  light-front form of the kinetic energy of the constituents, we derived our  basic Hamiltonian,
modulo Coulomb, spin-spin and and spin-orbit effects. The eigenstates of this Hamiltonian, were evaluated 
using different methods.

Now we are going to focus on the  ``residual" interactions, 
namely: \\
(iii) perturbative Coulomb interactions, \\
(iv) various forms of quasi-local operators descending from 't Hooft effective Lagrangian, or, more generally, from instanton-induced
zero  modes for light fermions\\
(v)  effects due to 
 the gauge fields of the instantons via nonlocal correlators of Wilson lines.
 

The traditional starting point is the  nonrelativistic Schroedinger equation in the CM frame.
As a compromise needed
for the use of  both a nonrelativistic approximation and the  't Hooft Lagrangian,
 we focus initially on the $strange$ quark channel, with a constituent mass  mass $m_s=0.55\, GeV$.  
As for any  compromise, it is not really accurate, yet it will provide
a preliminary information on the relative role of all  the interactions listed above. 

 Since the   't Hooft interaction must be flavor-asymmetric, we have to invent $another$ 
quark flavor  $s'$ with the same mass. (This idea is not ours, it originated in lattice studies
where it is used to eliminate two-loop diagrams. The pseudoscalar
$\bar s s'$ meson even has an established  name $\eta_s$.) 

We start with a $variational$ approach, using 
 simplified trial wave functions of two types
\be \psi_A\sim e^{-\alpha r^2}, \,\,  \psi_B\sim e^{-\beta r^{3/2} }
\label{eqn_A_B} \ee
to be referred to as trial functions A and B. The former (Gaussian) form leads to simple analytical expressions
for the mean kinetic energy, $\langle 1/r \rangle,  \langle r \rangle, \langle \delta^3(\vec r) \rangle$. However, 
 the trial  function B with  the power of the  distance in the exponent 
following from its semiclassical asymptotics, 
 turns out to be closer in shape to the numerical solution.
Some details about these variational functions can be found in Appendix \ref{sec_variational}.

Let us summarize the qualitative lessons we got from these variational studies. First, we
demonstrate that the contributions of both attractive  forces -- the Coulomb and  the 't Hooft ones -- 
are comparable for the strange quark mass. The light diquark binding and the r.m.s. size suggested by phenomenology and observed on the lattice,
can  be explained with the conventional values for the Coulomb and 't Hooft couplings.

 Second, we find the following distinction between these interactions: their contribution to
 the binding can change significantly if these values are changed. For example,
if the diquark size is reduced
 by a factor two, $ \langle 1/r \rangle$ increases by a  factor of 2, while  $\langle \delta^3(\vec r)  \rangle$ increases
by a factor $2^3=8$ and becomes dominant. 
So a reported balance between perturbative and nonperturbative contributions to the binding,
is in fact only valid for a strange quark mass, and is very sensitive to the actual quark masses.

Of course, in the current setting, there is no problem to solve the Schroedinger equation numerically.
For convenience, we represent the  't Hooft quasi-local term $-G_{qq} \delta^3(\vec r)$ by a smeared
delta function. In Fig.\ref{fig_diquarks_various_G} the resulting ground state
wave functions are compared, for  four values of the coupling $G_{dq} =0,10,20,30\, GeV^{-2}$.
As expected, an increase in the  negative potential near the origin, leads  to a  large wave function at small $r$.

In Table \ref{tab_diquark_G} we give the corresponding r.m.s. sizes and binding energies for the lowest four states. 
One can see that in order to get a binding of  $\sim -0.2 \, GeV$ and a size of $\sim 0.6 \, fm$, indicated by lattice studies,
the coupling needs to be $G_{qq} \sim 20 \, GeV^{-2}$. Recall  that by  Fiertzing the  't Hooft operator
from the $\bar q q$ to $qq$ channel, there is an additional factor of $1/(N_c-1)=1/2$. So  this value  corresponds
to  the coupling in mesons of $\sim 40 \, GeV^{-2}$.  Finally, we note that both the  't Hooft and Coulomb bindings
are only strong for the ground state, and their effect is strongly decreasing with $n$, as is listed  in this Table.

\begin{widetext}
\begin{table}[b!]
\caption{Root-mean-square sizes (fm) and additional binding energies (GeV) 
for the  lowest four states of $ss'$ diquarks, for four values of the $qq$ 't Hooft coupling }
\begin{center}
\begin{tabular}{|c|c|c|c|c|c|}
\hline
$G_{qq} \,GeV^{-2}$ & $R_{r.m.s} \, fm$ & $E_0-E_0(G_{qq}=0)$ & $E_1-E_1(G_{qq}=0)$ &$E_2-E_2(G_{qq}=0)$ &$E_3-E_3(G_{qq}=0)$ \\
\hline
0 & 0.67 & - & -& -& - \\
10 &    0.62 &          -0.082 & -0.044 & -0.032& -0.026 \\
20 &     0.56  &         -0.187& -0.089 & -0.063 & -0.051 \\
 30 &      0.50 &            -0.318 & -0.132 & -0.093 & -0.074 \\
\hline
\end{tabular}
\end{center}
\label{tab_diquark_G}
\end{table}%
\end{widetext}

\section{Diquarks on the light front  } \label{sec_dq_LFWF}
In order $not$ to use the nonrelativistic approach for the light quark systems,   in the previous papers of this series we have
advocated the use of the light front Hamiltonians. The confining forces were reduced by the  $^\prime$einbein trick$^\prime$. The resulting
Hamiltonian $H_{LF}=H_0+V$,
consits of $H_0$ which is the sum of a  harmonic oscillator in transverse momenta 
plus a Laplacian for longitudinal momenta, and of $V$ wich includes transverse
momenta in the numerator and longitudinal ones in the denominator.
One strategy of solving this Hamiltonian is, following our predecessors \cite{Jia:2018ary}, is
 to express $V$ as a matrix in the eigenbasis of $H_0$, 
  with its subsequent diagonalization. In paper \cite{Shuryak:2021hng} we have checked the wave functions
  obtained by this method with direct 3d numerical solutions, with very good agreement  between the two.

 Here we propose another method of solution for $H_{LF}$, relying on the approximate factorization of
the  transverse and longitudinal degrees of freedom. Treating the mean square of the transverse momentum as a parameter,
 we focus on the longitudinal wave equation following from  the relevant part of $H_{LF}$, which is proportional to
 \ba \label{eqn_long_Schr}
 &-&{d^2 \psi(x) \over dx^2} +A \bigg( {M_1^2+\langle \vec p_\perp^2 \rangle \over x}
 + {M_2^2+\langle \vec p_\perp^2 \rangle \over 1-x} \nonumber  \\
& - & 2(M_1^2+M_2^2+2\langle \vec p_\perp^2 \rangle)
 \bigg) \psi(x) \ea
The second derivative comes from the confining term (with parameter $a$ inherited from the
 ``einbine trick") following the substitution of  the longitudinal coordinate
 as $r_l\rightarrow i d/dx$. The ``potential"  is written for  two distinct masses, to keep it general.
  The mean squared transverse momentum is used as an  external parameter, together with the quark masses.
 The last term with the minus sign in the ``potential" is artificially subtracted here, and
 added in the remaining part of the $M_2$ for convenience (it is independent of the  longitudinal momentum fraction $x$.)
 
 The constant $A=a/\sigma_T$ in front of the potential contains the string tension $\sigma_T=(0.4\, GeV)^2$,
 and a parameter $a$ from the  ``einbine trick" we have used (which is fixed by minimizing the total mass squared).
 
 \begin{figure}[h!]
\begin{center}
\includegraphics[width=6cm]{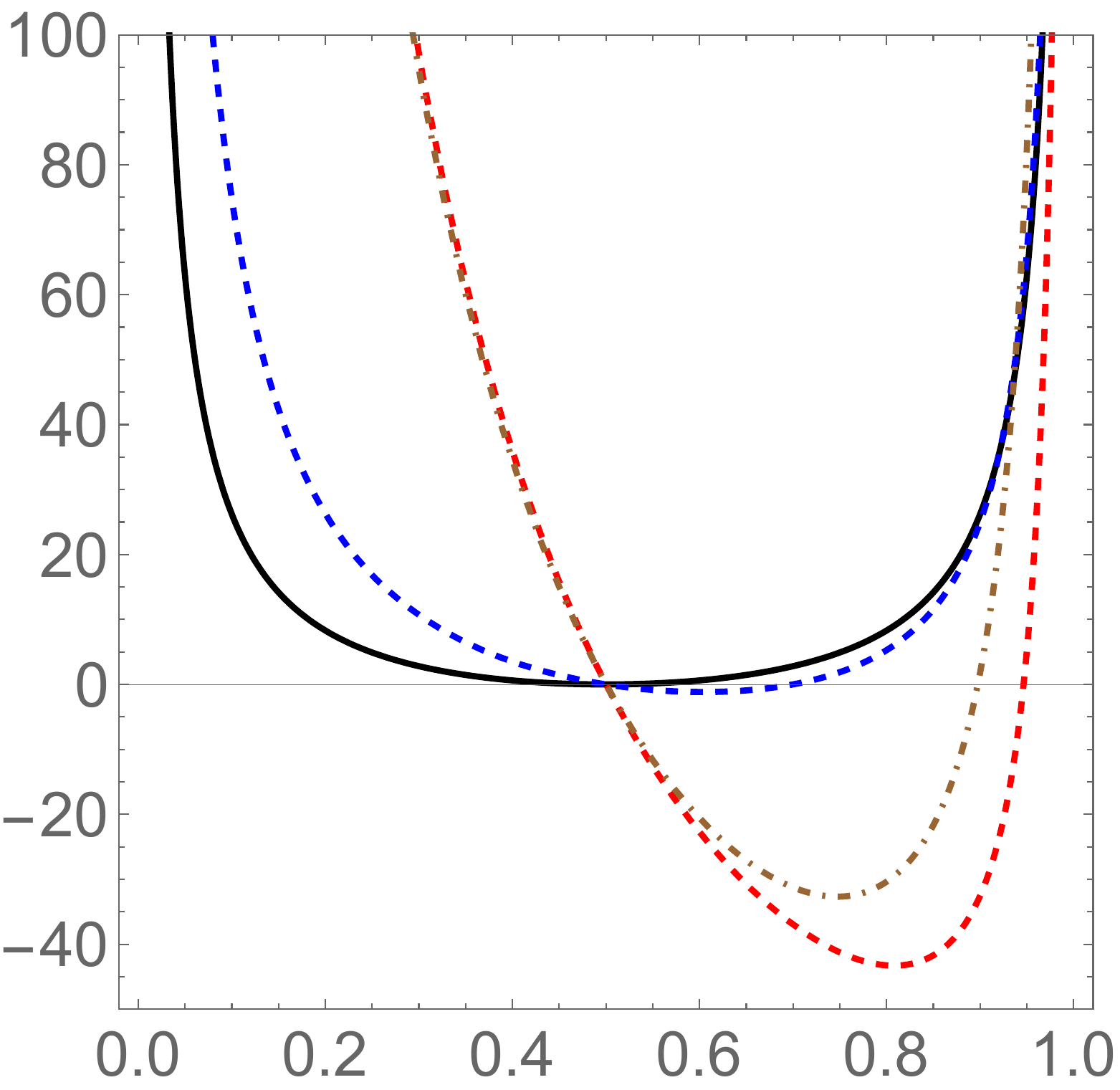}
\caption{Effective potentials in  the  longitudinal Hamiltonian (\ref{eqn_long_Schr}), for a $qq$ pair (black curve), for a $sq$ pair
(blue dashed curve), for a $cq$ pair (red dashed curve) and for a $c-(ud)$ charm-light diquark case (brown  dash-dotted curve).
}
\label{fig_long_4_pots}
\end{center}
\end{figure}

 In Fig.\ref{fig_long_4_pots} we show the shape of the corresponding effective potentials for four cases:\\
 (i) a pair of light constituent quarks $qq$ with masses $M_1=M_2=0.28\, GeV$ with comparable mean squared 
 transverse momentum $\langle \vec p_\perp^2 \rangle=(0.3\, GeV)^2$\\
 (ii) a strange-light pair (e.g. $K^*$ vector mesons)\\
 (iii) a charm-light pair (e.g. $D^*$ vector mesons) \\
 (iv) a charm-$ud$ diquark, an  approximation to $cud$ or $\Lambda_c$ baryon. We use here the diquark mass  $m_{ud}=0.5\, GeV$\\
 As one can see, the potential is small and symmetric in the $qq$ case (i), except near the edges of the
 physical domain, $x=0,$ and  $x=1$. Yet the potential becomes very asymmetric
 if the two masses are different.
 
\begin{figure}[htbp]
\begin{center}
\includegraphics[width=6cm]{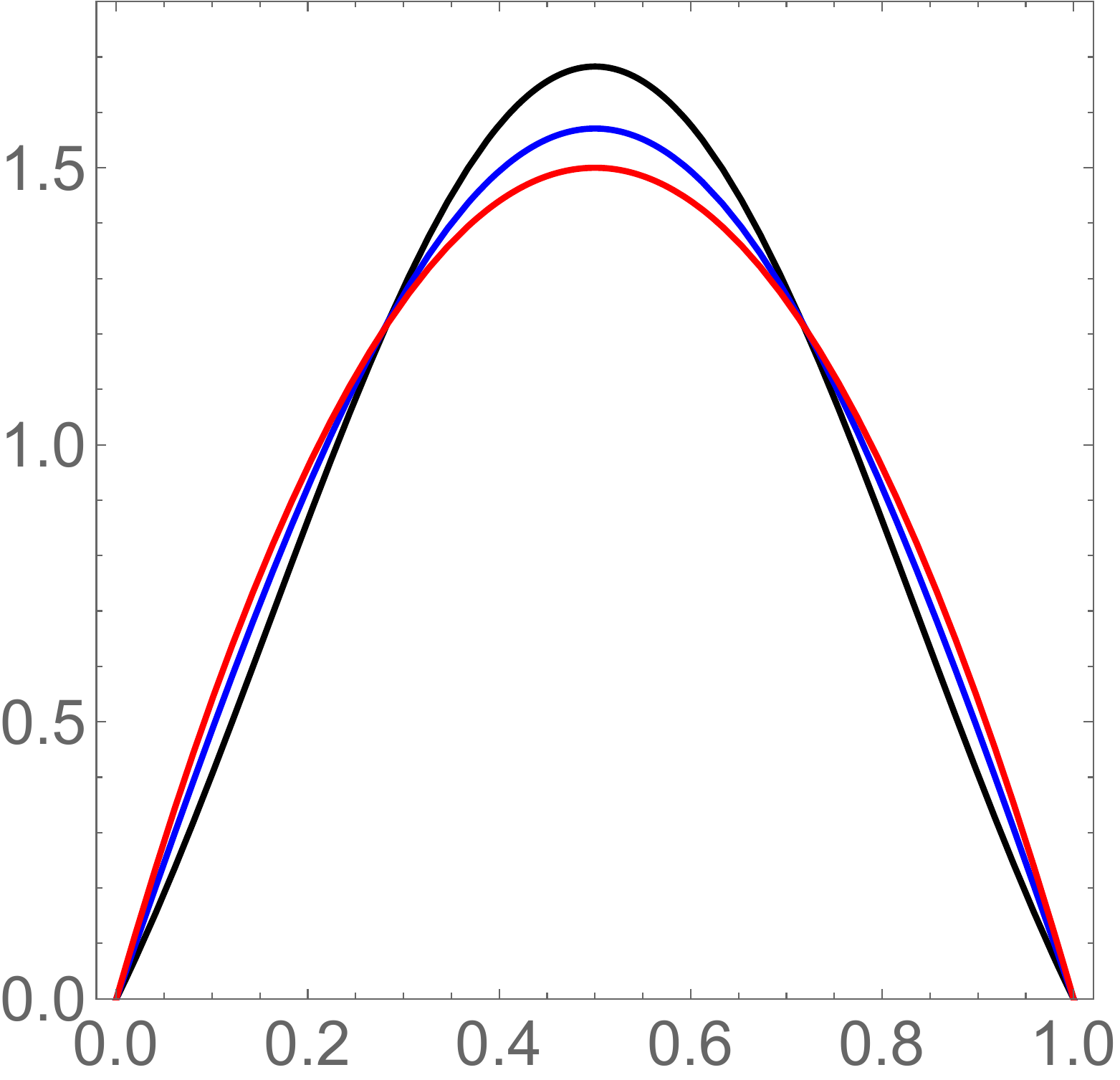}
\caption{Longitudinal wave function for a diquark  made of identical light quarks (black line, upper curve  at
the central point $x=1/2$).
For comparison we also show the lowest harmonic function $sin(\pi x)$ by a blue line, and 
the ``asymptotic" wave function $x(1-x)$ by the red line.}
\label{fig_qq_3psi}
\end{center}
\end{figure}

We  solved  the longitudinal wave  equation  following from (\ref{eqn_long_Schr}) for two light quarks (case (i)), and compared the solution 
(shown by a black line in Fig.\ref{fig_qq_3psi})e
to  two functions which are often used as simple approximations to these wave functions. 

(Note that in the approximation of constant $p_\perp$ distribution, these functions 
directly coincide with the Distribution Amplitudes (DAs). Therefore we have used here
a normalization traditionally used for  DAs, by putting to unity the integral of its $first$ power
rather than the  integral of the square, which is  more appropriate for the wave functions.)   

\begin{figure}[h]
\begin{center}
\includegraphics[width=6cm]{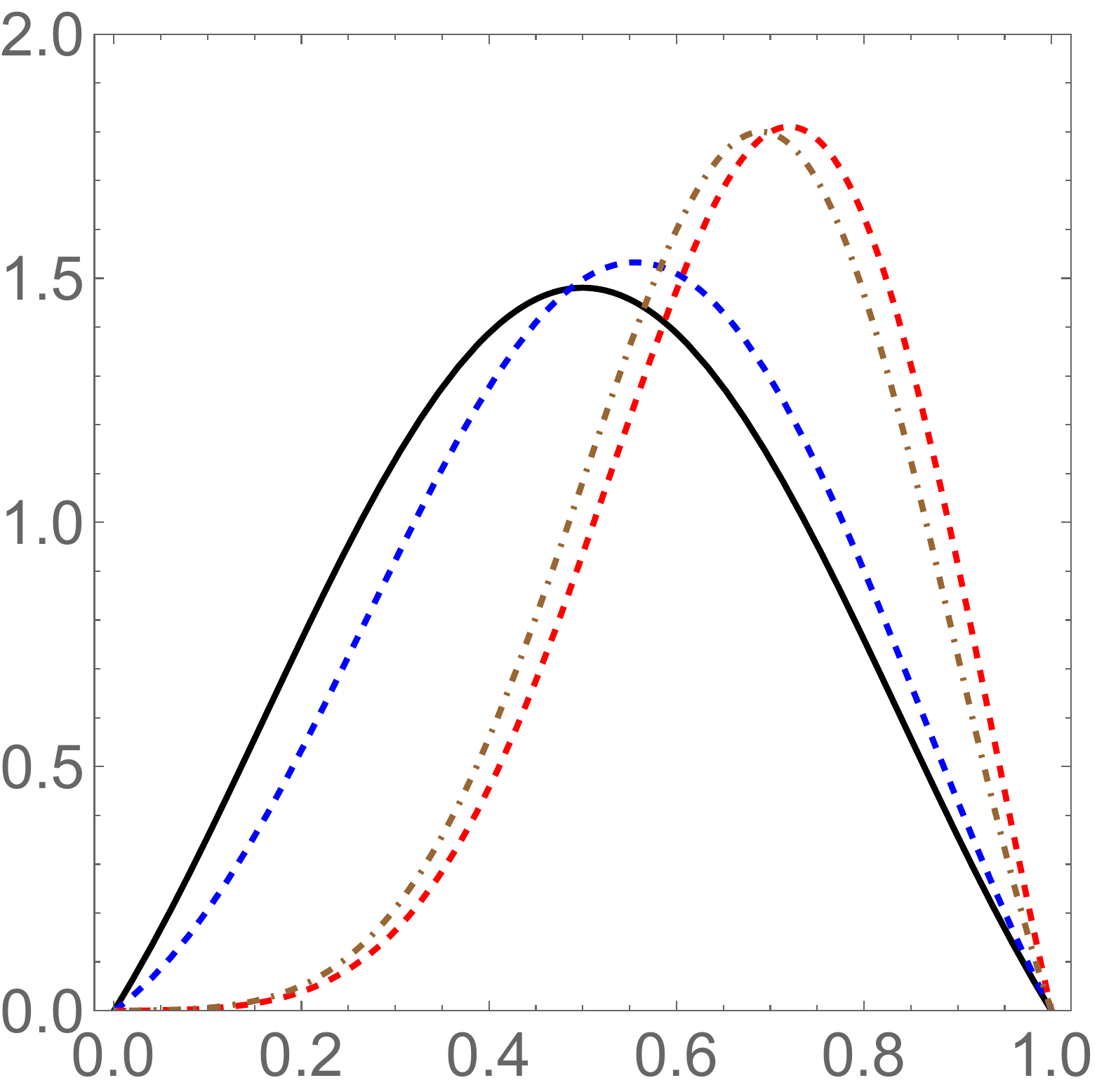}
\includegraphics[width=6cm]{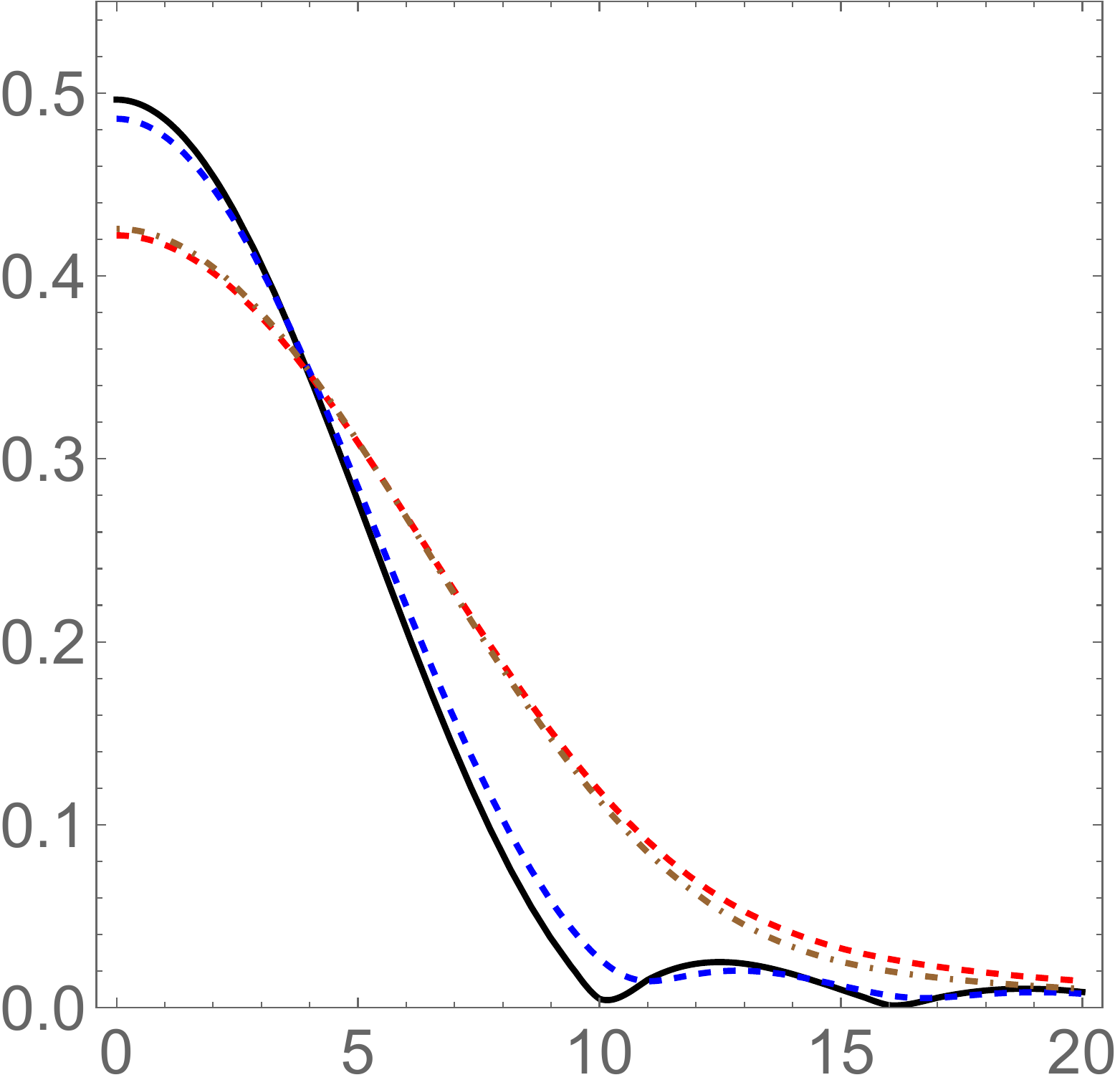}
\caption{The wave functions as a function of the momentum fraction $\psi(x)$ (upper plot) and their
Fourier Transform into coordinate space $\psi(P*r_l)$ (lower plot). The marking of the four curves 
in both plots are the same, and identical to those used 
fin Fig.\ref{fig_long_4_pots}: 
 for the $qq$ pair
 it is the black curve, for the $sq$ pair it is the
blue dashed curve, for the $cq$ pair it is the red dashed curve, and for the $c-(ud)$ charm-light diquark case it is the brown  dash-dotted curve.}
\label{fig_4psi}
\end{center}
\end{figure}

It is then straightforward  to solve the equation (\ref{eqn_long_Schr}) for the remaining three cases. All these ground state wave functions are  
shown in Fig. \ref{fig_4psi}, in the upper plot as numerical solution $\psi(x)$, and in the lower plot as its Fourier
transform $ \psi(P r_l)$.
The  asymmetric potentials lead to rather asymmetric wave functions, shifting toward larger $x$
of the first (heavier) particle. This makes sense, since the pair binding requires that the constituents 
move with the same relative velocity, which translates to a larger momentum fraction carried by the heavier
quark.

For completeness, let us give the (dimensionless) values of the ground state eigenvalues :
those  are  $$ 12.78,  14.23, -8.10, -0.23$$    for these four cases respectively.

The wave functions  dependence on the transverse
momenta is (near) Gaussian, and can be readily Fourier transformed. So, with the longitudinal wave functions 
Fourier transformed into the $coordinate$ representation,  we can 
calculate the matrix elements of any coordinate-dependent operators, such as 
the perturbative Coulomb.  However in this paper we would not do so. We will only only consider the quasilocal 't Hooft operator,
which can be done without any Fourier transform.

\section{Simplified baryons with Coulomb and 't Hooft interactions in the CM frame}~\label{SIMPLE}
In this section we proceed from two quarks in a diquark to three quarks in a baryon, including the  ``residual" 
interactions beyond confinement. As for diquarks, we start with a
  preliminary study of three nonrelativistic quarks
with equal masses in the CM frame, and proceed  in three steps: (i) a ``basic" 3-body problem, with only kinetic and confining energies; (ii)
adding Coulomb interaction; (iii) adding quasi-local 't Hooft interaction.

To  discuss the three-body Hamiltonians and wave functions, we first rrecall
how the three coordinate vectors $\vec x_i$ are redefined in terms of two Jacobi vectors 
$\vec \rho_i, \vec \lambda_i$ (see (\ref{eqn_rho_lambda}) for their longitudinal analog), with the kinetic energy 
\bea
\label{H0XX}
H_{kin}=-\frac 1{2M}\frac{\partial^2}{\partial\vec X^2}-\frac {1}{2m}\bigg(\frac {\partial^2}{\partial\vec\rho^2}+\frac {\partial^2}{\partial\vec\lambda^2}\bigg)
\eea
The first term, represents the center of mass motion, and is  factored out. The 
 remaining dynamics is performed in  the remaining six dimensions.

We start with the basic problem, with only the kinetic and confining energy. In this case
the problem is spherically symmetric in 6d and one can define the
hyperspherical radius
\bea
\label{RR2}
R^2=\vec\rho_{ij}^2+\vec\lambda_{ij}^2
\eea
(with any pair  combination $[ij]$, over which no summation will be assumed). 
 (\ref{H0XX}) is the 6-d Laplacian
 \be  H_{kin} =-\frac 12  \bigg({\partial^2 \over  \partial R^2} + {5 \over R} {\partial \over  \partial R} \  \bigg) +{L^2 \over 2 R^2} \ee
where $L$ includes  derivatives over the angles. The ground state wave function may be 
 assumed to depend only on the hyperradius $R$. 
 
 However, since we will consider a case in which {\em one quark pair} 1-2 will have a  't Hooft attraction,
 there is no such symmetry, and we expect a different dependence on $\rho=|\vec \rho |$ and $\lambda=|\vec \lambda |$.
We will use the following two-parameter Gaussian Ansatz
 \be \psi(\rho,\lambda)=  (2\alpha\beta/\pi)^{3/2}\,e^{-\alpha^2 \rho^2-\beta^2 \lambda^2}
 \ee
 The kinetic energy is then
 \be E_{kin}={3 \over 2m} (\alpha^2+\beta^2) \ee
The confining energy  is then
\be E_{conf}=\sigma_T \langle \sqrt{2}\rho + \sqrt{3/2}\lambda \rangle =\sigma_T { \sqrt{3} \alpha + 
   2 \beta \over \sqrt{\pi} \alpha \beta }
\ee
where the first term includes the  distance between the quarks 1 and 2, and the second term  the distance from quark 3
to the CM of the first pair. Here and below, the  angular brackets stand for the averaging over the wave function. 

Using our standard set of parameters, with the strange quark mass $m=0.55\, GeV$, and  the string tension $\sigma_T=(0.4\, GeV)^2$, 
we minimize these expressions and get variational wave function for the  ``basic" baryons, with
only the confining force written in this approximation. The minimum of the energy is at $\alpha_{min} =0.32,\, \beta_{min} =0.31 \,\, GeV^{-1}$, and the minimal values of the kinetic and confining energies are $E_{kin}\approx 0.38 \, , E_{conf} \approx 1.07\, GeV$. (The slight asymmetry is caused by the asymmetric geometry of the confining string.)

Now we include the  ``residual" interactions, and minimize the total energy.
The Coulomb interaction for the pair 12 includes the distance between these two quarks $r_{12}=\sqrt{2}\rho$,
with
\be E_{Coulomb}=-{2 \alpha_s \over 3} \bigg< {1 \over r_{12}} \bigg>=-{2 \alpha_s \over 3} {2 \alpha \over \sqrt{\pi}} \ee
The 't Hooft Lagrangian induces a quasi-local 4-fermion operator, which we regulate using a Gaussian form with
an fixed instanton size $\rho_0$,
\ba  E_{'t Hooft}&=&-G_{qq} \bigg<  \frac{e^{-\rho^2/\rho_0^2}}{\pi^{3/2} \rho_0^3}\bigg> \nonumber \\
&=&- G_{qq} \alpha^3 \bigg({ 2\over \pi+2\pi \alpha^2 \rho_0^2}\bigg)^{3/2} 
\ea
Using our standard set of parameters, the strange quark mass $m=0.55\, GeV$, the string tension $\sigma_T=(0.4\, GeV)^2$,
the Coulomb coupling from charmonium fit $4\alpha_s/3\approx 0.7$, the 
instanton size  $\rho_0\approx 0.3 \, fm\approx 1/(0.6\, GeV) $ and the quark-quark coupling $G_{qq}=20\, GeV^{-2}$, we minimize
the total energy, the sum of the terms mentioned above, over the parameters of the Ansatz.
With the Coulomb and 't Hooft terms acting only between particles 12, we found that the minimum is at
\be   \alpha_{min}= 0.430 \, GeV^{-1}, \,\,\,\beta_{min} = 0.306 \, GeV^{-1},
\ee
Note that parameter alpha has changed significantly as compared to the  ``basic" model above.

Using it, one finds the relative contributions of different terms in the Hamiltonian to be in this case (in GeV)
\ba E_{kin} &\approx & 0.76, \,\,E_{conf} \approx  0.93, \\
E_{Coulomb} & \approx &-0.17, \,\, E_{tHooft} \approx  -0.28 \nonumber \ea

The main lesson from this variational estimate,  is that again, we see that the instanton-induced 't Hooft effect is somewhat
larger than the Coulomb interaction, and that together they can generate significant diquark binding
comparable to the quark constituent mass $\sim 0.35\, GeV$. 

We conclude this section with the following comments:\\
(i)
The main lesson from this variational calculation is 
that the asymmetry induced by the ``residual" Coulomb and 't Hooft interactions is 
quite substantial, and therefore their perturbative account is $not$ justified.\\
(ii)
The Coulomb and 't Hooft attractions are  so far included only for $one$ pair
of quarks 1-2, out of three. If those would be multiplied by three, the
attraction can basically cancel the kinetic and confinement energy, leaving the total mass
close to its naive nonrelativistic value $3m$. \\
(iii) 
In our older paper  \cite{Shuryak:2003zi}  we developed a schematic model 
with certain  ``quasi-supersymmetry" between $masses$ of constituent quarks and light ``good diquarks". 
(Not between the number of states, as needed for true supersymmetry.)
In particular, it put certain mesons and baryons  into some (approximate)
multiplets. 
A similar meson-baryon symmetry has been developed in \cite{Brodsky:2015oia} based on 
a hybrid holographic approach.  Naively the same multiplets would include tetraquarks, as diquark-diquark states:
 yet those were not observed. It implies  that diquark-diquark interaction
 is strongly repulsive, violating this ``quasi-supersymmetry".

\section{Heavy-light baryons with diquarks} \label{sec_Lambda}
\subsection{Asymmetry induced in the longitudinal LFWFs by a heavy quark mass}  \label{sec_heavy_mass}
In our previous paper \cite{Shuryak:2022thi} we have studied {\em flavor-symmetric} baryons,
$qqq,sss,ccc,bbb$. Such choice was motivated by both the additional kinematical symmetry in each  case,
as well as the absence of an  instanton-induced flavor-antisymmetric 't Hooft interaction.  

The 3-body kinematics was discussed there in detail, and will not be repeated here. For completeness, we briefly recall 
our notations.
 We use the momentum representation and 6 Jacobi coordinates $\vec p_\lambda, \vec p_\rho$. Their longitudinal
momenta, normalized to total hadron momentum  called ``momentum fractions" $x_i,i=1,2,3$,  are expressed
in two coordinates $\rho,\lambda$
as follows 
\ba 
\label{eqn_rho_lambda}
 x_1 &=& 1/6 (2  + \sqrt{6} \lambda + 3 \sqrt{2} \rho); \nonumber \\
x_2 &=& 1/6 (2  + \sqrt{6} \lambda - 3 \sqrt{2} \rho) \\
x_3 &=& 1/3 (1 - \sqrt{6} \lambda) \nonumber \ea
The physical domain of the $\rho,\lambda$ variables  is an equilateral triangle, with corners corresponding to one of the
momentum fractions reaching one, and the others zero. 

The light front Hamiltonian considered in \cite{Shuryak:2022thi} included the kinetic energy of quarks and confining
term only. The latter term, with certain tricks,  is made proportional to  the 6-d quadratic form in coordinates.
After those are changed to derivatives over momenta, it is amenable to transverse and longitudinal
 Laplacians. (This is so for both the  $Y$ confining model, with three strings and a color junction, as well as for the  ``A Ansatz" with
 slightly different numerical coefficients.)
 Here we will focus on the ongitudinal momenta, so our main differential operator takes the form
 $\partial^2/\partial\rho^2+\partial^2/\partial\lambda^2$, defined on the equilateral triangle with corners
 at $$(\rho,\lambda)=(0, -\sqrt{2/3}), (1/\sqrt{2}, 1/\sqrt{6}), (-1/\sqrt{2}, 1/\sqrt{6}) $$
  In \cite{Shuryak:2022thi}, 
 its eigenfunctions were found analytically and numerically.

The kinetic part of Hamiltonian 
\be H_{kin}=\sum_i {\vec p_{i\perp}^2+m_i^2 \over x_i} \ee
depends nontrivially on the longitudinal momentum fractions. It is convenient to subtract and add its value
at the center point $x_1=x_2=x_3=1/3$ and call it a ``potential" $V$ plus a term depending only quadratically
on $\vec p_\perp$. The latter was used for the  ``transverse oscillator" part, defining
the basis set of functions. The potential $V$ was included either in the form of a
matrix in that basis, or found numerically from solving the Schroedinger-like equation. 

The confining (Laplacian) part of the Hamiltonian does not depend on masses, but the potential $V$ does.
If masses of the three quarks are the same (the case discussed in  \cite{Shuryak:2022thi}) it has a discrete symmetry
corresponding to maps of the triangle into itself (rotations by the angle $\pm 2\pi/3$). This symmetry is shared
by a Laplacian and thus the resulting wave functions. In this section we make the first step towards $unequal$
quark masses,  where  this symmetry is absent.

We are aiming first at heavy-light baryons of the type $qqQ$ (with light diquarks either flavor symmetric or antisymmetric),
and introduce two dimensionless parameters 
\be  A_Q= { \langle p_{Q\perp}^2\rangle +m_Q^2      \over  \sigma_T } , \,\,\,         A_q={ \langle p_{q\perp}^2\rangle +m_q^2      \over \sigma_T } 
\ee
The denominators  contain the string tension, while in the 
numerators we substituted the squared transverse momenta by their average. In this approximation the
longitudinal degrees of freedom split from the transverse ones, with the effective dimensionless potential
\be \tilde V(\lambda,\rho)=A_q (1/x_1+1/x_2-6)+A_Q(1/x_3-3) \ee
While it is still defined on the equilateral triangle, it no longer has  the triple $2\pi/3$ symmetry. Specifically, we note 
that these parameters have very different  magnitude for light and e.g. charmed quarks
 $$(\langle p_{q\perp}^2\rangle +m_q^2)/\sigma_T\approx  1.5, \,\,\, (\langle p_{c\perp}^2\rangle +m_c^2)/\sigma_T\approx  15 $$
For $b$ quarks the ratio is another order of magnitude larger. 
 So the triple symmetry is very strongly broken by the leading term with a large heavy quark mass. The effect
 is so strong, even for charm quark, that one can only keep the $m_c^2(1/x_3 -3) $ term in the kinetic energy.

We evaluated the matrix elements of the potential in the laplacian basis, and got the wave function of the $\Sigma_Q=Q(ud)_{1+}$ baryon. The  ground wave functions represented as a combination of (12) basis states 
\be \Psi_{\Sigma_Q(1)}=\sum_n C_n \psi_n(\rho,\lambda) \ee
with the coefficients equal to
\ba C_n^\Sigma=(0.520, 0.025, 0.736, -0.331, -0.167, -0.189, \nonumber \\
-0.105, -0.009, -0.024, 0.012, -0.014, -0.008 )   \nonumber  \ea
Note that several of coefficients are comparable, and the first coefficient is not even the largest.
It happens because the charm quark mass term creates such a  large perturbation, that
the lowest $\Sigma_Q(1)$ state is not even close to the lowest state of the Laplacian.

\subsection{Diquark pairing  in $\Lambda_Q$ baryons} \label{sec_Lambda_c}
Now we make the second step, to $\Lambda_Q=Q(ud)_{0^+}$ baryons with a flavor-asymmetric $ud$ diquark. In addition to what
was discussed in the preceding section, now there is also the   't Hooft determinantal interaction  between the $ud$ quarks. 
The symmetry $1 \leftrightarrow 2$ (or $\rho\leftrightarrow -\rho$) of the Hamiltonian and the LFWFs remains.

We now make use of a simplified and fully $local$ $^\prime$t Hooft interaction
\be H_{ud}=-G_{ud} \delta(\vec r_u-\vec r_d) \label{eqn_local} \ee
The matrix element of the $spatial$ delta functions is discussed in Appendix~\ref{sec_quasilocal}, see (\ref{eqn_mean_delta_of_rho}). Using
it in our set of basis functions,  we performed those integrals and obtained the Hamiltonian in a form of  a 12$\times $12 matrix.
Multiplying it by the coupling $G$, adding it to Hamiltonian detailed in the previous sections and getting the eigensystem,
we generated 12 states of the $\Lambda_c$ baryons. We have tuned the coupling so that the ground states have 
binding difference between ``good" and "bad" diquark fixed  by phenomenology (\ref{eqn_binding}).

\begin{figure}[t!]
\begin{center}
\includegraphics[width=6cm]{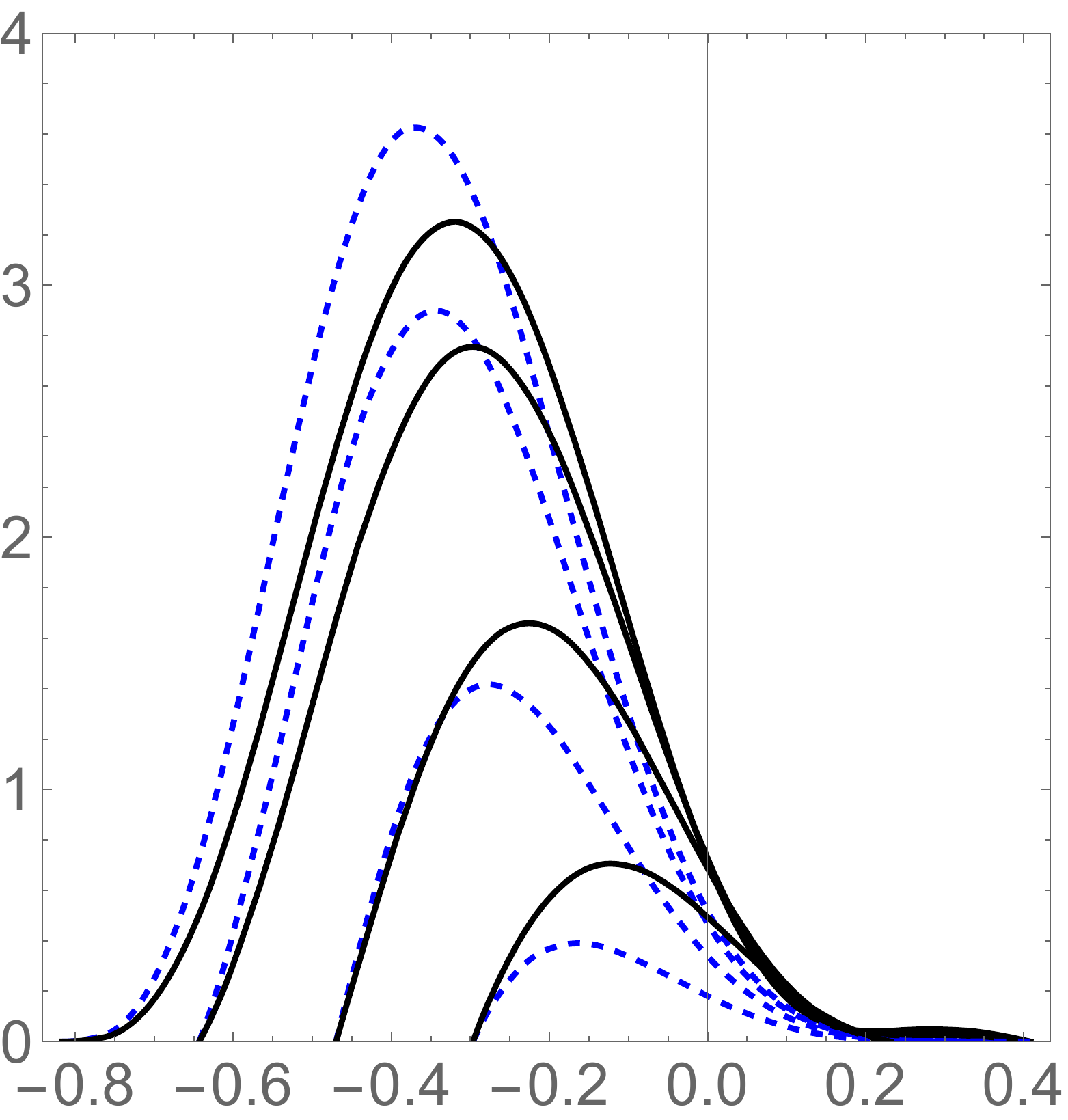} \includegraphics[width=6cm]{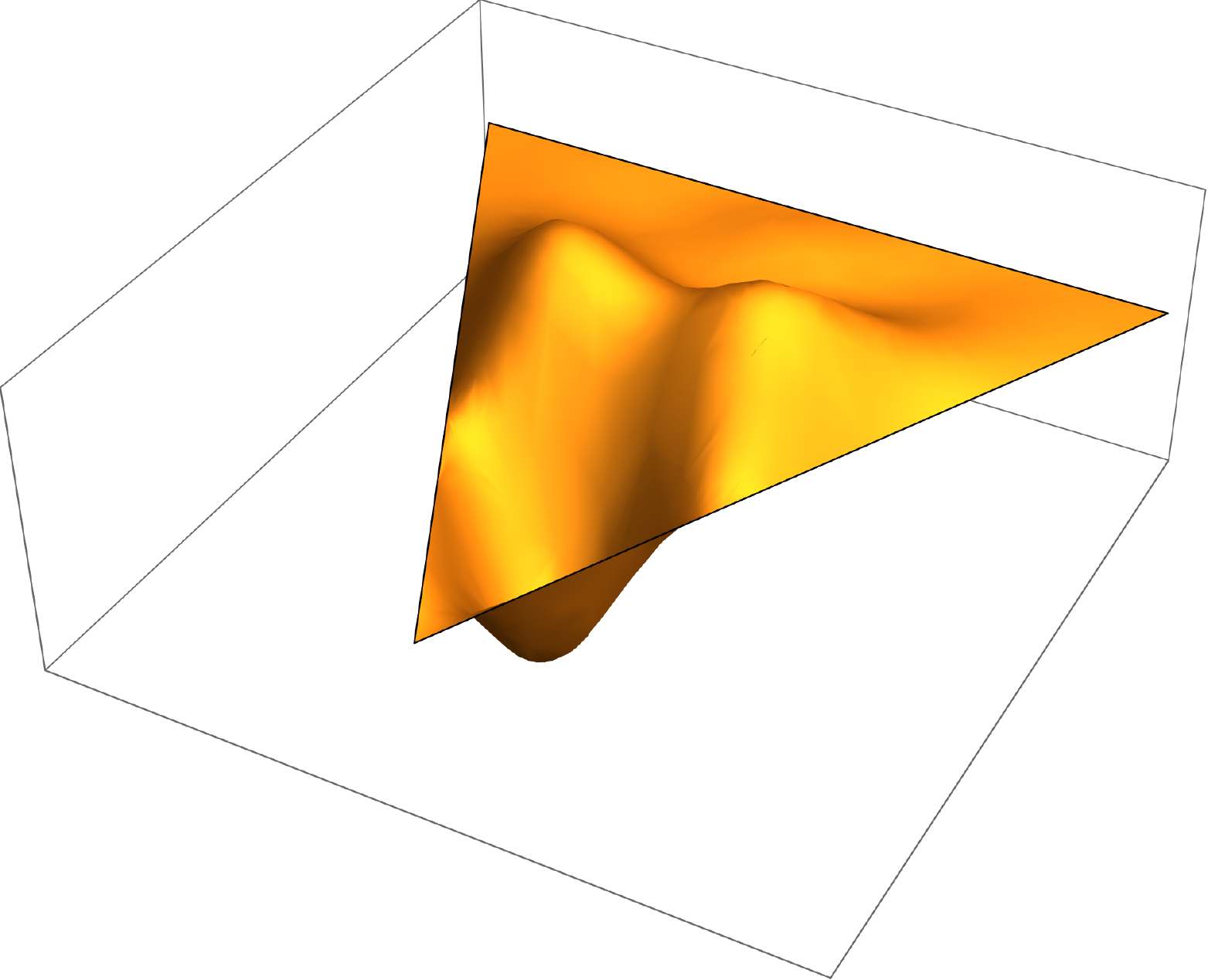}
\end{center}
\caption{Upper plot: the longitudinal wave functions 
$\Psi_\Sigma(\rho,\lambda)$ of $\Sigma_c$ (blue dashed) and  $\Psi_\Lambda(\rho,\lambda)$ of $\Lambda_c$ (black solid), 
as a function of $\lambda$. The four pairs of curves are for $\rho=0.0,0.1,0.2,0.3$, top to bottom.
 The lower 3D plot shows their $difference$,  $\Psi_\Lambda(\rho,\lambda)-\Psi_\Sigma(\rho,\lambda)$ .}
\label{fig_Sig_Lam}
\end{figure}

After this is achieved, we can compare the obtained $\Sigma_c$ and $\Lambda_c$ light front wave functions.
One way to do it is to give the coefficients of the decomposition in the Laplacian basis functions
as we did above for $\Sigma_c$ . Those are
\ba C_n^\Lambda=( 0.625, 0.025, 0.725, -0.249, -0.079, -0.089, \nonumber \\
-0.052, -0.004, 0.040, -0.020, -0.0272, 0.004)  \nonumber \ea
see Fig.\ref{fig_Sig_Lam} .  We can see from the upper plot, differences  at larger $\rho$ (lowest curves)
that are  as big as a factor of 2. However, a better representation of the shape $difference$
 is  given by the lowest plot

\subsection{Instanton-induced effects in heavy-light hadrons}

In the firts paper of this series~\cite{Shuryak:2021fsu}, we have addressed the instanton effects on Wilson lines 
(heavy quark potentials).  The novel point was the proposal of a   ``dense instanton liquid"  that also includes  instanton-antiinstanton molecules.
In the second paper of the series~\cite{Shuryak:2021hng} the instanton-induced t' Hooft
interaction was used for  light quarks, in the context of the pion LFWF.
 Since this interaction follows solely from the  near-zero fermionic modes, it is natural to limit its discussion
to the   $^\prime$dilute instanton liquid$^\prime$ as we did, with the hope of avoiding any confusion.

In this section we review some applications to heavy-light hadrons. The pioneering
study in the original instanton vacuum was  carried by  Chernyshev, Nowak and Zahed \cite{Chernyshev:1994zm},
on which this section is based. We will provide  some further discussion,
insuring connections to later papers and the remainder  of our series.

\noindent{\bf qQ interaction:}
The main point 
is that the instanton field strength (acting on a static quark $Q$), and its zero
mode (acting on a light quark $q$) are correlated. The appropriate setting is again  the
``dilute instanton liquid". 

If the instanton size is small, 
it can be written as a quasilocal  operator, to be included in a Lagrangian.
The interaction between a single light quark $q$ and a single static heavy quark $Q$  is
\bea
{\cal L}_{qQ} = &&-G_{Qq}
\left ( \overline{{\bf Q}} \frac{1+\gamma^0}2 {\bf Q}\,\,
\overline{{\bf q}} {\bf q} +
\frac{1}{4} \overline{{\bf Q}} \frac{1+\gamma^0}2 \lambda^a {\bf Q}\,\,
\overline{{\bf q}} \lambda^a {\bf q} \right )\nonumber\\
\label{qQ}
\eea
 The light quark effective vertex is based on the representation of
 the propagator as \be S_{ZM}(x,y)={\psi_0(x) \psi_0(y)\over m*}\ee with 
 some effective ``determinantal mass" characterizing the instanton ensemble.
 In the original ILM paper \cite{Shuryak:1982hk} this mass  was directly related 
 with the quark condensate $$m^*=\frac 23 \pi^2 \rho^2 |\langle \bar q q \rangle |\approx 170 \, {\rm MeV}$$ 
 (the number is for the empirical condensate value). Further development followed
 two directions: the gap equations in the mean field approximation (see references in 
 \cite{Pobylitsa:1989uq})  and numerical simulations of the instanton ensemble.
 The former expressions were used in~\cite{Chernyshev:1994zm} 
with  $$m^*=\sqrt{\frac n{2N_c}} \Sigma_0$$ and the  RILM 
 instanton density $n=1\,{\rm fm}^{-4}$. (Note that in~\cite{Chernyshev:1994zm} the factor of $1/2N_c$, 
 following from the averaging over the color moduli, was included in the
 definition of $n$). The explicit form of $\Sigma_0$ is quoted in Appendix~\ref{sec_G_tHooft}, with 
  $\Sigma_0\approx 240\, MeV$. The typical coupling in (\ref{qQ}) in the RILM is
 \be G_{Qq}=\bigg(\frac {\Delta M_Q\Delta M_q}{n} \bigg )
 \ee
with  the heavy quark mass shift
\bea
\label{DMQX}
\Delta M_Q=-\frac{4\pi^2}{3\rho}\,\frac{\pi^2\rho^4 n}{N_c}\,\bigg(J_0(\pi)+\frac 1\pi J_1(\pi)\bigg)\sim 70 {\rm MeV}\nonumber\\
\eea
and the light quark mass shift $\Delta M_q$ given in Appendix~\ref{sec_G_tHooft}. We recognize in (\ref{DMQX})
the packing fraction $\kappa=\frac 12 \pi^2 \rho^4 n$ in the RILM,  also used in our earlier papers.

Note that (\ref{qQ}) is dominated by the color-matrix (Coulomb-like) second term and
has a proper heavy quark spin symmetry.  The spin-dependent
correction 
 is subleading in $\Delta M_Q^{spin}\sim 1/m_Q$
\bea
{\cal L}_{qQ}^{spin} =  \frac{ \Delta M_q \; \Delta M_Q^{spin} }{n}
\; \frac 14 \;
\overline{\bf Q} \frac{1+\gamma^0}2 \lambda^a \sigma^{\mu\nu} {\bf Q}\,\,
\overline{\bf q} \lambda^a \sigma^{\mu\nu} {\bf q}\nonumber\\
\eea
For the charm quark
\bea
\Delta M_Q^{spin}=\frac{8\pi}{m_Q\rho^2}\frac{\rho^4 n}{N_c}
\int dx \,\frac{x^2{\rm sin}^2 f(x)}{(1+x^2)^2}\sim 3\, {\rm MeV}\nonumber\\
\eea
with the profile  $f(x)=\pi x/(1+x^2)^{\frac 12}$.

\noindent{\bf qq interaction:}
The instanton-induced 't Hooft Lagrangian has the form of a determinant in flavor indices
of certain $\bar q q$ operators, so the total numbers of quark legs is $2N_f$. Elsewhere
in our works we assumed that either there are only $u,d$ flavors, $N_f=2$ and the 
Lagrangian is of the usual four-fermion form, or that $N_f=3$ and the number of legs is 6,
with the  strange quarks contracted in the vacuum $\langle 
\bar s s \rangle $. 
However. we may  ask if there are situations in the  baryon sector  whereby the full
six-fermion operator may contribute. The operator, as any Lagrangian should be, is
flavor $SU(3)$ singlet: therefore if used as a ``baryonic current" operator aimed to $excite$  the vacuum into the $uds$ baryon, such baryon must be $SU(3)$ singlet as well. This condition
is not fulfilled for the usual $\Lambda,\Sigma$ baryons which are members of the $SU(3)$ octet.
There are known excited $\Lambda^*,\Sigma^*$  $SU(3)$ singlets, but those have nonzero
orbital momentum which the Lagrangian would not excite. Yet  we  think of averaging a 6-quark
't Hooft Lagranian over $\Lambda,\Sigma$ baryons, it does generate a   ``superlocal"
interaction (all three quarks at the same point, similar to Skyrme  force in nuclear physics).  
We do not pursue this issue quantitatively  and return to the 4-quark determinant. 

In the rest frame and using nonrelativistic spinors, we have
\bea
H^{\eta^\prime}_{qq}\approx -\bigg(\frac {\Delta M^2_q}{n} \bigg) \frac 12 (1-\tau_1\cdot\tau_2)\delta(\vec r_{12})
\eea
with $\vec r_{12}=\vec r_1-\vec r_2$, in leading order in $1/m_Q$. The sign corresponds to
the repulsive $\eta'$ channel, which is seen to flip in the pion channel by dropping 1. On the light front, the reduction of  the $\bar qq$ interaction in momentum space,  is detailed in Appendix~\ref{app_REDUCTION}.
For a meson  with a quark-antiquark pair and with  zero transverse momentum in and out 
($P_T=P_T^\prime=0$), the 2-particle interaction potential in momentum space is
\begin{widetext}
\bea
\label{DET5}
{\rm det}\overline {\bf q}_L {\bf q}_R\rightarrow (1-\tau_1\cdot \tau_2)
\bigg(m_{Q1}{\bf 1}_1+\frac 12 \sigma_1^-q_R\bigg)
\bigg(-m_{Q2}{\bf 1}_2+\frac 12 \sigma_2^+q_R\bigg)\nonumber\\
{\rm det}\overline {\bf q}_R {\bf q}_L\rightarrow (1-\tau_1\cdot \tau_2)
\bigg(m_{Q1}{\bf 1}_1+\frac 12 \sigma_1^+q_L\bigg)
\bigg(-m_{Q2}{\bf 1}_2+\frac 12 \sigma_2^-q_L\bigg)
\eea
with the transverse coordinates 
$q_{R,L}=q_1\pm iq_2$,  and $\sigma^\pm =\sigma_1\pm i\sigma_2$.
The contribution of (\ref{DET5}) to the light front Hamiltonian, is in the form of
a local 2-body interaction. In the singlet $U(1)$ or $\eta^{\prime}$ channel, it is of the form 
\bea
\label{HLFqq}
&&H^{\eta^\prime}_{LFqq}=-\bigg(\frac{\Delta M_q^2}{n}\bigg)  2(1-\tau_1\cdot \tau_2)\nonumber\\
&&\times\bigg[ \bigg(m_{Q1}{\bf 1}_1-\frac i2 \sigma_1^+\nabla_R\bigg)
\bigg(-m_{Q2}{\bf 1}_2-\frac i2 \sigma_2^-\nabla_R\bigg)+
\bigg(m_{Q1}{\bf 1}_1-\frac i2 \sigma_1^-\nabla_L\bigg)
\bigg(-m_{Q2}{\bf 1}_2-\frac i2 \sigma_2^+\nabla_L\bigg)\bigg]\delta(P^+x_{12}^-)\delta(b_{\perp})\nonumber\\
\eea
with $\nabla_{R,L}=\partial_1\pm i\partial_2$, and $P^+x^-\rightarrow id/dx$,
or equivalently
\bea
\label{HLFqqX}
&&H^{\eta^\prime}_{LFqq}=-\bigg(\frac{\Delta M_q^2}{n}\bigg) 4(1-\tau_1\cdot \tau_2)\nonumber\\
&&\times\bigg[ -m_{Q_1}m_{Q_2}\,{\bf 1}_1{\bf 1}_2
+\frac 12\big(\sigma_{1\perp}\cdot i\nabla_\perp\,m_{Q_2}{\bf 1}_2-m_{Q_1}{\bf 1}_1\,\sigma_{2\perp}\cdot i\nabla_\perp\big) 
-\frac 14\nabla^2_\perp \sigma_{1\perp}\cdot \sigma_{2\perp}
\bigg]
\delta(P^+x_{12}^-)\delta(b_{\perp})
\eea
\end{widetext}

After discussing the form of $qq$ effective Lagrangian, let us consider the magnitude
of its coupling. The mean field approximation assumes that the
instanton vacuum is  homogeneous. Hence, 
 in any expression the effective determinantal mass $m^*$ 
is treated as the same constant, as defined from the solution of the gap equation.
However, numerical studies of  instanton ensembles, show  significant deviations
from a  homogeneous vacuum. The parameter   $m^*$ is substituted by
a ``hopping matrix" $T_{IJ}$, with two-fermion and four-fermion operators proportional to
\bea {1 \over m^*_{qq}}=\bigg<{1 \over  T_{IJ} }\bigg>, \,\, 
 {1 \over (m^*_{\bar u u \bar d d})^2}=\bigg< {1 \over  T_{IJ}^2 }\bigg> \ea
The averages in the r.h.s. are subsumed over instanton and antiinstanton ensembles. Studies of those
averaging in~\cite{Faccioli:2001ug}  show that these  two definitions lead to 
different values of  the effective mass: while the former one is about $170\, MeV$ as
given by the mean fields, the second is much smaller, $m^*_{\bar u u \bar d d}\approx 90\, MeV$. This increases the
effective 4-quark coupling by about a factor 3.5.
Fits to the empirical pion
correlation function also agrees with this enhancement.

On top of light-light forces used in this work, there are other quasi-local forces induced by instantons, acting between heavy and light quarks.  For future
references, let us mention those.

{\bf $\bar {\rm {\bf Q}}$Q interaction:}
 As derived in \cite{Chernyshev:1994zm},
 to
order $1/m_Q$ and in the planar approximation this effective interaction
among the heavy quarks is
\begin{widetext}
\bea
{\cal L}_{QQ} = - \bigg (
\frac{ \Delta M_Q \Delta M_Q }{n} \bigg )
\left ( \overline{{\bf Q}} \frac{1+\gamma^0}2 {\bf Q}\,\,
\overline{{\bf Q}} \frac{1+\gamma^0}2 {\bf Q}   
+
\frac{1}{4} \overline{{\bf Q}} \frac{1+\gamma^0}2 \lambda^a {\bf Q}\,\,
\overline{{\bf Q}} \frac{1+\gamma^0}2 \lambda^a {\bf Q}
\right)\label{27}
\eea
\end{widetext}
The recoil effects are of first order in $1/m_Q$ and renormalize
$\Delta M_Q$. The spin effects are of second order in $1/m_Q$, 
and small

\bea
&&\Delta {\cal L}_{QQ}^{spin} = \bigg (
\frac{ \Delta M_Q^{spin} \; \Delta M_Q^{spin} }{n} \bigg ) \;\nonumber\\
&&\times\frac{1}{4} \overline{{\bf Q}} \frac{1+\gamma^0}2 \lambda^a
\sigma_1^{\mu\nu} {\bf Q}\,\,
\overline{{\bf Q}} \frac{1+\gamma^0}2 \lambda^a \sigma_2^{\mu\nu} {\bf Q}
\label{28}
\eea

Note however, that since in this case there are no light quarks, and following  our
first paper \cite{Shuryak:2021fsu}, we do not   need well-isolated zero modes. Hence, 
we should include contributions of instanton-antiinstanton molecules. If so,
the original estimate of this interaction in \cite{Chernyshev:1994zm}
should be increased by a factor $n_{dense\,ILM}/n_{dilute\, ILM}\sim 7$.

{\bf Qqq interaction:}
For heavy baryons the induced $qqQ$ interaction in leading order is in mean field
\cite{Chernyshev:1994zm}
\begin{widetext}
\bea
{\cal L}_{qqQ} = -2\bigg ( \frac{\Delta M_Q \Delta M_q^2}{n^2} \bigg )
\bigg ( &&\overline{\bf Q} \frac {1+\gamma^0}2 {\bf Q} \,\,
\left({\rm det} \overline {\bf q}_L {\bf q}_R \,\,+{\rm det}\overline
{\bf q}_R {\bf q}_L \,\,\right) +\nonumber\\ &&
       \frac 14 \,\,\overline{\bf Q} \frac {1+\gamma^0}2 \lambda^a {\bf Q} \,\,
\left({\rm det} \overline {\bf q}_L\lambda^a {\bf q}_R \,\,+{\rm det}\overline
{\bf q}_R\lambda^a {\bf q}_L \,\,\right) \bigg )
\label{29}
\eea
with the short-hand notation  for 2-flavors
\bea
{\rm det} \overline {\bf q}_L\lambda^a {\bf q}_R+{\rm det} \overline {\bf q}_R\lambda^a {\bf q}_L=
\frac 14
\big((\bar q \lambda^aq)\bar q q+(\bar qi\gamma^5\lambda^a \tau^Aq)(\bar qi\gamma^5\tau^A q)
-(\bar q\lambda^a\tau^A q)( \bar q\tau^A q)-(\bar q i\gamma^5\lambda^a q)( \bar qi\gamma^5 \lambda^a q)\big)\nonumber\\
\eea
\end{widetext}
The  $1/m_Q$ spin correction is

\bea
{\cal L}_{qqQ}^{spin} =&& - \bigg ( \frac{ \Delta M_Q^{spin} \; \Delta
M_q^2}{n^2 } \bigg )
 \,\,\overline{\bf Q} \frac {1+\gamma^0}2 \lambda^a
\sigma_{\mu\nu} {\bf Q} \;\nonumber\\
&&\times\bigg ( \; {\rm det} \overline {\bf q}_L\lambda^a
\sigma_{\mu\nu} {\bf q}_R \,\,+{\rm det}\overline
{\bf q}_R\lambda^a \sigma_{\mu\nu} {\bf q}_L \; \bigg )\nonumber\\ \label{30}
\eea

Since there are two light quark propagators, we should  use the corrected
$m^*_{\bar u u \bar d d}$ instead of the mean field $m^*$ value. Again,
this increases the effective coupling by about a factor of
 $(m^*_{\bar u u \bar d d}/m^*_{\bar q q})^2 \sim 3.5$.
  
{\bf QQqq interactions:}
The same enhancement due to deviations from the mean field, 
should also be present in this case. This carries to the 
 exotics, such as the $cc\bar u \bar d$ tetraquark recently discovered 
  at LHCb. For the tetraquarks,  the induced interaction is

\begin{widetext}
\bea
{\cal L}_{qqQQ} = &&-2
n \bigg ( \frac{\Delta M_q}{n} \bigg )^2
\bigg ( \frac{\Delta M_Q} {n} \bigg )^2 \nonumber \\
&&\times\bigg (  \overline{\bf Q} \frac {1+\gamma^0}2 {\bf Q} \,\,
\overline{\bf Q} \frac {1+\gamma^0}2 {\bf Q} \,\,
\left({\rm det} \overline {\bf q}_L {\bf q}_R \,\,+{\rm det}\overline
{\bf q}_R {\bf q}_L \,\,\right) \nonumber\\
&& +\frac 14 \,\,\overline{\bf Q} \frac {1+\gamma^0}2 \lambda^a {\bf Q} \,\,
\overline{\bf Q} \frac {1+\gamma^0}2 {\bf Q} \,\,
\left({\rm det} \overline {\bf q}_L\lambda^a {\bf q}_R \,\,+{\rm det}\overline
{\bf q}_R\lambda^a {\bf q}_L \,\,\right)  \nonumber \\
&& +\frac 14 \,\,\overline{\bf Q} \frac {1+\gamma^0}2 \lambda^a {\bf Q} \,\,
\,\,\overline{\bf Q} \frac {1+\gamma^0}2 \lambda^a {\bf Q} \,\,
\left({\rm det} \overline {\bf q}_L {\bf q}_R \,\,+{\rm det}\overline
{\bf q}_R {\bf q}_L \right)\,\,\nonumber\\
&&+\frac 18 d^{abc} \,\,\overline{\bf Q} \frac {1+\gamma^0}2 \lambda^b {\bf Q} \,\,
\,\,\overline{\bf Q} \frac {1+\gamma^0}2 \lambda^c {\bf Q} \,\,
\left({\rm det} \overline {\bf q}_L\lambda^a {\bf q}_R \,\,+{\rm det}\overline
{\bf q}_R\lambda^a {\bf q}_L \,\,
\right)\bigg)
\eea
\end{widetext}
The overall sign is consistent with the naive expectation, that the $n$-body
interaction follows from the $(n+1)$-body interaction by contracting a light
quark line, resulting in an overall minus sign (quark condensate).

{\bf QQqqq interactions:}
for pentaquarks the quasi-local Lagrangian reads
\begin{widetext}
\bea
{\cal L}_{qqqQQ} = &&+4n \bigg ( \frac{\Delta M_q}{n} \bigg )^3
\bigg ( \frac{\Delta M_Q} {n} \bigg )^2 \nonumber \\
&&\times\bigg (  \overline{\bf Q} \frac {1+\gamma^0}2 {\bf Q} \,\,
\overline{\bf Q} \frac {1+\gamma^0}2 {\bf Q} \,\,
\left({\rm det} \overline {\bf q}_L {\bf q}_R \,\,+{\rm det}\overline
{\bf q}_R {\bf q}_L \,\,\right) \nonumber\\
&& +\frac 14 \,\,\overline{\bf Q} \frac {1+\gamma^0}2 \lambda^a {\bf Q} \,\,
\overline{\bf Q} \frac {1+\gamma^0}2 {\bf Q} \,\,
\left({\rm det} \overline {\bf q}_L\lambda^a {\bf q}_R \,\,+{\rm det}\overline
{\bf q}_R\lambda^a {\bf q}_L \,\,\right)  \nonumber \\
&& +\frac 14 \,\,\overline{\bf Q} \frac {1+\gamma^0}2 \lambda^a {\bf Q} \,\,
\,\,\overline{\bf Q} \frac {1+\gamma^0}2 \lambda^a {\bf Q} \,\,
\left({\rm det} \overline {\bf q}_L {\bf q}_R \,\,+{\rm det}\overline
{\bf q}_R {\bf q}_L \right)\,\,\nonumber\\
&&+\frac 18 d^{abc} \,\,\overline{\bf Q} \frac {1+\gamma^0}2 \lambda^b {\bf Q} \,\,
\,\,\overline{\bf Q} \frac {1+\gamma^0}2 \lambda^c {\bf Q} \,\,
\left({\rm det} \overline {\bf q}_L\lambda^a {\bf q}_R \,\,+{\rm det}\overline
{\bf q}_R\lambda^a {\bf q}_L \,\,
\right)\bigg)
\eea
\end{widetext}
with the short-hand notation  for three flavors
\bea
{\rm det} \overline {\bf q}_L{\bf q}_R=&&
\begin{pmatrix}
\overline {\bf u}_L{\bf u}_R& \overline {\bf u}_L{\bf d}_R &\overline {\bf u}_L{\bf s}_R \\
\overline {\bf d}_L {\bf u}_R& \overline {\bf d}_L {\bf d}_R & \overline {\bf d}_L {\bf s}_R \\
\overline {\bf s}_L {\bf u}_R& \overline {\bf s}_L {\bf d}_R & \overline {\bf s}_L {\bf s}_R 
\end{pmatrix}
\eea
and
\bea
{\rm det} \overline {\bf q}_L\lambda^a {\bf q}_R=&&
\begin{pmatrix}
\overline {\bf u}_L\lambda^a {\bf u}_R& \overline {\bf u}_L\lambda^a {\bf d}_R &\overline {\bf u}_L\lambda^a {\bf s}_R \\
\overline {\bf d}_L {\bf u}_R& \overline {\bf d}_L {\bf d}_R & \overline {\bf d}_L {\bf s}_R \\
\overline {\bf s}_L {\bf u}_R& \overline {\bf s}_L {\bf d}_R & \overline {\bf s}_L {\bf s}_R 
\end{pmatrix}+u\leftrightarrow d\leftrightarrow  s\nonumber\\
\eea

The flavor composition of the light quarks should of course be $uds$, but the net color
 inside the pentaquark need  not be a singlet. There 
should also be a significant enhancement over mean field estimates, but
$\langle 1/T_{IJ}^3\rangle$  is in so far not evaluated.

\section{ Diquark  pairing in the nucleons} \label{sec_N}
The role of the  instanton-induced quasi-local interaction (diquark pairing) in the wave functions of  $\Delta$ and $N$ 
was already discussed in the paper by one of us \cite{Shuryak:2019zhv}. However several principal and technical
tools were different. In particular, the light front hamiltonian
 $H_{LF}$ was different (constructed a la mesonic Hamiltonian of Vary et al \cite{Jia:2018ary}),
and the set of basis functions was completely different. 

As in the preceding section, we start with baryons without quasi-local instanton-induced 't Hooft
interaction, namely $\Delta^{++}(3/2)=uuu$, and proceed similarly by 
expressing the potential $V$ as a matrix in the Laplacian basis,  and diagonalize $H_{LF}=H_0+V$.
If we use the basis of 12 such functions, the spectrum (of squared masses) is

\begin{figure}[t]
\begin{center}
\includegraphics[width=6cm]{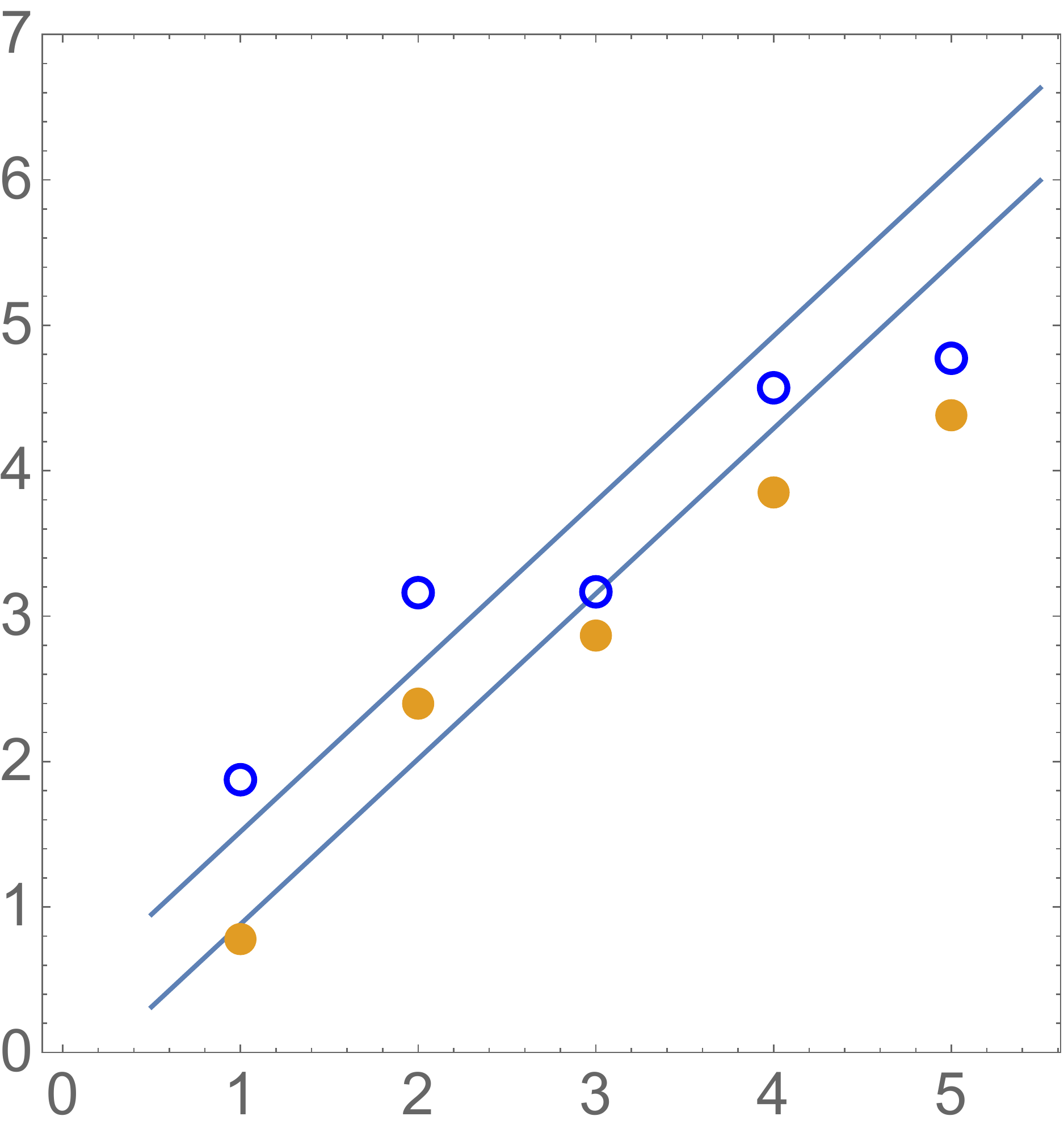}
\includegraphics[width=6cm]{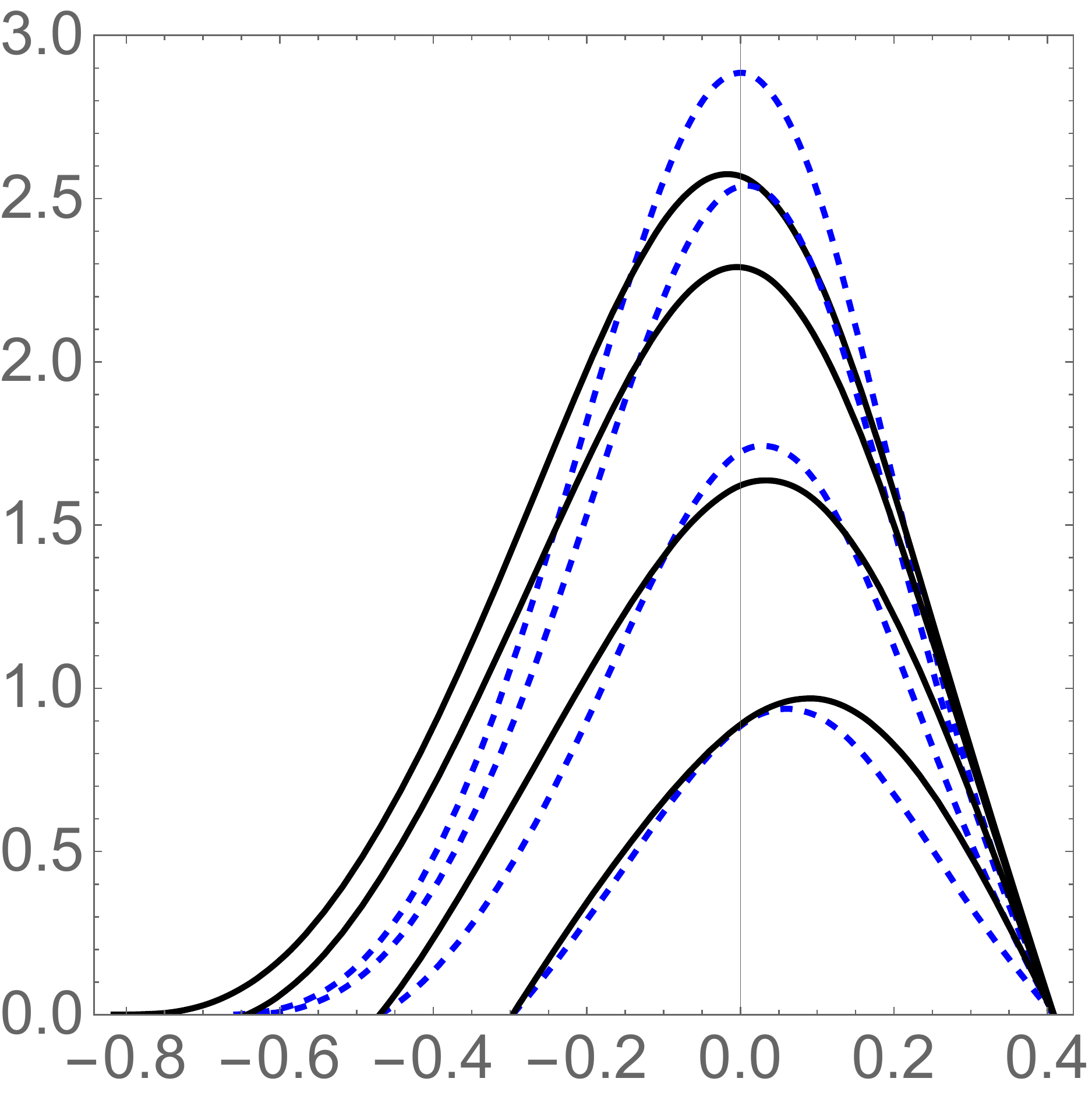}
\caption{Upper: Squared masses of the Delta (open points) and N (closed) resonances versus their successive quantum number $n$. The two straight lines shown for comparison, are the Regge trajectories fitted to the experimental values of $M^2(J)$ ,
versus the total angular momentum $J$, with the  slope $\alpha'=0.88\, GeV^2$.\\
Lower: LFWFs for the lowest Delta (dashed lines) and N (solid lines). The plots are shown
versus the Jacobi coordinate $\lambda$, for  fixed $\rho=0,0.1,0.2,0.3 $, top to bottom.   }
\label{fig_Delta_N}
\end{center}
\end{figure}

Pairing in the proton $p=uud$ takes place in two $(ud)$ channels, which we denote as $(13)$ and $(23)$. 
For that, it is more convenient to use alternative  Jacobi coordinates $\rho_\pm,\lambda_\pm$,  
rotated from the original  $\rho,\lambda$ by the ``triple symmetry" matrices  of the equilateral triangle
\ba M_\pm = \begin{bmatrix} 
	cos(2\pi/3) & \pm sin(2\pi/3) \\
	\mp  sin(2\pi/3)  & cos(2\pi/3) \\
		\end{bmatrix}
\ea	
The hamiltonian now has two pairing terms, and each can be written as a matrix in our basis 
in appropriate coordinates using the same form (\ref{eqn_rho_lambda}), adding those to the light front
hamiltonian $H_{LF}$ and diagonalizing
it, we obtain the spectrum and the wave functions. For one choice of the  't Hooft coupling,
the results for the squared masses of the lowest $\Delta,N$ baryons are shown in Fig.\ref{fig_Delta_N}.

Recall  that these masses are calculated from limited basis set,
with only 12 longitudinal eigenfunctions of the Laplacian. Also note, that 
neither the perturbative Coulomb nor the  spin-dependent  interactions are included. The
Delta-N splitting is due only to the  't Hooft operator treated in a quasi-local approximation.

The lower part of the plot shows the light front wave functions of the lowest mass $\Delta$ and $N$ baryons.
The attractive and quasi-local interaction makes the wave function of the nucleon $N$  wider than that of  the isobar $\Delta$, 
i.e. greater both at the left and right side of the plot (corresponding to $x_d\rightarrow 1$ and $x_d\rightarrow 0$.).
This widening effect is similar to that observed for  mesons. The LFWF of vector mesons is relatively narrow,  while that of the pion is 
nearly flat. Furthermore, as $d$ participates in two pairings while each $u$ only in one, this effect is more
pronounced for the $d$ quark.

\section{Baryon formfactors} \label{sec_ff}
In the non-relativistic formulation, the formfactors are defined as overlap integrals.
This carries to the light front, with the formfactors as overlap of the  LFWFs. 
In particular the helicity preserving Dirac formfactor is~\cite{Drell:1969km,West:1970av,Lepage:1980fj}
\be 
\label{LFFF}
F_1(q^2)=\langle \vec P+\vec q, \uparrow | J^+/2P^+| \vec P, \uparrow \rangle 
\ee

To evaluate (\ref{LFFF}), we  select the momentum transfer to be $q$ in the transverse x-direction,
and select the struck quark to be number 3 ($d$ quark). More specifically, the transverse momenta in the struck 
baryon   (with prime) are related to those in the non-struck one (without prime) by
\ba k^{1^\prime}_x&=&k^1_x-x_1q,\,\,k^{2'}_x=k^2_x-x_2 q,\nonumber \\
k^{3'}_x&=&k^3_x+(1-x_3)q 
\ea
The three transverse momenta $\vec k^i$ and longitudinal fractions $x_i$ need to be 
re-expressed in terms of two Jacobi momenta, in our notations $\vec p_\lambda,\vec p_\rho$ and
$\rho,\lambda$, on which LFWFs depend. 

\begin{figure}[t]
\begin{center}
\includegraphics[width=6cm]{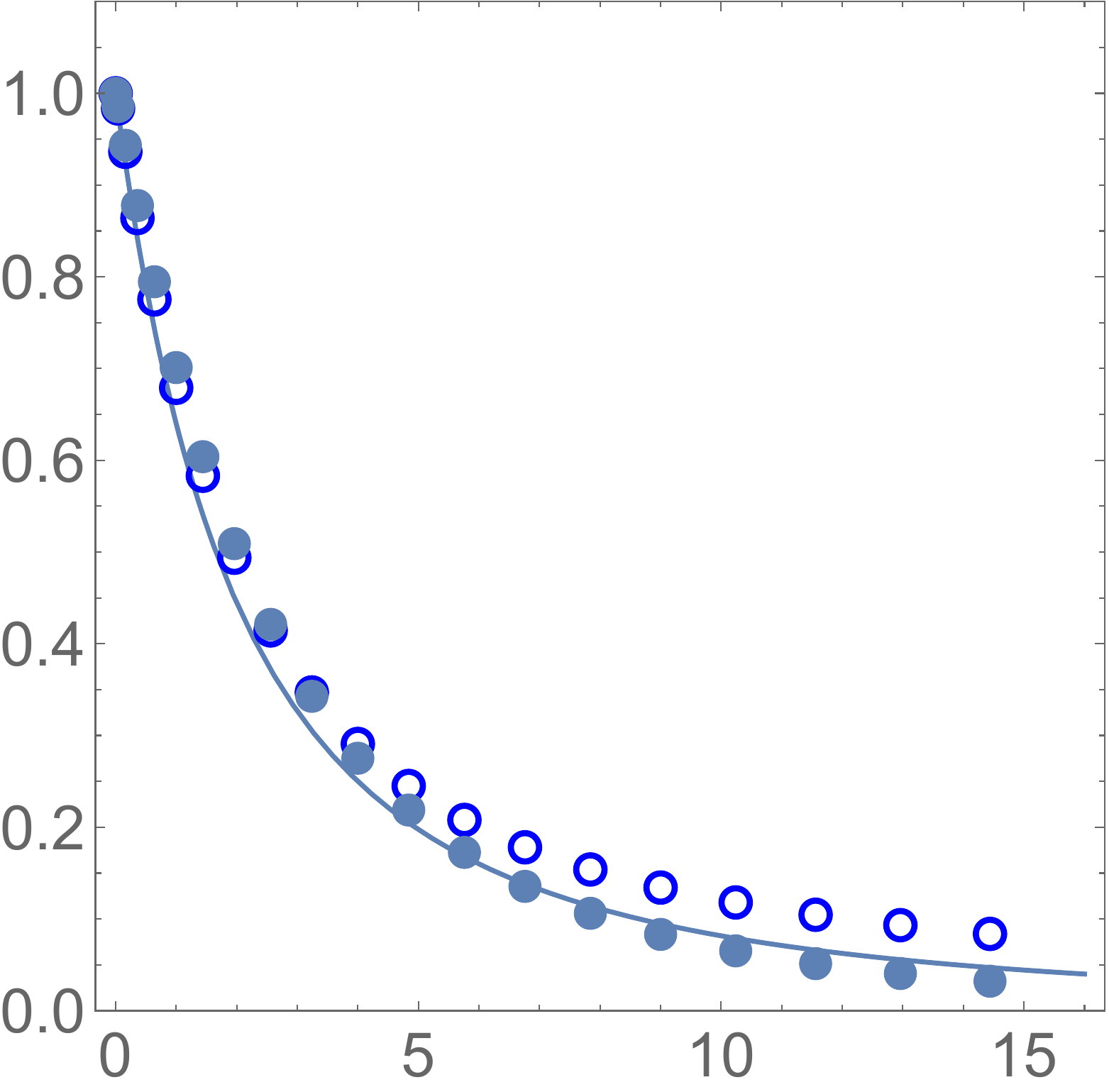}
\caption{The formfactors for `Neumann" (open points) and ``Dirichlet" (closed points)  wave functions defined
in the text, versus the momentum transfer $q^2$.A line, shown for comparison,
corresponds to the dependence $1/(1+ q^2/4)^2$.}
\label{fig_ff_example}
\end{center}
\end{figure}

For  illustration, let us take  the example of a LFWF with a Gaussian transverse momentum
dependence. For the struck LFWF it takes the form
\begin{widetext}
\be exp\big(-A\sum_1^3 (\vec k^{i'})^2\big) \rightarrow 
exp \big(-A \big[\vec{p_\lambda}^2 + \vec{p_\rho}^2
- {2\over 3} (\vec{p_\lambda}\cdot \vec q) (\sqrt{6} + 3 \lambda) - 
 2 (\vec{p_\rho} \cdot \vec q)  \rho + 
  \vec q^2 (2 + 2 \sqrt{6} \lambda + 3 \lambda^2 + 3 \rho^2)/3 \big]\big)   \label{eqn_factor_for_ff}     \ee
 \end{widetext}
Selecting the momentum units such that $A=1$, and convoluting it with various longitudinal wave
functions (defined  as always on the equilateral triangle),  we can see how the formfactor depends on their shape. In Fig.\ref{fig_ff_example} we present results of LFWF convolution, for two extreme cases: ``Neumann" wave function flat (constant) on the physical triangle,
and  ``Dirichlet"  wave function  $\psi\sim x_1 x_2 x_3$ satisfying linear  boundary conditions
on all sides of the equilateral triangle. As  expected, the  Neumann wave function   with sharper edges, produce a larger formfactor
at large $q^2$, although the overall difference is not that large. Note that
this methodical example (not expected to be realistic)  is well reproduced 
by the dipole form  $1/(1+C^2 q^2)^2$,  by which the nucleon formfactors were
originally fitted decades ago.

\begin{figure}[t]
\begin{center}
\includegraphics[width=6cm]{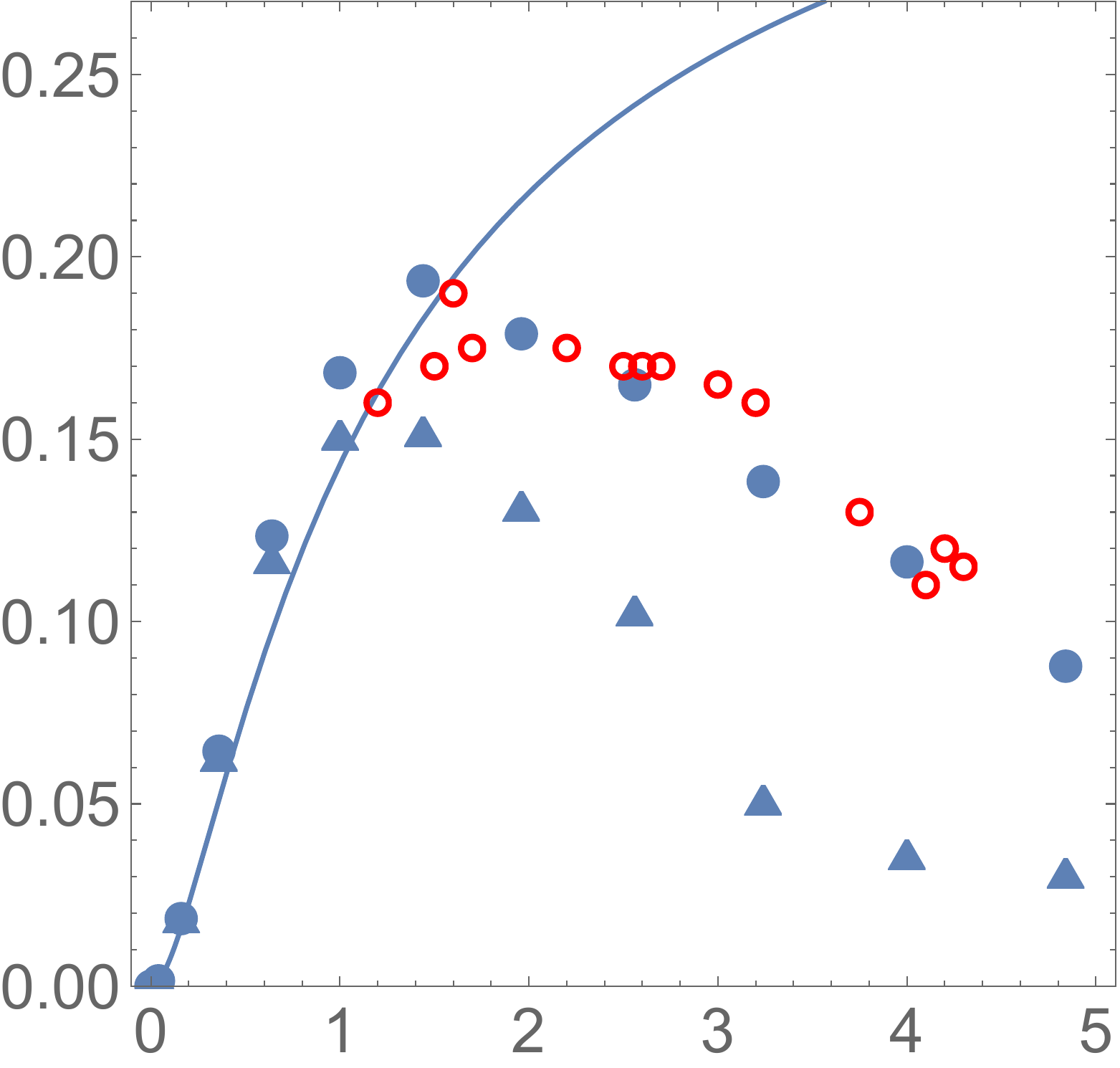}
\caption{$Q^4 F_1^d(Q^2),\, (GeV^4)$ versus the
momentum transfer $Q^2\, (GeV^2)$. The triangles and closed points correspond
to  the Delta and Proton LFWFs, respectively. The red circles are extraction from
the experimental  data on the $p$ and $n$ formfactors mentioned in the text.
The solid line shown for comparison, 
corresponds to the dipole form factor $Q^4/(1+ Q^2/m_\rho^2)^2$.}
\label{fig_Q4_F1d}
\end{center}
\end{figure}

We now show the formfactors calculated with the longitudinal wave functions
for the Delta and Proton, following from our analysis in the preceding section.  
In order to see  better the most interesting region of large $q^2$,  we plot
$Q^4 F_1^d(Q^2)$ in Fig.\ref{fig_Q4_F1d}.

Experiments are of course done with protons and neutrons, but using them one
can extract separate formfactors for $u$ or $d$ quarks. This was done e.g. in
\cite{1209.0683}, and the red circles in Fig.\ref{fig_Q4_F1d} are from Fig.8 of this work.
(For clarity we do not show the datapoints  in the range  $Q^2<1\, GeV^2$, as the error bars for these points
are $\pm 0.02$ on average.)  From the plot, we see that  this formfactor 
does not appear to reach a constant limit at large $Q^2$,  with the measured points slowly decreasing
towards the right hand side.   Old  dipole parametrization 
$Q^4F=Q^4/(1+ Q^2/m_\rho^2)^2$ asymptotes  a constant at large $Q^2$ from below.

Remarkably, our longitudinal proton wave functions convoluted with (\ref{eqn_factor_for_ff}) 
reproduces such a trend, and  (with the parameter $A=4\, GeV^{-2}$  ) they follow the shape indicated by the data rather well. The calculated formfactor for the case when the struck quark is $u$ has a similar shape. Unfortunately, according to \cite{1209.0683}, the experimental trend is different, the constant at $Q^2\rightarrow \infty$ is approached from below. By the Drell-Yan relation this flavor difference is also seen in the PDFs of $u$ and
$d$ at $x\rightarrow 1$. Flavor asymmetry must be related with the asymmetry of the
spin-orbit part of the wave function, which in our approximations is so far ignored.

Note that the corresponding formfactor for Delta (triangles) is significantly softer, as one would expect from the size of the wave function.
Recall that the large difference between the Delta and Proton formfactors (so well seen
in this plot) is completely due to the
't Hooft quasi-local pairing $ud$ interaction. While we do not have Delta targets
for experiments, perhaps its formfactor can be calculated on the lattice, or in other models.

Let us now make a more general comment on possible improvements of the LFWFs
calculated in this work, and in particular their consequences for formfactors at large $Q^2\rightarrow \infty$ and PDFs at large $x \rightarrow 1$. We treated light quarks as 
 ``constituent quarks" with fixed mass $M\sim 400 \, MeV $ everywhere, including the 
 ``cup potential" $\sim M^2/x$ diverging at kinematical edges. As a result, our
  LFWFs vanish at these edges in a smooth way. However it is known that $M^2$
  decreases with virtuality of the quark and vanishes if it is highly virtual. The instanton-based theory of chiral symmetry breaking shows how it is related to the instanton zero modes
  and describe smooth transition from on-shell constituent quarks to near-massless
  quark-partons. We are planning to include this effect in subsequent works.

Completing the section on formfactors and trying to avoid any confusion, let us comment on the
relation between our results and those in the literature on ``hard regime" $ Q^2\rightarrow \infty$ 
limit. These terminology is used in literature in very different settings. One is physics of heavy
boson or quark production $W,Z,H,t$ or jet observables at colliders: here $Q^2\sim (100\, GeV)^2$ and pQCD 
is fully accountable for those.

A completely different situation is with $exclusive$ processes, such as elastic scattering and formfactors.
Specific powers of $1/Q^2$ and powers of $\alpha_s(Q^2)$ follow from the lowest orders pQCD diagrams~\cite{Brodsky:1973kr}.
Furthermore, in some cases (e.g. the pion formfactor) even the constant in the hard limit can be expressed in
terms of $f_\pi$, so the pQCD asymptotic prediction is fully known. It is further  known for decades, that 
in the ``semi-hard" domain of current experiments $ Q^2 < 10\, GeV^2$ the formfactors are $not$ dominated
by pQCD mechanisms.
 In our paper on formfactors \cite{Shuryak:2020ktq} we included the instanton contributions in the hard blocks.
 While important in  the ``semi-hard" domain, 
 at large  $ Q^2$ those become exponentially small $\sim exp(-Q\rho)$ at $Q\rho \gg 1$.
In this series of papers, we used the  't Hooft Lagrangian as a $quasi-local$ operator. This  means  the
opposite regime, where the 
distance scales considered are $large$ compared to the instanton size $\langle r \rangle \gg \rho$. 

\section{Bridging the gap between hadronic and partonic dynamics} \label{sec_chiral_sea}
\subsection{The matching scale} \label{sec_matching_point}
Before building a bridge, one should have a good assessment of both sides of the river. Therefore let us start
with a brief summary of what we know on the two extremes. 

Since the 1970's we know that
{\em hard processes} defined at some high scale $Q^2 \gg 1\, GeV^2$  can be described as a set of independent
``partons", $g,q,\bar q$.  The probabilities to find those in a target (or beam) hadrons are known as 
 PDFs $q(x,Q^2)$. Due to  high resolving power in this regime,
the  pointlike quark-partons and gluons can emit each other with ``splitting functions" 
 following directly from the QCD Lagrangian. So PDFs and structure functions at different Q are related by 
 perturbative   DGLAP ``evolution".  Combining these  with fits to experimental data on various partonic processes,
has mushroomed into a large body of work. For a recent summary see e.g. reviews by CTEQ collaboration such as \cite{1912.10053}. 

(While the partonic PDFs definitely represent a very solid end of the bridge, they are not constructed without certain approximations. 
Subsequent gluon emissions are assumed to be incoherent, i.e. the  DGLAP equations are
 {\em probabilistic kinetic equations}. As noted in \cite{Kharzeev:2017qzs}, this assumption
 generates an entanglement entropy.  However, hadrons are pure states,
  and their consistent treatment should be based on 
 their complete LFWFs, without entanglement.)

  By evolving DGLAP to sufficiently low normalization point,  one finds a scale at which gluons can no longer be
emitted. The lowest glueball masses  have a mass scale $M_{glueballs}\sim 2 \, GeV$, and so a crude estimate for ithis limiting scale
is  the gluon effective mass   $M_{eff}(gluon)\sim 1\, GeV$. We expect the lower end of the DGLAP evolution to be located at
$Q^2\sim M_{eff}(gluon)^2$.

On the other side (lower $Q^2$), 
chiral symmetry breaking puts special emphasis on the lightest mesons, the Nambu-Goldstone modes -- the pions --
and the condensate $\sigma$. Instead of the QCD Lagrangian, one has a chiral effective Lagrangian and its
higher order descendants.  Its $upper$ cutoff scale can be identified with the original cutoff of the Nambu-Jona-Lasinio (NJL) model 
$\Lambda_\chi\sim 1\, GeV$. Later its mechanism was related to $instantons$ \cite{Shuryak:1982hk}, and this cutoff with the typical instanton size $\rho\sim 1/3\, fm $. 

So, our preliminary assessment suggests a nice plan for a bridge, with a hope that its ``arcs"  -- the chiral and 
the pQCD ones --
will join relatively smoothly  at the $1\, GeV$ scale. In this section we are going to investigate 
if the PDFs evaluated from both sides do indeed join there. (Needless to say, there are many other observables
for which one may need more sophisticated strategy than a jump from one theory to another. Here, we would like
to emphasize the  efforts by many,  to include both
perturbative and nonperturbative effects, among which our own discussions of meson formfactors in \cite{Shuryak:2020ktq}, and 
spin-dependent forces in  \cite{Shuryak:2021fsu} .)

\begin{figure}[t]
\begin{center}
\includegraphics[width=8cm]{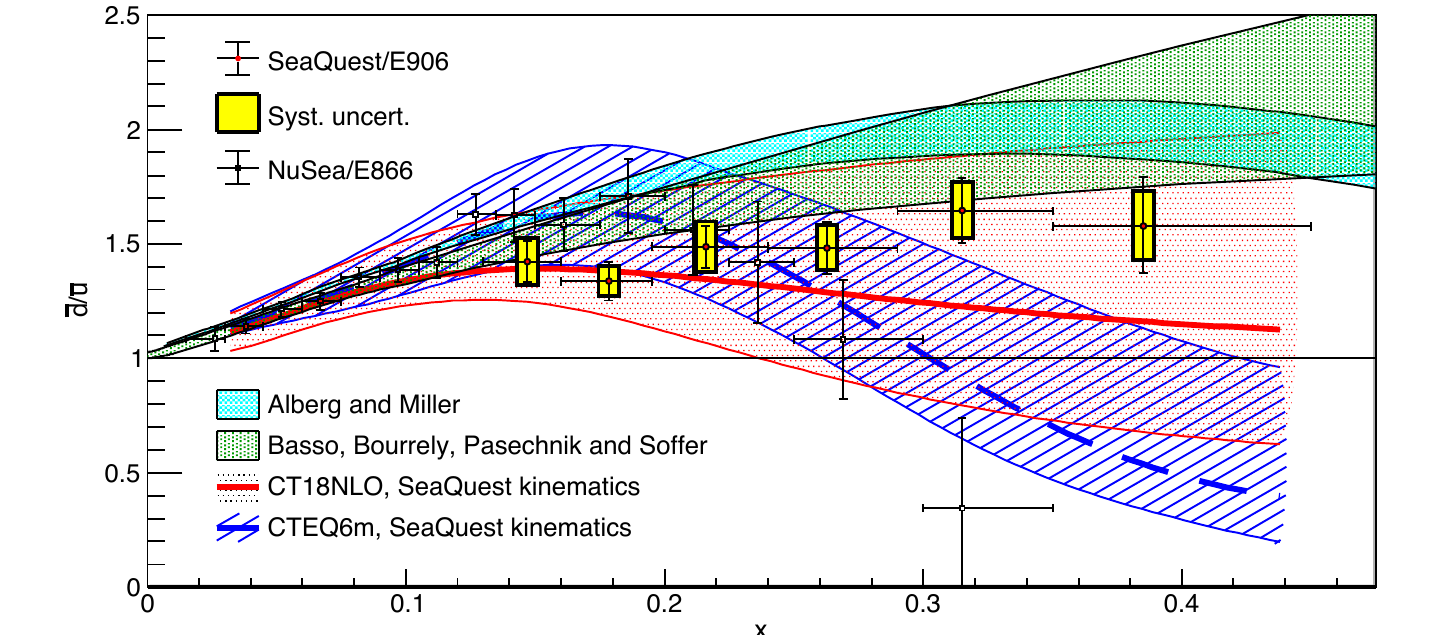}
\caption{Flavor asymmetry of the antiquark sea as the ratio of PDFs of $\bar d/\bar u$ versus $x$~\cite{2103.04024}.}
\label{fig_flavor_asymmetry}
\end{center}
\end{figure}

 
In  Fig.\ref{fig_sea} we illustrate 
the processes affecting the PDFs . Perturbatively,
an extra $\bar q q$ pair can be mediated by virtual gluons, see (a,b). Since
 gluons are ``flavor blind",  they produce $\bar u u, \bar d d$ in equal number.
 Another process shown in Fig.\ref{fig_sea} (c,d) uses the instanton-induced 't Hooft four-fermi interaction in the production channels. As first noted  in \cite{Dorokhov:1993fc}, it is ``maximally flavor antisymmetric" due to the Pauli principle for instanton zero modes of fermions. Also, since there are
 two $u$ and one $d$  valence quarks, in the first order in this mechanism the simple prediction for antiquark flavors would be
\be {\bar d \over \bar u} \rightarrow 2 \ee
And for $\Delta^{++}$ with valence $uuu$ only, the sea should  consists of only $\bar d$, without $\bar u$!

The flavor asymmetry of the sea is a very convenient tool  to discriminate what part of the ``sea" (multiparticle/multiparton sectors
of the LFWFs)  come from chiral (step 2) processes and which from pQCD (step 3), because it can only come
from the former.

\subsection{Phenomenology of the nucleon antiquark sea}  \label{sec_flavor_assymetry}

  Let us now briefly describe what is known about the flavor asymmetry of the antiquark sea. First discovered as
    ``violation of Gottfried sum rule" (an assumption that the sea is produced entirely by gluons) three decades ago,
    it is still being developed. For reference, at $4 \, GeV^2$ scale, the NMC collaboration found
    \be \int_0^1 dx \big(\bar d(x)-\bar u(x)\big)= 0.147\pm 0.039  \label{anti_difference}\ee
Indeed, experiments have  shown a surprisingly strong  violation. The  NuSea experiment \cite{hep-ex/0103030}  gives for this 
 ratio $\approx 1.5$ at  $x\sim 0.2 $. The more recent  SeaQuest experiment \cite{2103.04024},
 has found that the asymmetry persists to larger momentum fractions, at least up to $x\sim 0.4 $, see Fig.\ref{fig_flavor_asymmetry} (copied from \cite{2103.04024}).  Also shown in this figure, are
 the CT parameterizations of  the previous data, the so called MMHT14 PDF polynomial parameterization \cite{Basso:2015lua}, and some
theoretical predictions. Ref~\cite{Alberg:2017ijg} (green strip)  uses a model based on the ``pion cloud" of the nucleon
( see also~\cite{Alberg:2021nmu}). 

In Fig. \ref{fig_d_minus_u} we show the difference $ \bar d(x)-\bar u(x)$ as a function of $x$ from NuSea/866 experiment.
It shows that the effect is localized at $x<0.2$,  but strongly
$grows$ towards small $x$.

Having briefly reviewed the experimental situation, let us outline the related theory efforts. 
The LF wave functions of the baryons ($\Delta$ and $N$)  with the $five-quark$ sector 
was studied in  \cite{Shuryak:2019zhv}.  This paper included the four-quark 't Hooft interaction
 to  first order, and obtained antiquarks PDFs from rather complicated wave functions.
 Its overall shape and scale reproduced the data shown in Fig.\ref{fig_d_minus_u}. Yet
 there were visible oscillations, coming perhaps from the rather limited functional basis set used.

\subsection{The $q\bar q$ pair production, to  the first order in 't Hooft Lagrangian} \label{sec_tHooft_sea}

The first study of the five-quark sector of the baryon wave functions has been done by one of us in \cite{Shuryak:2019zhv}.
In it the diagrams Fig.\ref{fig_sea} (c) or (d) were used to calculate the matrix of basis matrix elements relating 3- and 5-q states, and the Hamiltonian was then diagonalized. This procedure  includes diagrams of all orders.
However, the set of basis states used in that paper was based on a nonlinear map of momentum fractions $x_i$, from 5 to 4,
and the procedures was rather complicated. 

In fact, there is no need to follow this path, as the (modified) Jacobi
coordinates provide a linear map.  The physical domain of the
5 momentum fraction with a condition $\sum_1^5 x_i=1$ is the 4-d manifold called $pentachoron$  (or 5-cell or or
4-simplex), which is one dimension higher than  tetrahedron. For the case of 5 quarks such map is detailed in Appendix \ref{sec_Jacobi5}.
The four coordinates $\alpha,\beta,\gamma,\delta$  play the same role as $\rho,\lambda$ for three quarks. 
We conjecture that the eigenstates of the Laplacian on the 5-simplex can also be worked out analytically using some set of standing waves, 
see~\cite{Shuryak:2021mlh} for discussion and for the ground state in this form.

 The PDF is the integral of the squared wave function projected onto a $single$ variable of the set. For  $N$
constituents with  $N-1$ Jacobi coordinates, the integral has dimension $N-2$,
by tracing over the coordinates called generically $"\rho"$  
 \be PDF("\lambda")= \int d^{N-2}"\rho" |\psi ( "\rho","\lambda")|^2
\ee 
If for a crude estimate we take the wave function be flat over the manifold, then $PDF(x)\sim (1-x)^{N-2}$.
For $N=3$ the power is linear, while for $N=5$ it is a cube. 

In this work we carry  all the way to the evaluation of the 5-q LFWF in Jacobi coordinates.
 Instead,  we provide some estimates
of the probability of the process to {\em lowest order} in the  't Hooft vertex. It follows the same spirit as
DGLAP treatment of extra gluons and sea quarks. More specifically, the interaction among the emitted quarks is ignored, motivated by
the observation that the kinematical domain for the newly produced quarks correspond to $x$ much smaller than
those of the valence quarks. So, their production is treated as in free space, by a diagram with free phase space integration.

Let us denote by $A_1$  the amplitude  of the $q \bar q$ pair production in the first order in the   't Hooft vertex, corresponding to diagrams Fig.\ref{fig_sea} (c) or (d). Since it uses a process with fermionic zero mode
of the instanton, it should be 
proportional to a small instanton  packing fraction
$\kappa=(\pi^2/2)\rho^4 n_{I+\bar I}\sim 1/10$ of the original Instanton Liquid Model which sets the order of magnitude
of the effect. 

 All u- and d-quark emissions yield the final states
\bea
u\rightarrow A_1^2\,u+A_1^2 (d+\bar d)\nonumber\\
d\rightarrow A_1^2\,d+A_1^2 (u+\bar u)
\eea
If we note that the probability for a quark to do nothingh is $1-A_1^2$, then the proton composition after one interaction is
\bea
(2+A_1^2) u+ (1+2A_1^2) d+A_1^2 \bar u+2A_1^2 \bar d
\eea
which is valence quark preserving with $u_v=u-\bar u=2$ and $d_v=d-\bar d=1$. 
The parton distribution for the neutron follows by isospin symmetry.
In this schematic description of the proton and neutron sea contributions  to first order in $A_1^2$, the Gottfried sum rule reads
\bea
\label{GSRX}
\frac 13+\frac 23 (\bar u-\bar d)=\frac 13 -\frac {2A_1^2} 3 \rightarrow 0.227
\eea

More specifically, the simplified and ultra-local't Hooft vertex
\bea
G_{tHooft} ({\rm det}(\overline q_R q_L)+{\rm det}(\overline q_L q_R))
\eea
reduced to $u,d$ flavors, is characterized by the  coupling  
\be   G_{tHooft}  \approx 35 \, GeV^{-2} \label{eqn_GT}
\ee
see the related discussion in Appendix \ref{sec_G_tHooft}.
For the process 
\bea
u (K_u)\rightarrow u(K_{\underline u})+ d(K_d) + \bar d(K_{\bar d})
\eea
we define the 4-momenta in the infinite momentum frame as
\bea
K_u&=&(P,P,0_\perp)\nonumber\\
K_d&=&(xP+\frac{\vec k_\perp^2}{2xP}, xP, \vec k_\perp)\nonumber\\
K_{\bar d}&=&(zP+\frac{p_\perp^2}{2zP}, zP, p_\perp)\nonumber\\
K_{\underline u}&=&((1-x-z)P+\frac{(\vec p_\perp+\vec k_\perp)^2}{2(1-x-z)P},\nonumber\\
&&(1-x-z)P, -(\vec p_\perp+\vec k_\perp))
\eea
Assuming the produced $\bar d$ in an unpolarized $u$, we calculate the $probability$ of the process
as a square of the amplitude defined by old-fashion perturbation theory
\bea
\label{DND}
dN_{\bar d/u}=&&\frac {G^2_{tHooft}}{2}\frac {\overline{|V_{u\rightarrow ud\bar d}|^2}}{(E_u-E_{\underline u}-E_d-E_{\bar d})^2}
\nonumber\\
&&\times \frac 1{2E_{u}}\frac 1{2E_{\bar d}}\frac 1{2E_{\underline u}}\frac 1{2E_{d}}\frac{d^3K_{\underline u}}{(2\pi)^3}\frac{d^3K_d}{(2\pi)^3}
\eea
with the energy denominator, and the bar on the matrix element refers
to spin averaging over the't Hooft vertex. In the vertex there are terms proportional to constituent quark masses squared $m_q^2$ 
and to  momenta squared $\vec p^2$. As we will show in more detail in Appendix ~\ref{THOOFTDGLAP}, 
the momenta are regulated by the instanton formfactors and therefore $\vec p^2\sim 1/\rho^2$. For simplicity
we have ignored all terms with masses because $(\rho m_q)^2\sim 1/4$ can be considered small.
With this in mind,  the spin averaging gives
\be
\overline{|V_{u\rightarrow ud\bar d}|^2}=\frac 12 16(K_u\cdot K_{\underline u})\,(K_d\cdot K_{\bar d})
\ee
with (\ref{DND}) taking the form
\begin{widetext}
\bea \label{eqn_dN}
dN_{\bar d/u}=\frac{G_{tHooft}^2}4\,\bigg[\frac{\frac{(\vec k_\perp+\vec p_\perp)^2}{2(1-x-z)}
[\frac{\vec k_\perp^2}{2x}+\frac{\vec p_\perp^2}{2z}+(1-\frac 1{x+z})\frac{(\vec k_\perp+\vec p_\perp)^2}{2(1-x-z)}](x+z)}{(\frac{\vec k_\perp^2}{2x}+\frac{\vec p_\perp^2}{2z}+\frac{(\vec k_\perp+\vec p_\perp)^2}{2(1-x-z)})^2}\bigg]
\frac 1z \frac{dx}{x}\frac{dz}{(1-x-z)}\frac{d^2\vec k_\perp}{(2\pi)^3}\frac{d^2\vec p_\perp}{(2\pi)^3}
\eea
\end{widetext}
The sea distribution of $\bar d$ in an unpolarized constituent quark $u$,  is given by
\bea
N_{\bar d/u}(z,Q^2)=\int_{p_{\perp}}  \int_{\vec k_\perp}\int_x  \,\frac{dN_{\bar d/u}}{dz}
\eea
with  the integrals carried sequentially in the ranges:
$0\leq x\leq 1$  and $0\leq p_{\perp}^2\leq Q^2$. The sea distribution $\bar u$  in an unpolarized constituent quark $d$,
is identical with $dN_{\bar d/u}=dN_{\bar u/d}$.
For its evaluation see Appendix~\ref{THOOFTDGLAP}.  Using the value of the coupling 
of the  't Hooft operator (\ref{eqn_GT}),  
one finds the probability of the process $A_1^2\approx 0.11$, which sets the scale of the ``primary sea"  
produced by $1\rightarrow 3$ instanton-induced process. (The regulator on which the dependence
is logarithmic, is taken to be $\epsilon=0.01$.)

\begin{figure}[htbp]
\begin{center}
\includegraphics[width=6cm]{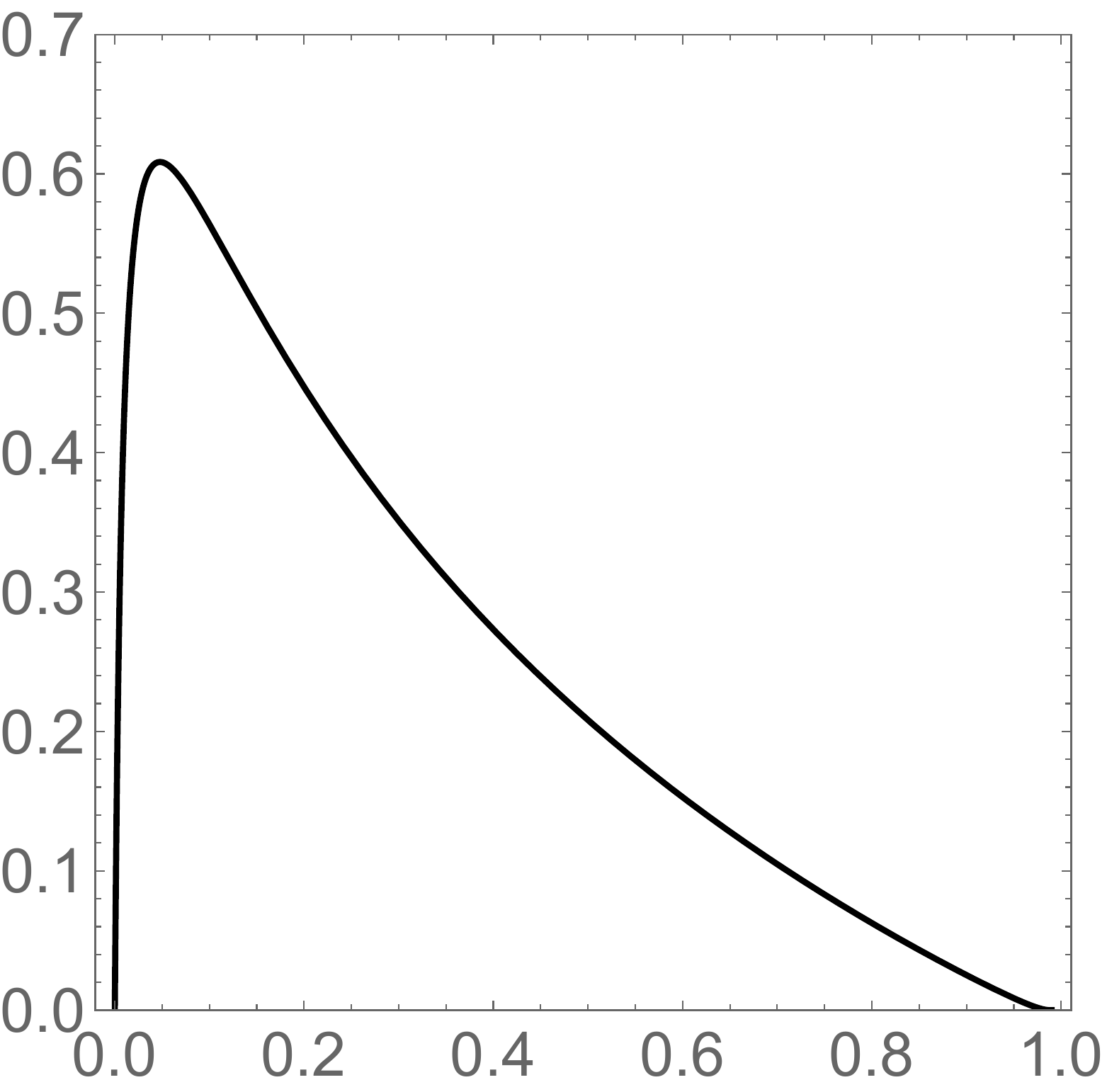}
\caption{The  ratio of the  ``produced" sea PDF to the one which initiated it  $\bar d^N(x)/u^N(x)$ as a function of $x$ following from~(\ref{EVOL}).}
\label{fig_dbar_to_u}
\end{center}
\end{figure}

The next issue we address is the $shape$ ($x$-dependence) of the PDFs of these sea (anti)quarks produced by the
processes in Fig.~\ref{fig_sea} (c,d).
Those can be obtained by convolution of this squared amplitude, treated as a ``splitting function"
with the original (valence) distributions $d^{N}(y),u^{N}(y)$
(e.g. those calculated from the LFWFs in the 3-q sector above).
The unpolarized  sea $\bar u, \bar d$ distributions in the nucleon are then
\bea
\label{EVOL}
\bar u_N(x,Q^2)&=&\int_x^1\frac {dy}y\,d^{p}(y,Q_0^2)N_{\bar u/d}\bigg(\frac{x}y, Q^2\bigg)\nonumber\\
\bar d_N(x,Q^2)&=&\int_x^1\frac {dy}y\,u^{p}(y,Q_0^2)N_{\bar d/u}\bigg(\frac{x}y, Q^2\bigg)\nonumber\\
\eea
where $q^p(x)$ is the unpolarized flavor $f=u,d$ distribution in a proton $p$, at the low
resolution point which we argued above is $Q_0^2\approx (2/\rho)^2$. In Fig.\ref{fig_dbar_to_u}
we plot the $ratio$ of the produced sea PDF to the one which initiated it, $\bar d^N(x)/u^N(x)$. 
The sea is strongly shifted to small $x\sim 1/10$. 
 At
 large $x$, if  the initial PDF  $u^N(x)\sim (1-x)^a$ has a certain power $a$, the produced one has power $a+1$.
 (While both these features are clearly steps in the right direction, the reader is perhaps aware that the observed sea PDF
 have at small $x$ negative power singularities, and a much larger difference between powers of $(1-x)$ in $u^p(x)$ and
 $\bar d^p(x)$, being approximately 3 and 7, respectively. These features however are known to be generated by
 subsequent DGLAP evolution from $Q_0$ to the the scale $Q$ at which experiments are done.)

\subsection{The sea induced by the ``pion cloud"} \label{sec_pion_sea}
The ``pion-induced"
contributions to the PDFs, see Fig.\ref{fig_sea} (e,f) (written in  DGLAP-like form),  were proposed in \cite{Eichten:1991mt} three decades ago.
  Note that the $\pi^0$ diagrams (e) for the specific in and out quark pairs have a factor of $(1/2)^2$, while the  $\pi^+$ diagram (f) 
  leads to flavor transition $u\rightarrow u\bar d $ with a larger factor $1$. The former changes the flavor of the recoil quark, but
  the latter does not. Because these two diagrams have different probabilities, they together lead
also to  a flavor asymmetry  of the sea.  Note  that these pion-induced diagrams can also be
 considered to be higher order iterations of the 't Hooft vertex, in different channels.

 Here we evaluate the contribution of the pion-induced antiquark production,  following~\cite{Eichten:1991mt}:
Let the probability of the pion-generated pair production process be  $P_\pi$, then all valence $uud$ quarks together produce a sea with probability $P_\pi(7/4 \bar u+11/4 \bar d)$, or
\be {\bar d \over \bar u}={11 \over 7}\approx 1.57
\ee
which is in good agreement with the ratio  reported experimentally. Using the absolute observed magnitude of
 $\bar d-\bar u$ (\ref{anti_difference}),  one finds that in order to explain its integrated magnitude one would need  
  $P_\pi\approx 0.2$.  

The expression for $P_\pi$ from the chiral Lagrangian~\cite{Eichten:1991mt} (20, 
 contains the following dimensionless combination of pion constants (in their notations)
\be  {g_a^2 m_q^2 \over 8\pi^2 f_\pi^2}\approx 0.09 \ee 
times certain integral being $O(1)$ (mildly depending on upper cutoff $\Lambda_{chiral}$. 
 As a result, $P_\pi^{th}$ given by the pion
diagrams gives about $half$ of the empirical effect. 

In summary, we conclude  that the first-order in the  't Hooft interaction, and the iterated (pion) diagrams give comparable
contributions to integrated flavor asymmetry of the antiquark sea. Unfortunately, at this time it is not
possible to make a more quantitative evaluation  including both.

\begin{figure}[t!]
\begin{center}
\includegraphics[width=6.5cm]{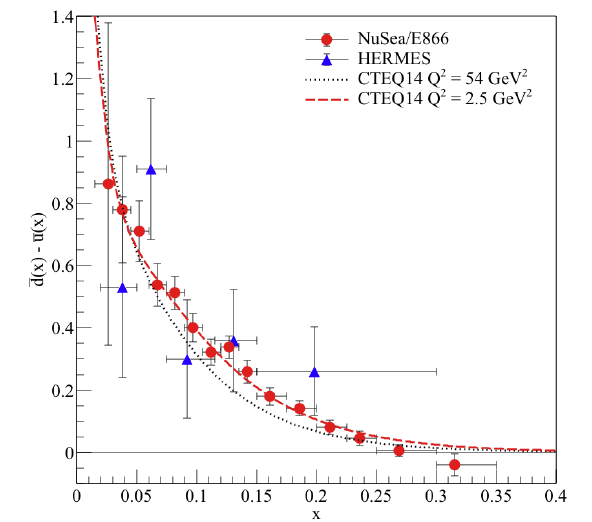}
\includegraphics[width=6cm]{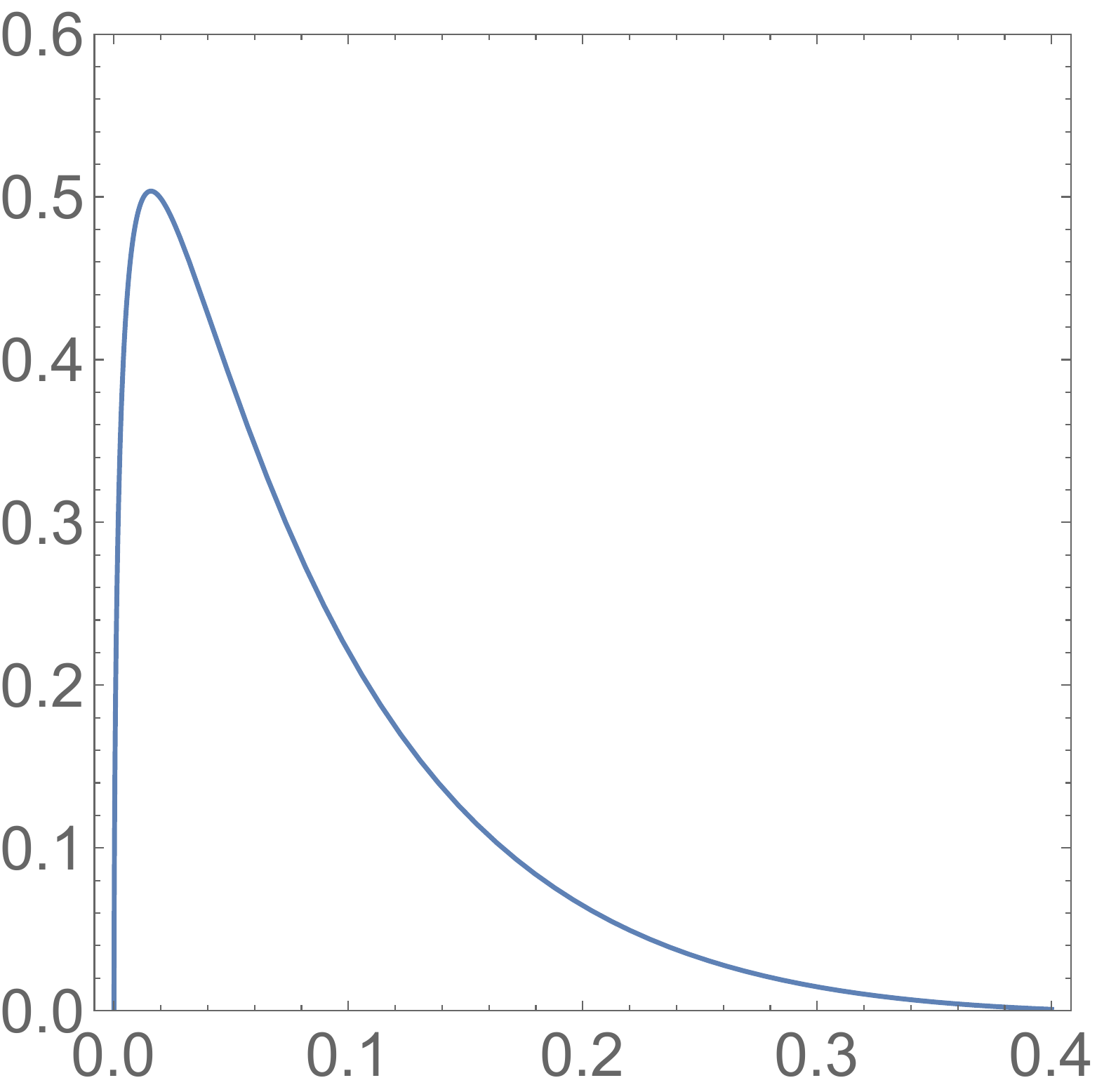}
\caption{Upper: the difference of sea antiquark PDFs $\bar d(x) -\bar u(x)$ from experiments. Lower:
the calculated shape of sea antiquark distribution (arbitrary units) described in the text.
}
\label{fig_d_minus_u}
\end{center}
\end{figure}

The next question is how the  antiquarks produced by the  intermediate pions are distributed in $x$. For that we adopt 
the expression (18) in~\cite{Eichten:1991mt}, using the 
convolution of the quark PDFs in the nucleon $q^N(y)$ with the ``splitting function" $P(z)$, followed by a convolution with the  pion PDF $q^\pi(x/y z)$. 
Unlike Eichten et al, however, we do not use here the PDFs fitted from experiments at some high $Q^2$, since
 our intensions here is to build another --chiral -- ark of the bridge.  So we take $q^N(y)=12 y (1 - y)^2$ (approximately
corresponding to the wave functions derived for three  quarks with the quasi-local attraction above).  We also take the symmetric 
PDF for the pion $q^\pi(x)=6 x (1-x)$ corresponding to a two-quark semicircular wave function.
 Convoluting those with  the ``splitting function" $P(z)$,  we get the shape of the pion-induced antiquark PDF shown in Fig.\ref{fig_d_minus_u} (lower).
As one can see  by comparing it to the upper experimental plot, it does reproduce the observed shape quite well.

Finally, let us briefly discuss the issue of {\em  flavor asymmetry} of valence quark distribution. The empirical PDFs
are such that
$(u_v(x)/2d_v(x)>1$ at large $x\rightarrow 1$. As was explained in section \ref{sec_N}, the quasi-local pairing interaction
makes the LFWFs ``flatter" (larger at large x) and, since the $d$ quark in the proton participate in two of those, one finds
the opposite, $(u_v(x)/2d_v(x)<1$ at $x\rightarrow 1$. 
However, chiral processes leading to the sea quark production work in the  opposite  direction, towards the one observed.
Let us single out the last diagram (f) of \ref{fig_sea} with an intermediate $\pi^+$. (Its contribution is larger than 
diagram (e) with $\pi^0$ by a factor 4, the flavor factors.) It interchanges the flavors of the leading quarks: if the original one is $u$ (as shown in this figure) then the recoil one -- the one at larger x -- is $d$.

\subsection{Evolution down to the matching point} \label{sec_DGLAP}
The PDFs describing the experimental and lattice data are professionally 
fitted to certain analytic forms, and connected to each other via
 perturbative DGLAP evolution.  For definiteness, we will rely on a sufficiently  complete global fit  called CTEQ18, in  Ref~\cite{1912.10053}, Appendix C.  It is defined 
 at $their$ lowest scale
 $$Q^2_{CTEQ}\approx 1.7\, GeV^2$$
The reason we need to discuss it
at the end of this paper,  is that we are  going to   evolve it further down,
 to the matching point discussed in section \ref{sec_matching_point}.

\begin{figure}[h!]
\begin{center}
\includegraphics[width=6cm]{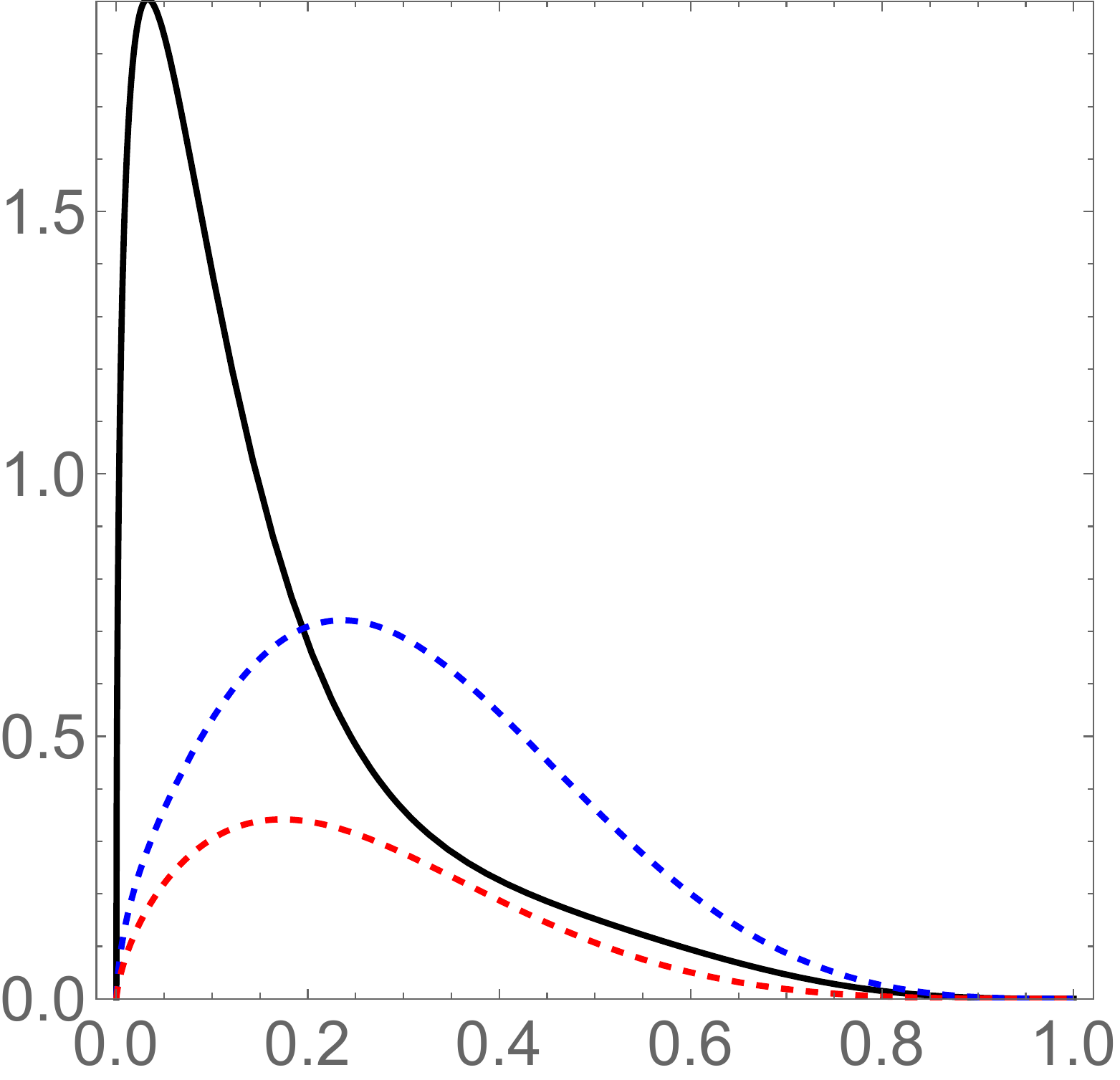}
\caption{CTEQ18 PDFs of the gluons $x g(x,Q^2_{CTEQ})$ (black solid curve), valence up $ x u_v(x,Q^2_{CTEQ})$
and down  $ x u_v(x,Q^2_{CTEQ})$ quarks (blue and red dashed curves.}
\label{fig_CTEQ}
\end{center}
\end{figure}

 We would not repeat those expressions here, just show their plot  for valence quarks and gluons in the proton
 in Fig.\ref{fig_CTEQ}. Unlike many other similar plots, we have not reduced the gluons by any artificial  factor, to fit it better in the plot. Our aim is to remind the reader that, even at this scale $Q^2_{CTEQ}$, the proton contains significant
 amount of glue. In fact, as is obvious from the plot, they are dominant at $x<0.2$. Integrating these curves,
 one gets the corresponding momentum fractions at scale $Q^2_{CTEQ}$, $\langle x_g\rangle=0.385, \langle x_{uv}\rangle= 0.325,  \langle x_{dv}\rangle =
 0.134$. So, at this scale gluons are by no means subleading. 
 
 We have recalled these details to stress once more that the scale $Q^2_{CTEQ}$ {\it is not low enough}
 to match to the hadronic spectroscopy. Indeed, it operates in terms of constituent quarks and have no gluons.
So, what happens with the gluons when one  performs  DGLAP evolution $downward$, say to our ``matching scale" $1\, GeV^2$?

\begin{figure}[htbp]
\begin{center}
\includegraphics[width=6cm]{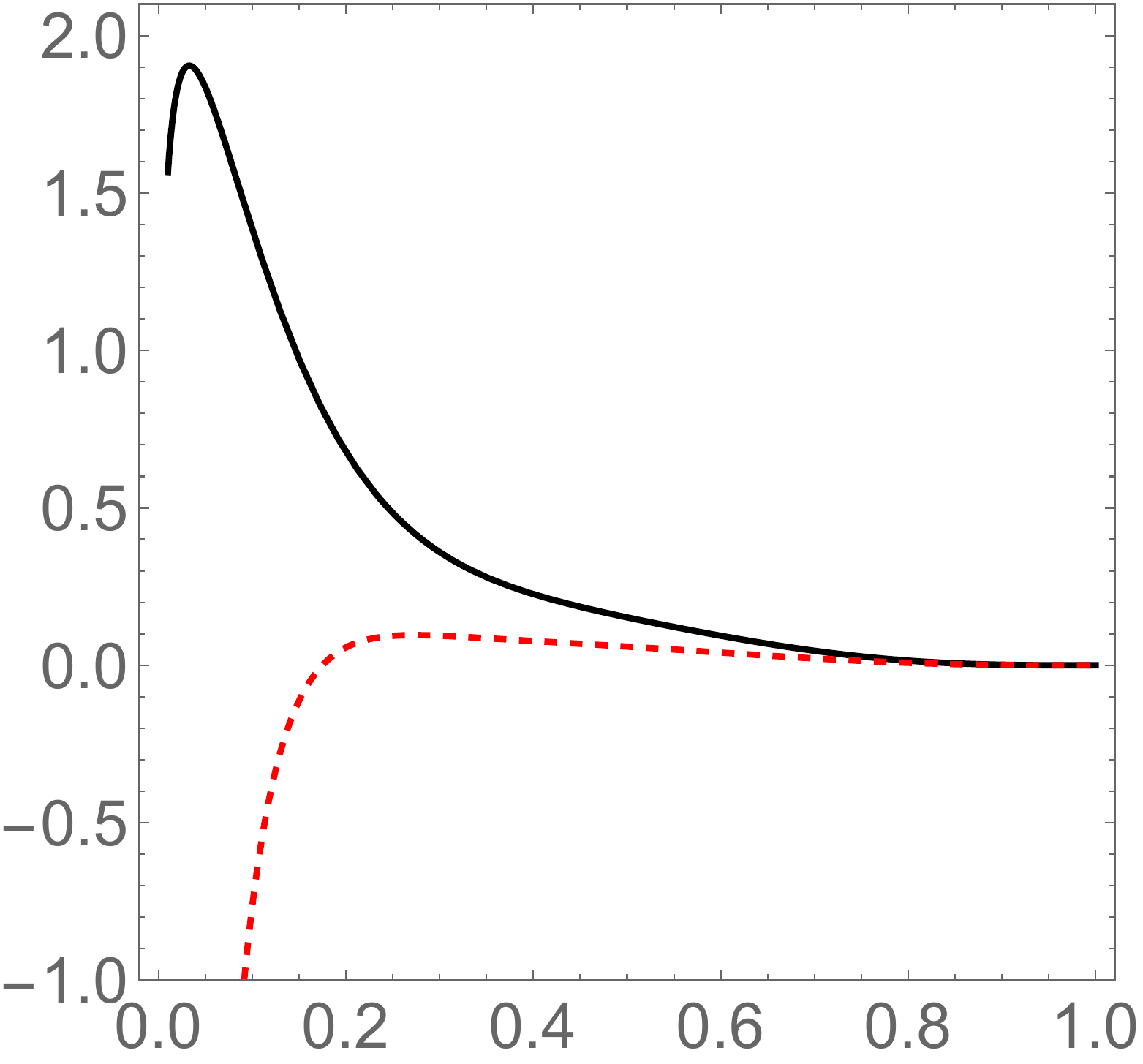}
\caption{Black solid line is the CTEQ18 gluon distribution $x g(x,Q^2_{CTEQ})$, and red dashed
line is its version evolved downward by DGLAP to $Q_{matching}^2=1\, GeV^2$. }
\label{fig_glue_disappear}
\end{center}
\end{figure}

Despite the fact that the amount of corresponding "DGLAP evolution time" is not long, $log(Q^2_{CTEQ}/Q_{matching}^2)\approx 0.52$, 
dramatic changes take place for the gluons. 
Using the lowest order splitting function $$P_{gg}(z)=6\bigg({z \over 1 - z} + {1 - z\over z} + z(1 - z)\bigg)$$
convoluted with $CTEQ18$,  we have evolved the gluons downward to our ``matching scale": the results are shown in  Fig.\ref{fig_glue_disappear}. As expected, we  see the gluons (and in particular $\langle x_g\rangle$) practically disappear! (Except at small $x$ where
$g(x)$ gets negative, which of course makes no sense.)
Oviously, the same downward evolution increases $u,d$ momentum fractions roughly to $2/3,1/3$.

Furthermore, the same perturbative downward evolution practically erases the quark-antiquark sea at the matching scale.
However, what is left there is the sea generated by chiral dynamics, as discussed in the
two preceding subsections, generating observable flavor asymmetry of the antiquarks.
Note further that the mean
momentum fractions $\langle x_{\bar q} \rangle$ generated there, are
at about 2-3 \% level, and of course they are  complemented by the gluon-generated sea at
higher scale. 

The use of DGLAP in this section is by necessity very  crude. The analysis can be improved, 
 to make our ``bridging the gap" goal smooth. It should eliminate artifacts (like negative PDFs) authomatically. Some modifications
are rather obvious, e.g. the inclusion of quark and gluon masses in the splitting functions.
Others  may  include the transition from kinetic equations to  LFWFs.

\section{Summary} \label{sec_summary}

This is the concluding paper of the series of five papers, and it is fitting to briefly overview here the main goals and results of the whole program, with some specifics about
each of them. 

When starting this program we had two general goals:\\ (a) One was to bring the quark models used
in hadronic spectroscopy to the light front.\\
(b) The other was to connect the obtained LFWFs to partonic observables as deduced
from various hard scattering processes.

The first goal is basically accomplished. Somewhat unusual, the  LF Hamiltonians create
technical problems, but those were solved. We have shown how one can include the confining forces
and solve the corresponding Schroedinger equation for mesons and for baryons (``on equilateral triangle"). This 
construction reproduces the  masses of multiple lowest states in each channel,
with agreement with empirical Regge trajectories. 
We also were able to include the ``residual" 
quasi-local binary attraction in a nonperturbative way, describing ``good diquark" correlations in a nucleon.
Of course, a lot of work remains to be done, such as the  inclusion of the spin-depend
potentials and the mixing between the various spin-orbit components of the wave functions. Clearly, 
 this can be done in a relatively straightforward way.

The second goal is accomplished only partly. By adding the 5-quark sector to baryons 
via certain approximate methods of chiral dynamics, we found a reasonable magnitude
of the antiquark sea, as well as its distribution in $x$. The observed  flavor asymmetry of the sea is
 explained. 
 
 However matching the experimental data on the PDFs, DAs, GPDs etc at high 
scale,  to those  we calculated from the LF wave functions at low scale, can so far be done
only at  the level of average quantities, e.g.  $\langle x_q\rangle$ but is not yet quantitative
for their $x$-dependence. We attempted to bring downward the DGLAP to a  scale as low as 1 $GeV^2$, 
where  we see that $\langle x_g \rangle\rightarrow 0$. However, we need  to tweek the  DGLAP
evolution to accomodate switching off the gluons in a consistent manner. Also 
this probabilistic description is only justified when $x$ of the produced partons is small
compared to that of their parents, for otherwise we need to develop a coherent Hamiltonian description for the  quark-gluon sector.

This is now a good place, to remind the reader for the specific content of these five papers.
We started in the first paper \cite{Shuryak:2021fsu} with discussion of the physical origin of the confining and spin-dependent potentials  for heavy quarkonia
in a  traditional  setting, in which they are defined via some correlators involving  Wilson lines. Specifically, we
focused on instanton-induced effects. Following our earlier paper on mesonic formfactors \cite{Shuryak:2020ktq},
we did so in a novel ``dense instanton liquid" which includes both instantons forming the quark condensate (a subject of
studies in the previous four decades) and close instanton-antiinstanton molecules. We have shown that such a vacuum model
reproduces the phenomenological confining potential up to distances of $r\sim 0.8 fm$. The spin-dependent potentials
are defined via Wilson lines with added magnetic field strengths. The
perturbative and instanton-induced effects are both short-range and were shown to have comparable  magnitude
for charmonia, with instanton effects dominating for light quark systems.

The second paper \cite{Shuryak:2021hng} starts the derivation and usage of the light-front Hamitonians $H_{LF}$
for the description of meson light-front wave functions (LFWFs). Here we developed the ``einbine trick", by means of which
a potential linear in coordinates turns into a quadratic one. Then, writing the coordinates as derivatives over momenta,
we net  Laplacian-like confining terms, while the kinetic energy $\sim (p_\perp^2+m^2)/x$ is treated as 
a certain potential energy.  The masses of the obtained states were shown to be close to the expected Regge trajectoris, with
novel LFWFs to follow.  In this paper we also managed to put the instanton-related Wilson lines from Euclidean time
into the light cone, by analytic continuation from Euclidean angle to Minkowskian rapidity (the hyperbolic angle).
We have derived the spin-dependent terms of $H_{LF}$ following from Wilson lines (nonzero fermionic modes),
and 't Hooft effective Lagrangian (zero modes). The latter was shown to generate massless pion: as a benefit we
have its LFWF.

The third paper \cite{Shuryak:2021mlh} was also devoted to mesons on the light cone, focusing on spin-spin  and spin-orbit forces. Starting with heavy quarkonia (bottomonium), we compared the traditional Schroedinger equation in the CM frame and spherical
symmetry, to $H_{LF}$-based approach in which the symmetry is just axial. We do get the correct spectrum of bottomonia,
in agreement with  its Regge trajectory.
 Using the fact that $H_{LF}$ has
full relativistic kinematics, in which there is no principal distinction between heavy and light quarks, we extended
the latter to strange and light mesons.  We focused on spin and orbital momentum mixing onthe  LF, in which both
are represented just by their longitudinal projections. We studied the role of the tensor forces and mixing in vector mesons,
generating their quadrupole moments (both in the CM and on the LF frames). At the end, we studied the  relations between LFWFs 
and the PDFs and distribution amplitudes (DAs) of the mesons.

In the fourth paper \cite{Shuryak:2022thi} we proceeded to three-quark baryons, with heavy and light quarks. However,
in this paper  we  restricted our analysis to $flavor-symmetric$ baryons ($bbb,ccc,sss,uuu$) in which the  't Hooft four-fermion
effective Lagrangian does $not$ operate. One novel feature was the detailed calculation of the instanton-induced 
three-static-quark potentials, which were also compared with available lattice data for the same geometries. 
Our conclusion is that all of them seem to favor the model we call ``Ansatz A", half the sum of binary two-quark potentials.
Another feature of this work, separating it from others in literature on baryon LFWFs, is that we used  (modified) Jacobi
coordinates and thus have as many coordinates as necessary, without spurious center-of-mass motion. The longitudinal
momentum fractions $x_1,x_2,x_3$ are then defined on an equilateral triangle in two Jacobi coordinates. The natural
basis functions are therefore those of a Laplacian on such triangle. We  were able to give analytic form for this set.
Solving the full Hamiltonian requires numerical approaches: one of them uses matrices in terms of basis functions,
another is a direct numerical solution of 2d Schroedinger-like eqns (provided the transverse and longitudinal motion
can be approximately factorized).

Now we summarize the main content of this paper, the fifth in the series. It is devoted to two very different issues.
The first is the  flavor-asymmetric ``good diquarks" $ud,us,ds$ with $J^P=0^+$ quantum numbers. Multiple
phenomenological and lattice results show that those are rather deeply bound, in comparison to two
constituent quarks or ``bad diquarks" with other $J^P$ values. We calculated the LFWFs including the pairing correlations
induced by the instanton-induced 't Hooft operator, in its quasi-local form. (We showed how to do so without
fullly Fourier transforming our momentum wave functions into  coordinate representation.) We did so for
the heavy-light baryons $\Lambda_c=cud$ with a single diquark,  and the nucleon with its two pairing $ud$ channels. The masses and,
most importantly LFWFs of those,  were compared to states without ``good diquarks", $\Sigma_c$
and $ \Delta$ respectively. These differences of
LFWFs due to quasi-local pairing are found to be rather significant.

The second issue is in fact the underlying reason for why all those paper were written. 
Two important subfields of hadronic physics -- the $spectroscopy$ (done in the rest frame with the wave functions and constituent quarks), and the {\em partonic physics} (done in terms of density matrices PDFs and pointlike quarks and gluons on the LF). 
The previous four papers make the first step, exporting the spectroscopy to the light front. Here we made the second step,
by adding to  the 3-q  baryons ``the sea", again using the  't Hooft four-fermion operator, but now in 1-to-3 channel. 
One way to do that would be to use Jacobi coordinates for 5-q systems, which we detailed. However,
we do not follow this path to
calculate the 5-q LFWFs. Rather, we have proceeded {\em a la} perturbative DGLAP evolution, treating this Lagrangian to the lowest order,
 and evaluated the appropriate ``splitting function" and probabilty.  
The higher orders are approximated by ``pion cloud" contribution, already known in the literature. We show that
these effects do account for flavor asymmetry of the antiquark sea, both in magnitude and in $x$-dependence.

The final point of this paper is ``matching" the valence quark and sea PDFs to phenomenological ones.
We think that the matching scale should be $Q^2\approx 1\, GeV^2$, being both the upper scale of chiral (instanton) physics,
 and the lowest scale at which the gluon components of the PDFs disappear. We show that these  three
subsequent steps do indeed provide a bridge between spectroscopy and partonic PDFs, in so far at the semi-quantitative level. 

\appendix 
\section{ Variational study of $s s'$ diquarks } \label{sec_variational}
Starting from diquarks as a spherically symmetric two-quark system in 3d, 
and $quadratic$ confinement, one is in an oscillator setting, with a Gaussian  wave function for the ground state 
\be \phi_0(r)=\frac{e^{-{ r^2 \over 2\beta^2}}}{\pi^{3/4} \beta^{3/2} }
\ee 
where $r$ is the $relative$ distance between quarks. The r.m.s. distance is $R_{r.m.s.}=\sqrt{3/2} \beta$.

The simplest interaction between $u$ and $d$ quarks  is the local form of the  't Hooft Lagrangian
\be \label{eqn_local_Hooft}
V_{'t Hooft} = - {G_{'t Hooft} \over 2} \delta^3(\vec r) \ee
where the coupling constant is the one in mesonic (pion or $\eta'$) channels. The $-1/2$ stems 
from the Fiertz transformation in the diquark channel, see \cite{Rapp:1999qa} for details. Averaging it over simplified
wave function one gets
\be \langle V_{'t Hooft} \rangle = -{G_{'t Hooft} \over 2 \pi^{3/2}\beta^3} \ee 

The Coulomb interaction in the diquark channel, has also half of the strength compared to the mesonic channel
$V_C(r)=-(2/3)\alpha_s/r$. Its average using the same wave function is
\be  \langle V_C\rangle = -  { 2 \alpha_s\over 3} {2 \over \sqrt{\pi} \beta } \ee
which for the same  $R_{r.m.s.}$ give about $-0.15 \, GeV$.
(The additional nonperturbative component -- instanton gauge fields -- will be evaluate later ). 

\begin{figure}[h]
\begin{center}
\includegraphics[width=6cm]{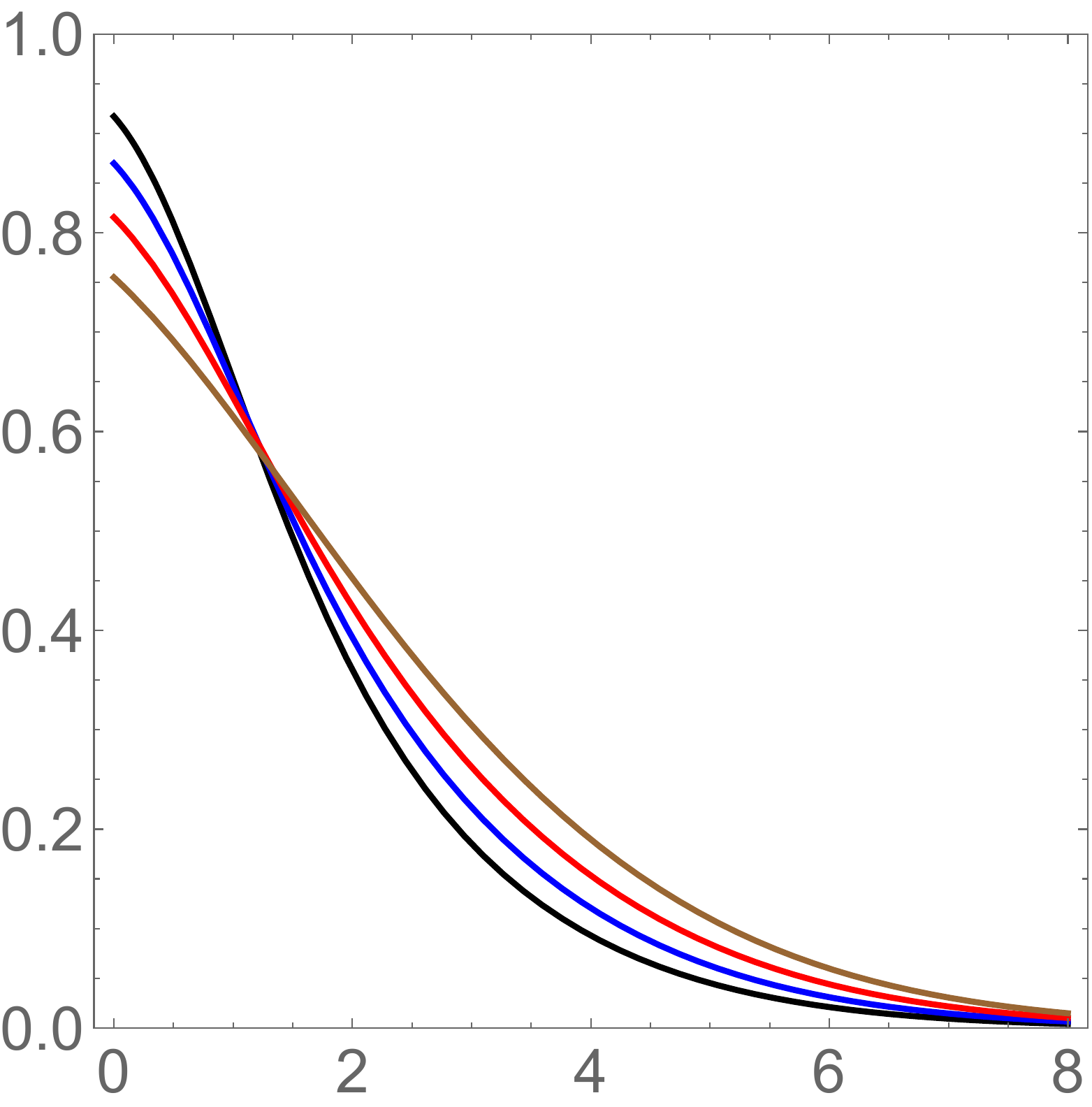}
\caption{Comparison of the shapes of the wave functions $\psi_{dq}(r)$ vs $r \, (GeV^{-1})$ for $ss'$ diquarks, for $G_{qq} =30,20,10,0\, GeV^{-2}$,
black, blue, red and brown curves, respectively.}
\label{fig_diquarks_various_G}
\end{center}
\end{figure}

The wave functions dependence on the strength of 't Hooft coupling is shown in Fig.\ref{fig_diquarks_various_G}. The diquarks get more compact as the
pairing strength grows. This effect is nonlinear in binding, and  grows  stronger.

\section{From the wave functions in momentum representation to local and Coulomb interactions} \label{sec_quasilocal}
The generic  two-body interaction, assumed to be between $u$ and $d$ quarks, is  of the form
\be \label{eqn_generic_coordV}
\langle V \rangle=\int d^3\vec r_u d^3\vec r_d |\psi_u(\vec r_u)\psi_d(\vec r_d)|^2 V(|\vec r_u-\vec r_d|)  \ee
with a potential depending  on the relative coordinate $r_{ud}=| \vec r_u -\vec r_d |$,
while the WFs depend on the individual coordinates.  The average 
 of the local potential  (\ref{eqn_local_Hooft}) takes the form
\bea 
&&\langle \delta^3(\vec r_u -\vec r_d) \rangle = \nonumber\\ 
&&\int d^3 r_u d^3 r_d  |\psi(\vec r_u)|^2 |\psi(\vec r_d)|^2\delta^3(\vec r_u -\vec r_d) =\int d^3 r |\psi(\vec r)|^4 \nonumber\\
\eea
familiar in few-body physics, with the 4-th power of the single-body wave functions in the CM frame.

In order to use a more general potential, it us customory to 
 proceed to  the momentum representation via Fourier transform $\psi(\vec r) \rightarrow \psi(\vec p)$, and introduce the
 so called {\em overlap function}
\be S(\vec q) =\int {d^3 p \over (2\pi)^3} \psi^*(\vec p) \psi(\vec p+\vec q)
\ee
In these  notations,  the interaction (\ref{eqn_generic_coordV}) can be rewritten as a convolution 
of overlap functions squared with the Fourier transform
of the potential
\be \langle V \rangle=\int {d^3 q \over (2\pi)^3} |S(\vec q) |^2 V(q) \ee
For color Coulomb interaction between quarks $V(q)=(2/3)\alpha_s/q^2$. This approach
is so-to-say $t$-channel description of scattering.

Now, the difficulty  we have is related to the fact that
the   LFWFs are defined in {\em momentum representation}, 
while the binary interactions is given in {\em coordinates}. However, since the 
 't Hooft interaction
can be approximated by a local form (\ref{eqn_local}), we  can 
avoid the cumbersome Fourier transforms, and use an alternative $s$-channel description without oscillating exponents.
 
Let us explain it first using the simplest example of a meson. In this case the wave function is a function 
of the relative coordinates $\vec r=\vec r_1- \vec r_2$ already, so the  local interaction is just 
proportional  to the coordinate wave function at the origin $-G |\psi(\vec r=0) |^2$. (For example,
such an approximation is in fact exact for perturbative spin-spin interactions, in atoms, nuclei and baryons.)
The point is that it has a simple expression in terms of the wave function in momentum representation
\be \psi(\vec r=0) = \int {d^3p \over (2\pi)^3 }\Psi(p) \ee

The problem discussed in section \ref{sec_Lambda}  uses binary local interaction only between particles 1 and 2.
In terms of  Jacobi coordinates, it is proportional to $\sim \delta(\vec r_\rho)$, while particle 3 (related to the
coordinate $\vec r_\lambda$ is not affected. Its matrix element in the momentum representation 
has $two$ integrals over $\vec p_\rho, \vec p^\prime_\rho$ but only $one$ over $\vec p_\lambda$, as the momentum of particle 3 does not change 
\bea
 \label{eqn_mean_delta_of_rho}
 &&\langle \delta(\vec r_\rho) \rangle =\nonumber \\
 &&\int  {d^3 p_\lambda \over  (2\pi)^3 } \bigg( \int {d^3 p_\rho \over  (2\pi)^3 } \Psi(\rho,\lambda) \bigg)
  \bigg( \int {d^3 p_\rho' \over  (2\pi)^3 } \Psi^*(\rho',\lambda) \bigg) \nonumber\\
\eea

\section{The longitudinal basis functions on the equilateral triangle} \label{sec_basis_triangle}
The analytic form for these functions were found in \cite{Shuryak:2022thi}, and they were also
obtained numerically  using a Mathematica 2d solver. In parts of this paper we used them as a basis,
expressing the nontrivial parts of the Hamiltonian as matrices. Therefore, it is helpful to show
the shapes of the lowest ones (actually 12 lowest) that we used as our reduced basis, see Fig.\ref{fig_b}.

\begin{widetext}
\newpage
\begin{figure}[b]
\begin{center}
\includegraphics[width=5cm]{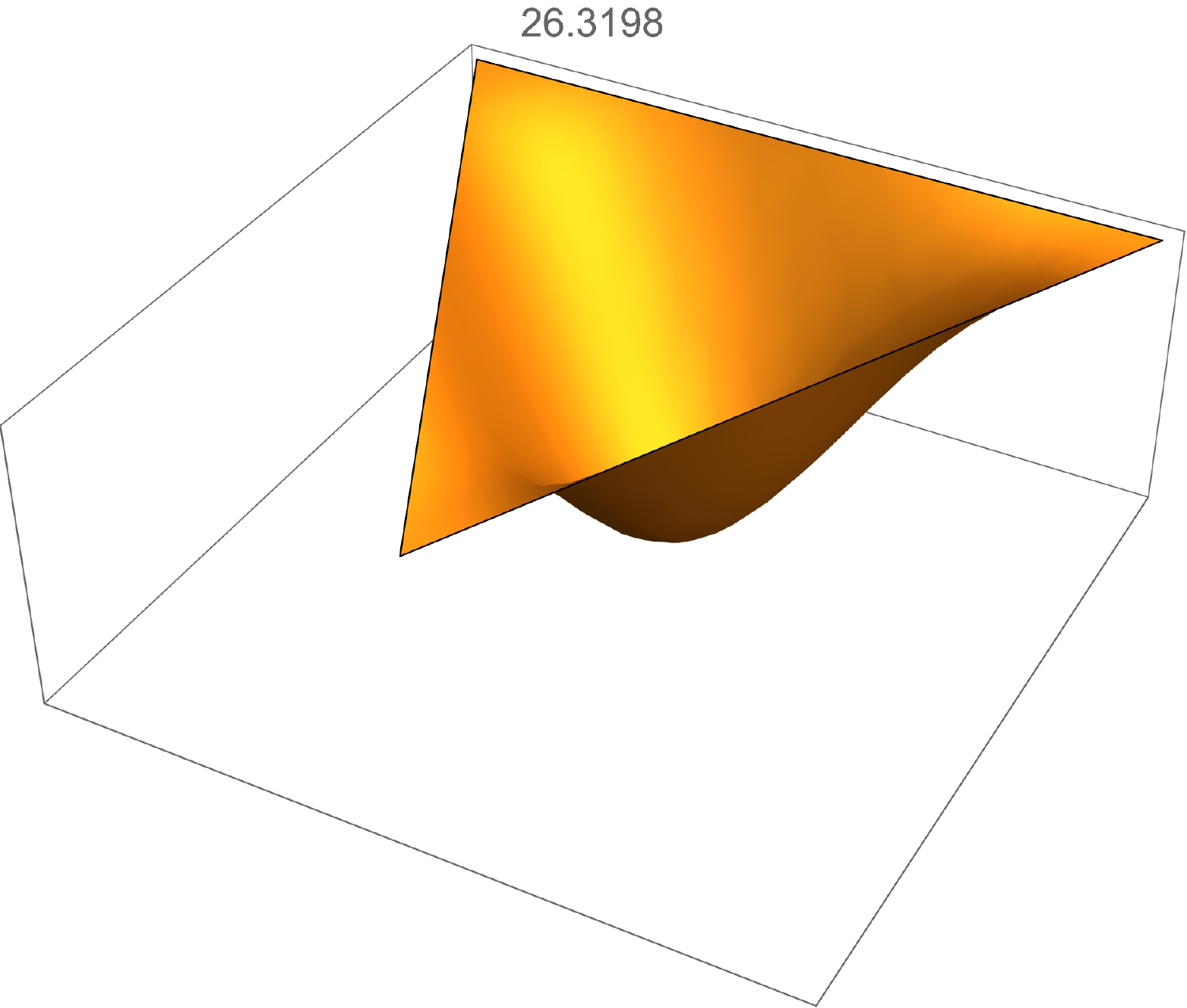}\includegraphics[width=5cm]{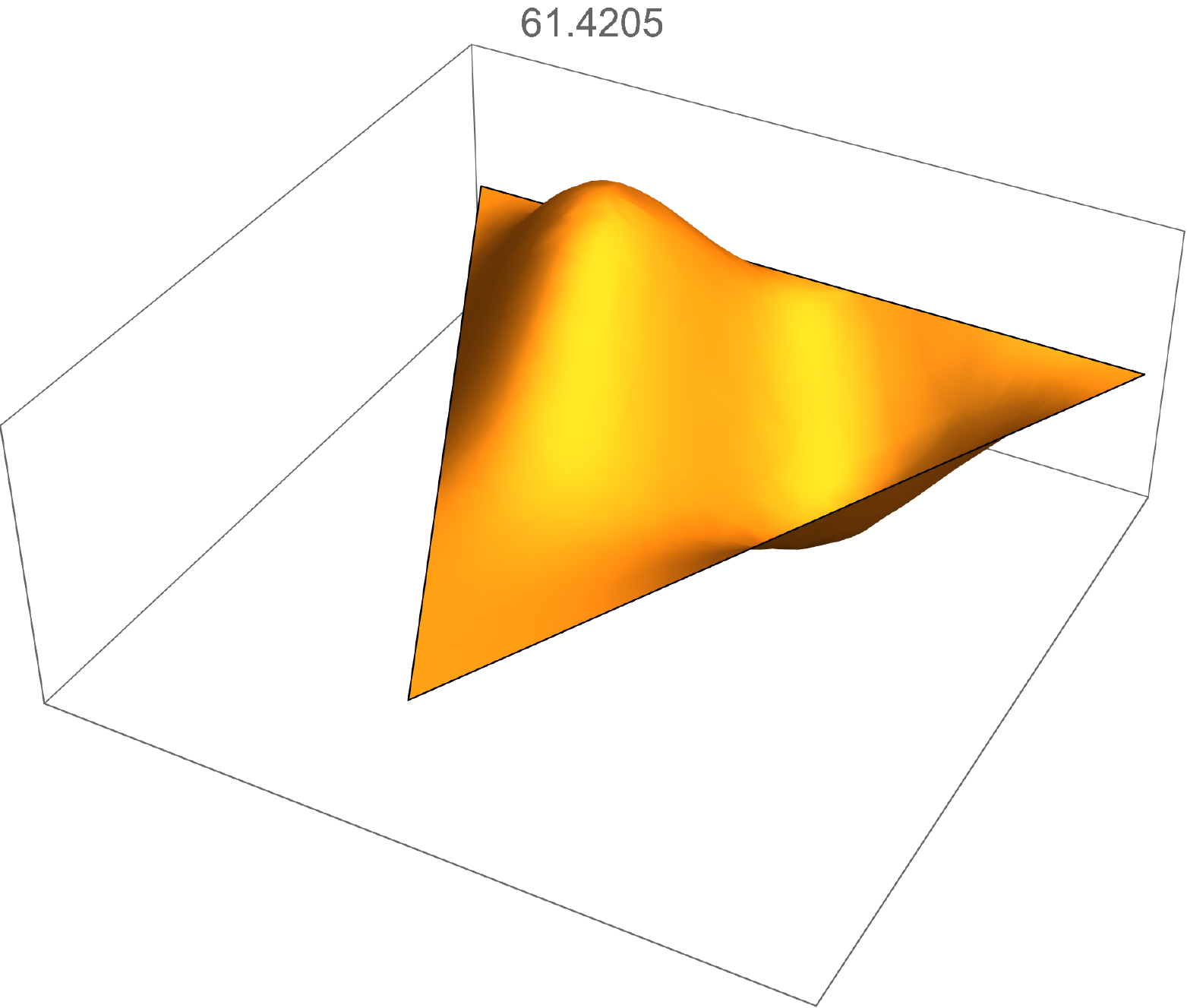}\includegraphics[width=5cm]{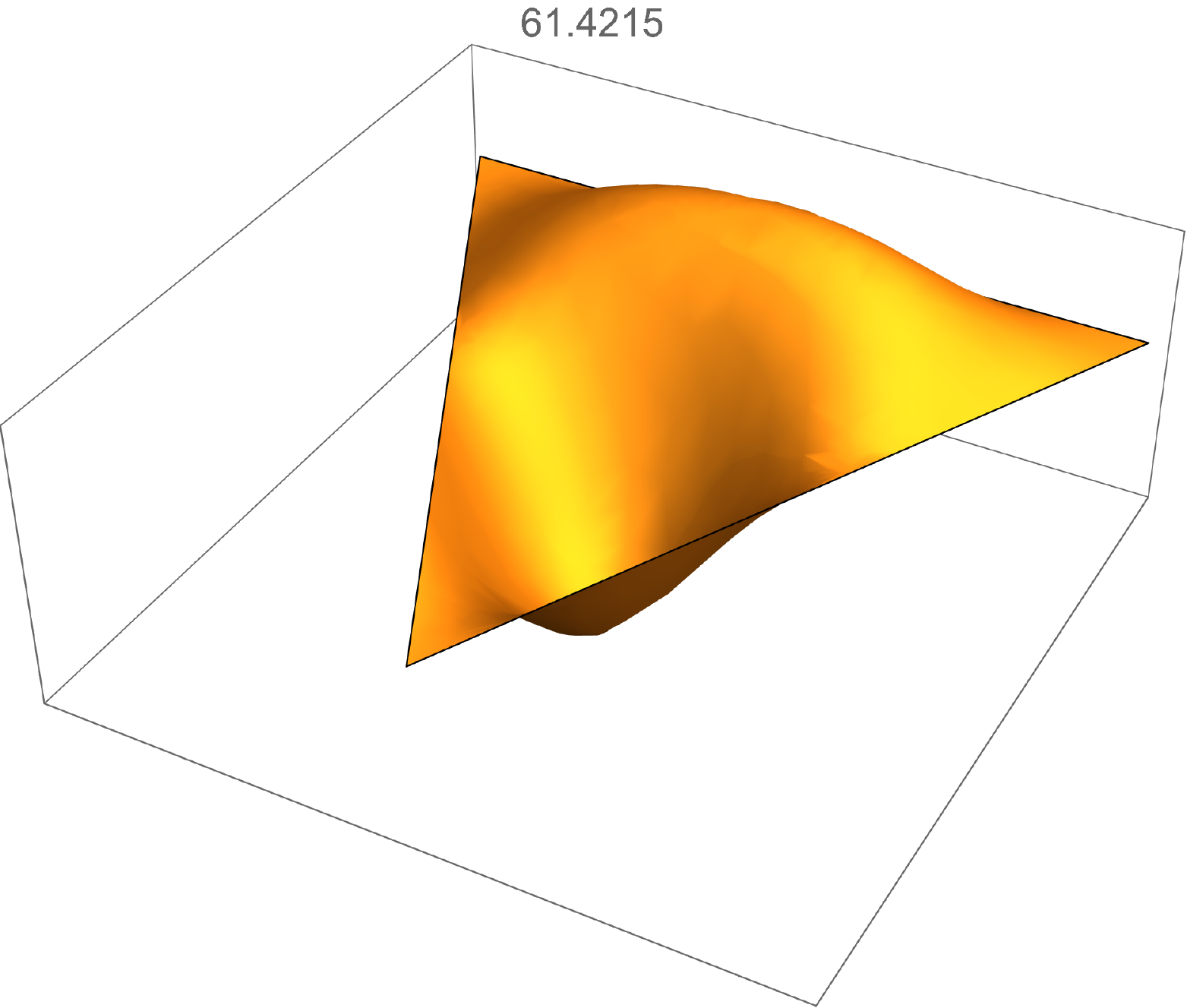}
\includegraphics[width=5cm]{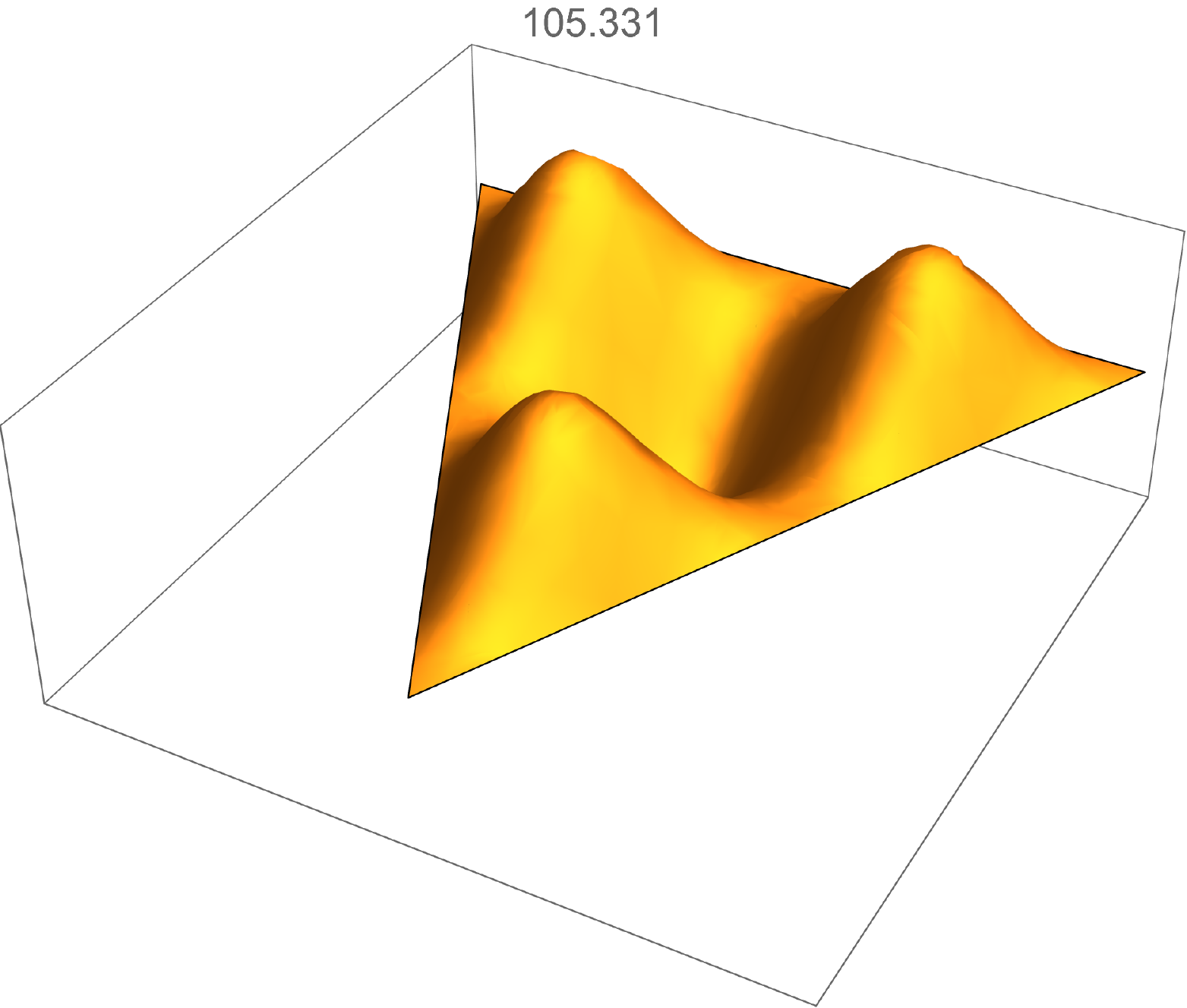}\includegraphics[width=5cm]{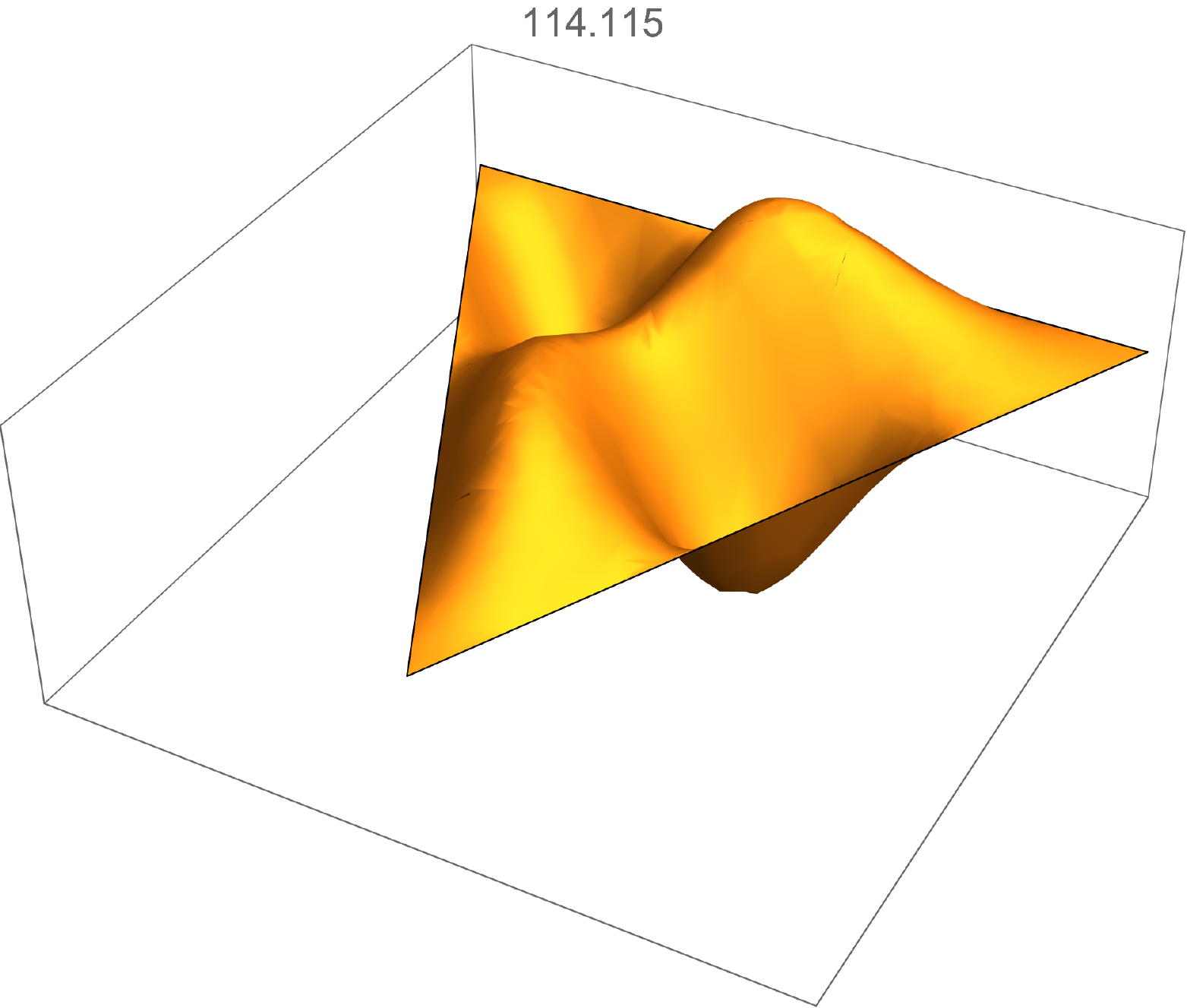}\includegraphics[width=5cm]{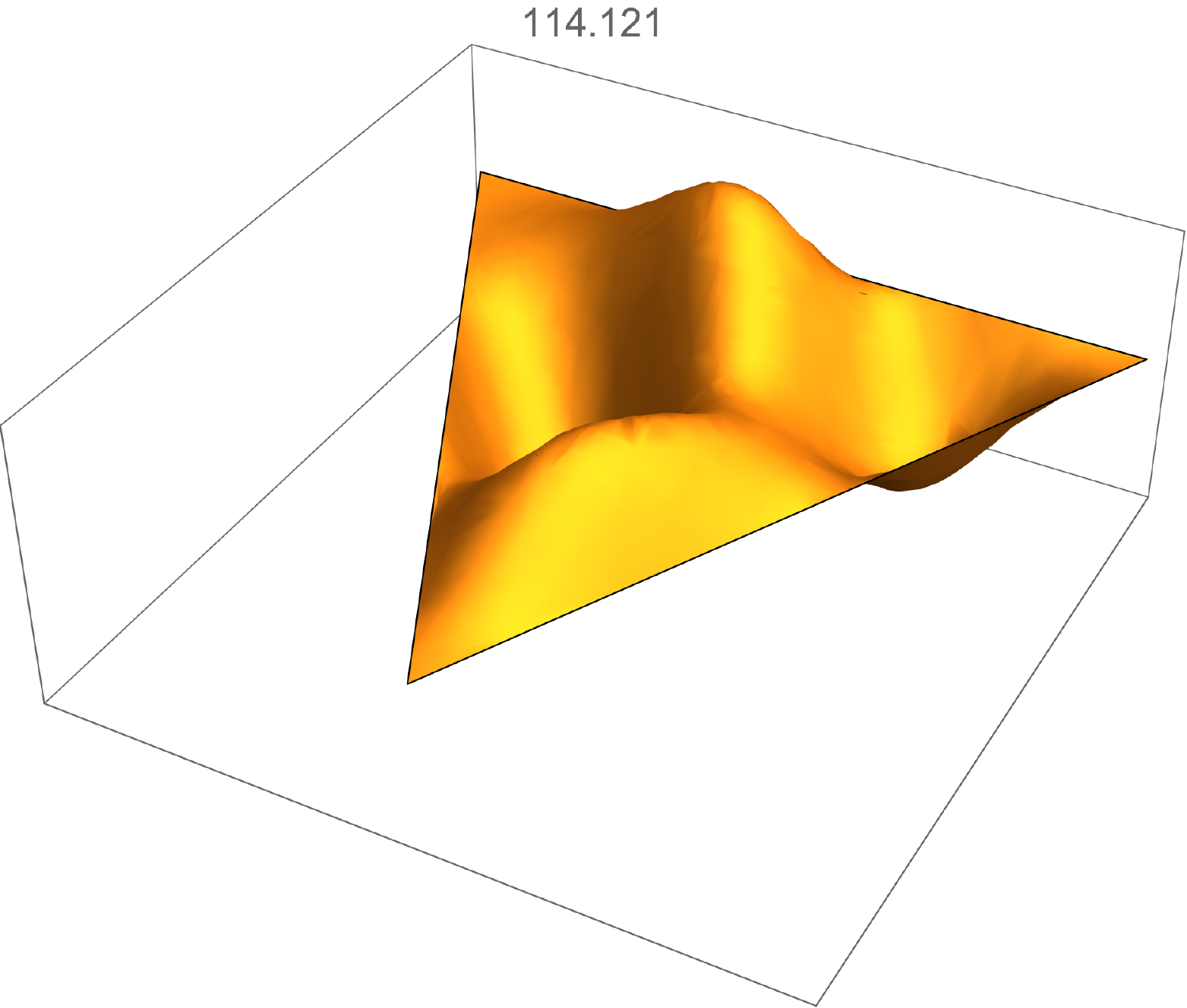}
\includegraphics[width=5cm]{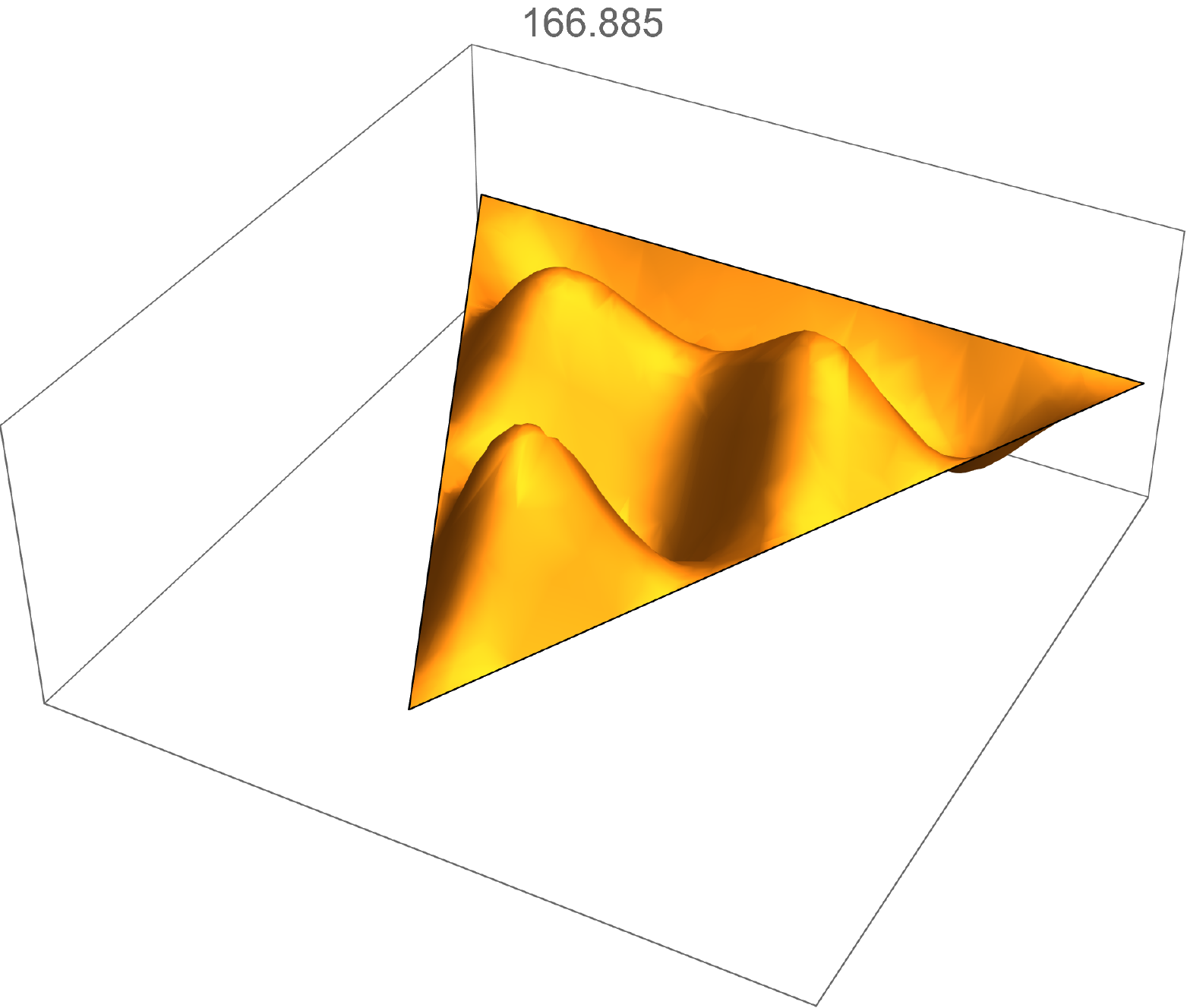}\includegraphics[width=5cm]{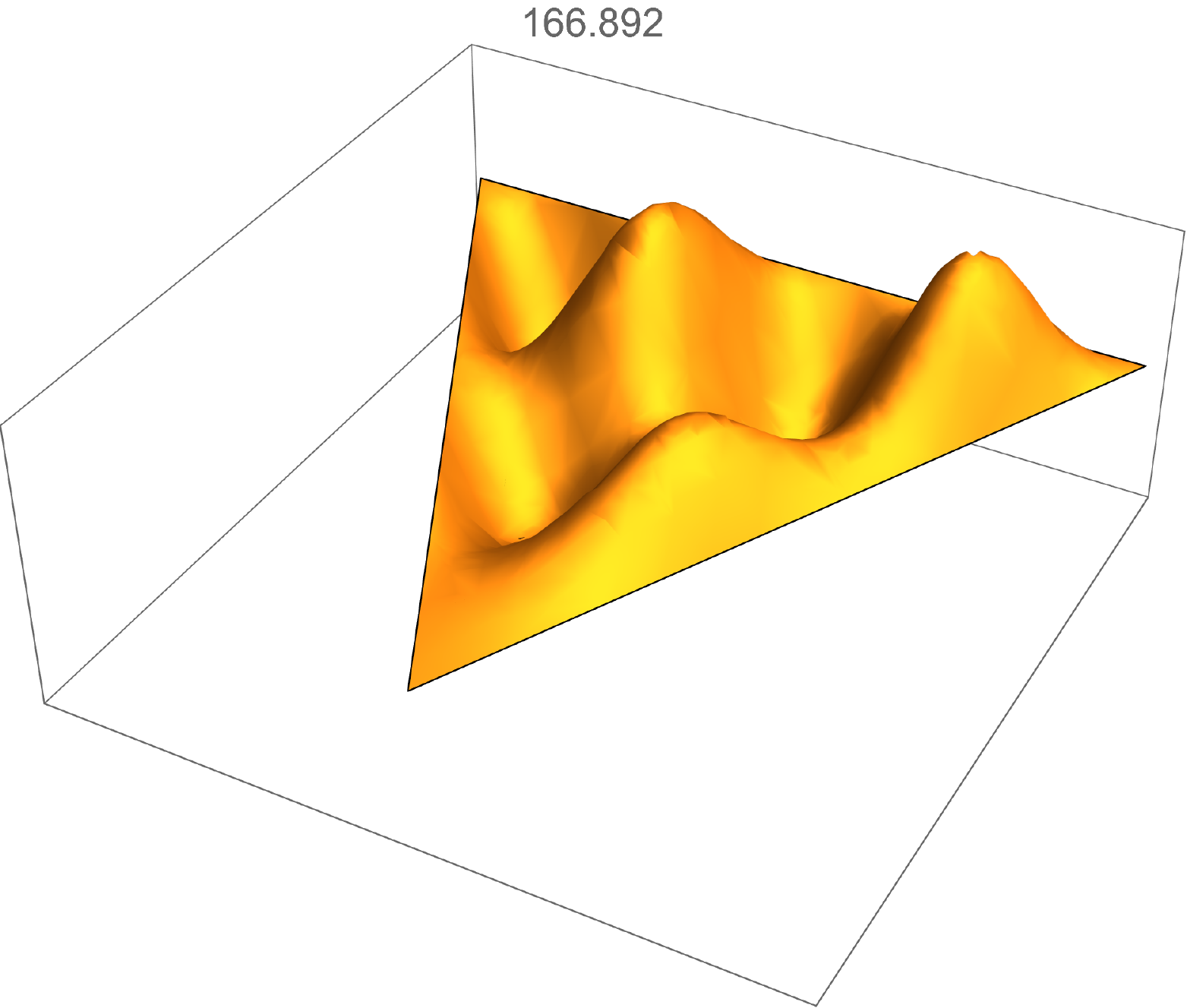}\includegraphics[width=5cm]{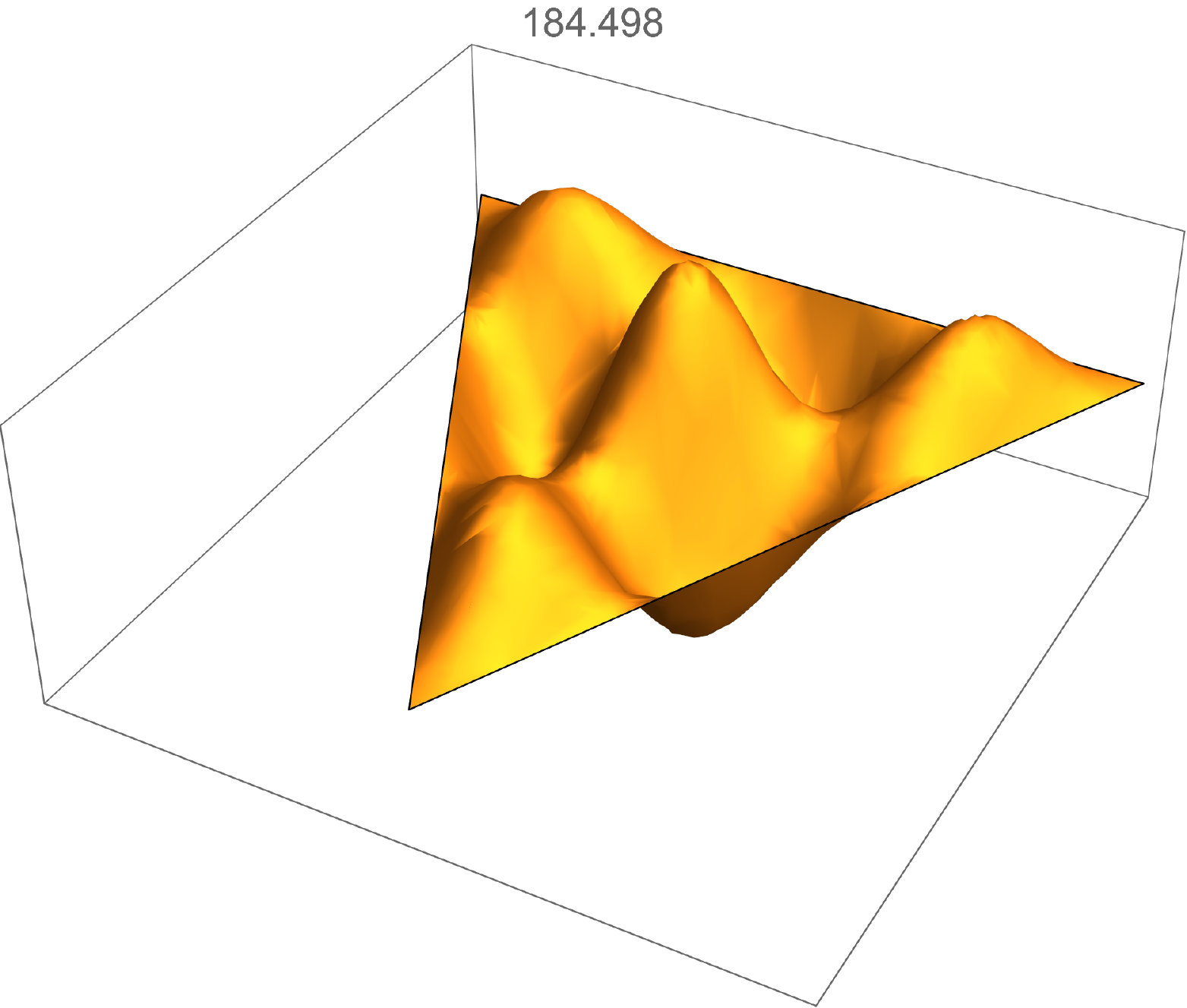}
\includegraphics[width=5cm]{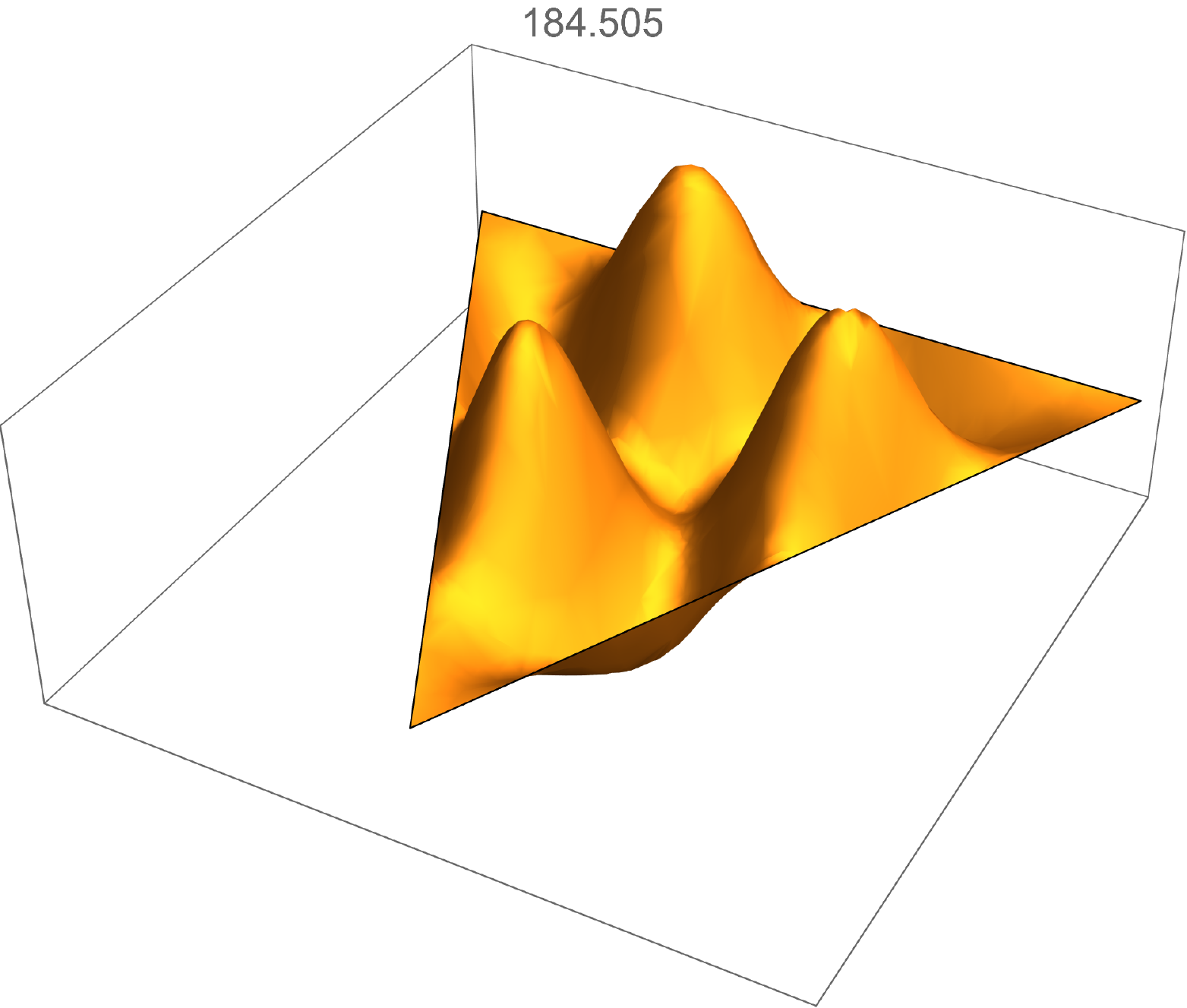}\includegraphics[width=5cm]{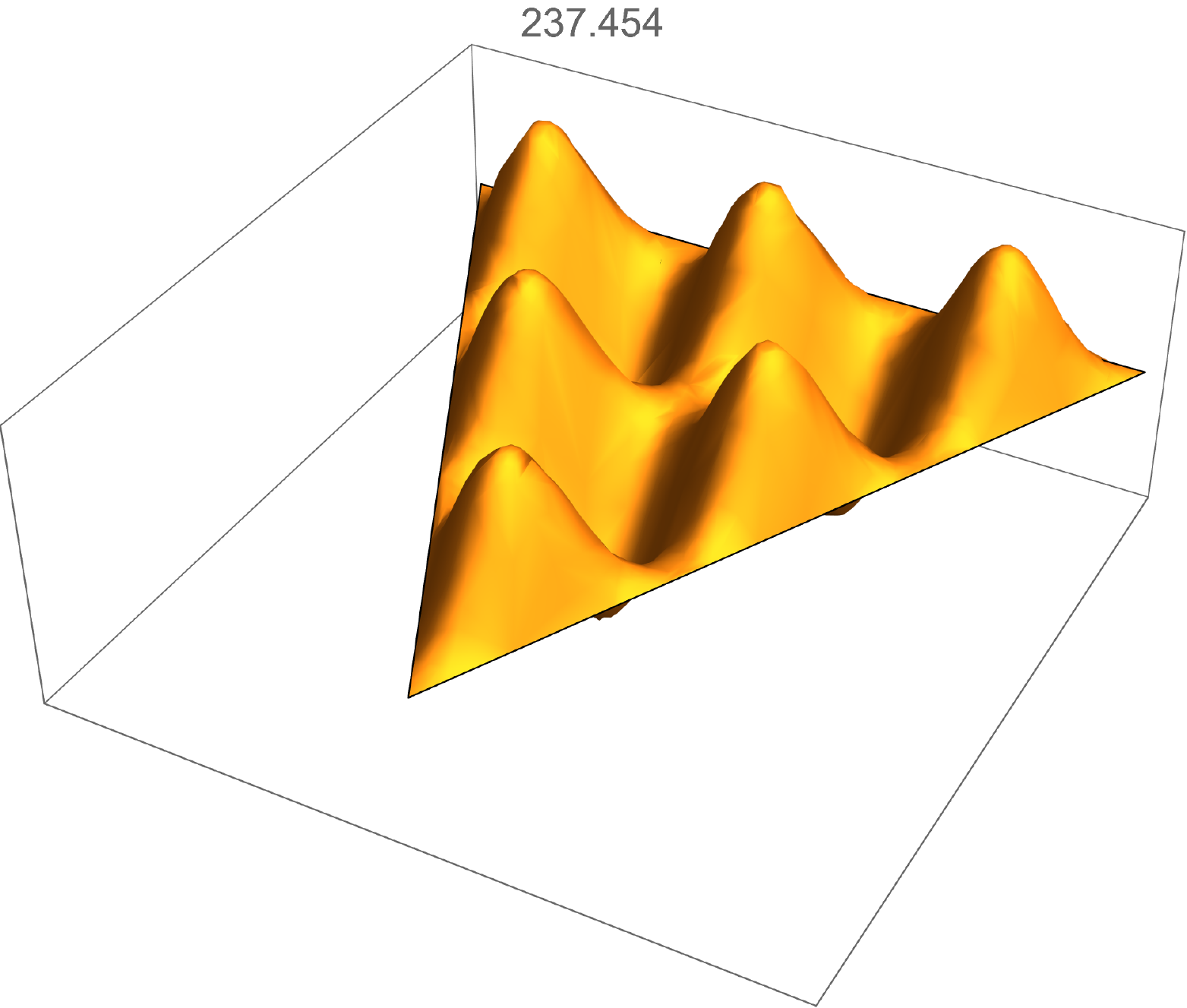}\includegraphics[width=5cm]{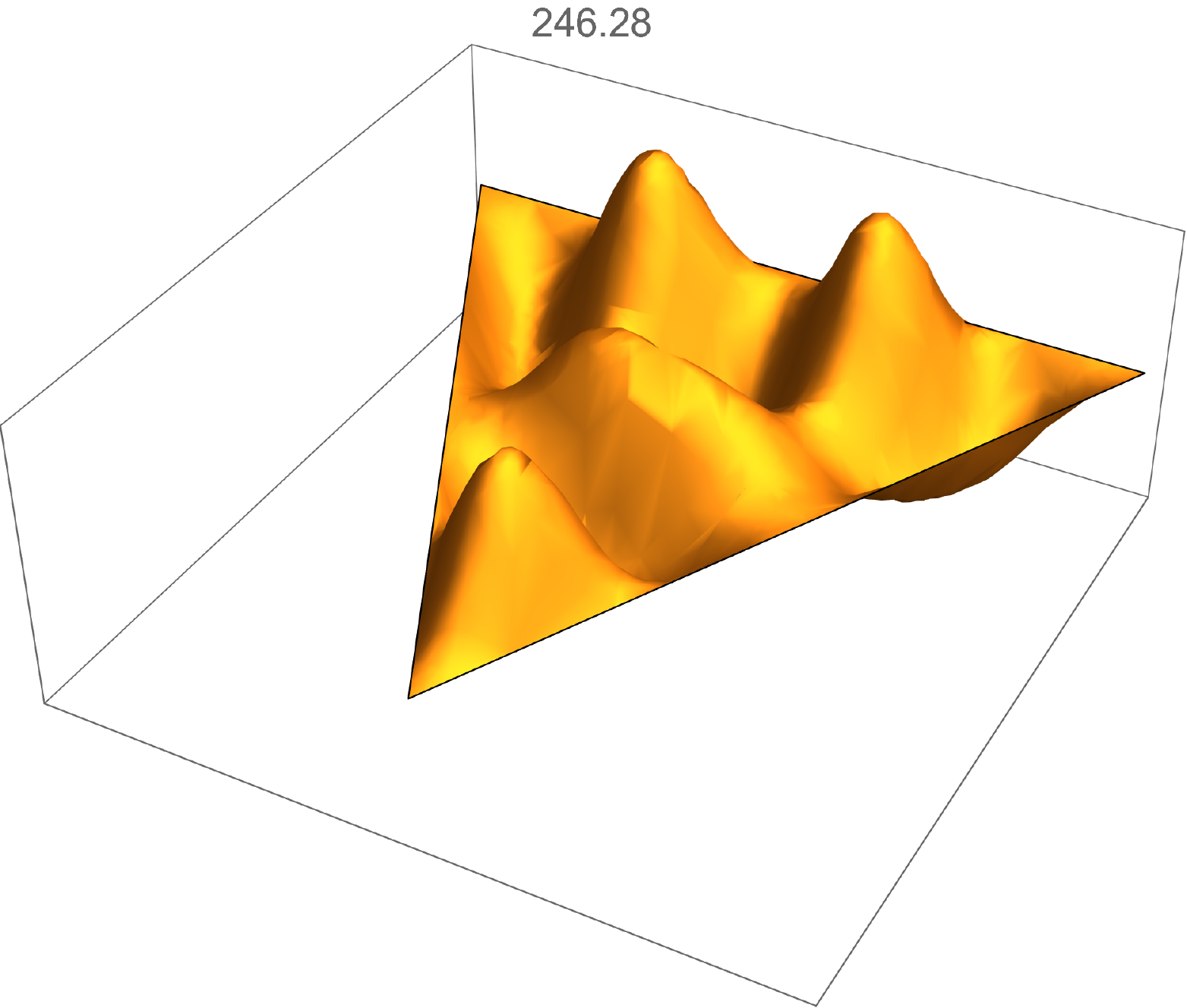}
\caption{Twelve lowest eigenfunctions of the Laplacian on the equilateral triangle, used as our basis functions. The numbers
on the top of each plot are their respective eigenvalues.}
\label{fig_b}
\end{center}
\end{figure}
\end{widetext}

\section{Modified Jacobi coordinates for five-body systems} \label{sec_Jacobi5}
The tandard form of the Jacobi coordinates $\vec r_i,i=1..5$  is
\ba \vec r_1&=&\vec x_1-\vec x_2, \,\,\, \vec r_2={1\over 2} \vec x_1+{1\over 2} \vec x_2-\vec x_3 ,\nonumber \\
 \vec r_3&=&{1 \over 3} \vec x_1+{1 \over 3} \vec x_2+{1 \over 3} \vec x_3 -\vec x_4 ,\nonumber\\
  \vec r_4&=&{1 \over 4} \vec x_1+{1 \over 4} \vec x_2+{1 \over 4} \vec x_3 +{1 \over 4} \vec x_4-\vec x_5, \nonumber \\
   \vec r_5&=&{1 \over 5}(\vec x_1+\vec x_2+\vec x_3+\vec x_4+\vec x_5)
\ea 
with the determinant of this matrix -- the Jacobian -- being 1. The last coordinate $\vec r_5$ is the location of the center of mass, if
those are coordinates, or 1/5 of the total momentum if $x_i$ are momentum fractions, and is in any case redundant.

As for  two  and three bodies, the simplest ``star-like" potential simplifies to purely a diagonal form 
\be \sum_1^5 \vec x_i^2= {1 \over 2} \vec r_1^2+{2\over 3} \vec r_2^2+{3 \over 4}  \vec r_3^2+{4 \over 5} \vec r_4^2
+5 \vec r_5^2 \ee
Further modification of the Jacobi coordinates is done by a simple rescaling of $\vec r_i$
\ba \vec r_1&\rightarrow& \sqrt{2} \vec \alpha,\, \vec r_2\rightarrow \sqrt{3/2}\vec \beta,\,  \vec r_3\rightarrow 
 \sqrt{4/3} \vec \gamma, \nonumber \\
 \vec r_4 &\rightarrow& \sqrt{5/4} \vec \delta, \vec r_5\rightarrow 
 \sqrt{1/5}\vec \sigma
\ea
which makes the ``star-like" potential  a sum of squares of greek-letter coordinates.

Another important potential, the ``Ansatz A", is half the sum of ten binary potentials. If it  is assumed  quadratic, 
it is proportional to the same sum of squares in terms of greek-letter coordinates
\ba V_A&=&{1\over 2} \sum_{i>j} (\vec x_i -\vec x_j)^2 \\
&=&  {5\over 2} \big(\vec \alpha^2+\vec \beta^2+\vec \gamma^2+\vec \delta^2 \big)  \ea
except that the $\vec \sigma$ term is missing (which is unimportant as it is  constant anyway). 

In sum: in proper coordinates, the confining five-body potentials $V_{star}$ and $V_A$, 
can  be reduced to the sum of greek-letter coordinates squared. In our approach  to the quantization  of
$\vec x_i \rightarrow i \partial /\partial \vec p_i$, these operators take the form of Laplacians.
The appropriate basis function are eigenfunctions of the Laplacian on the four-dimensional 5-simplex,
the descendants of our basis on an equilateral triangle for three bodies. 

\section{The coupling constant of instanton-induced 't Hooft vertex} \label{sec_G_tHooft}

$^\prime$t Hooft standard interaction for two light flavors reads
${\bf q}=({\bf u}, {\bf d})$
\bea
\label{DET1}
{\cal L}_{qq} = 2\bigg ( \frac {\Delta M^2_q}{n} \bigg ) \,\,
\left( {\rm det }\,\overline {\bf q}_R {\bf q}_L \,\,+\,\,{\rm det}\,
\overline {\bf q}_L {\bf q}_R\right)
\label{26}
\eea
The  light constituent quark mass, for the original parameters of the (dilute) random instanton liquid model (RILM) is~\cite{Chernyshev:1994zm,Kock:2020frx} 
\bea
\label{SQRTN}
\Delta M_q\sim \sqrt{\frac{n}{2N_c}}\bigg(\frac{4\pi^2\rho^2}{\Sigma_0}\bigg)\sim 420\,{\rm MeV}
\eea
with  the zero mode integral
\bea
\Sigma_0=\bigg(\int \frac{d^4k}{(2\pi)^4}k^2\varphi^{\prime 4}_I(k)\bigg)^{\frac 12}\sim (240\,{\rm MeV})^{-1}\nonumber\\
\eea
In singular gauge, the Fourier transform of the  fermionic zero-mode profile is
\bea
\varphi_I^\prime(k)=\pi\rho^2\bigg(I_0(z)K_0(z)-I_1(z)K_1(z)\bigg)^\prime_{z=k\rho/2}
\eea
The prime is a z-derivative,  and $I,K$ are modified Bessel functions. 
This function  with the canonical value $\rho=1/3\, fm$,  gives a dependence of the constituent quark mass on the virtuality
$k$ in good agreement with phenomenology and lattice studies.

Note that (\ref{SQRTN}) is not {\it analytic} in 
the packing fraction $n$, as expected from the spontaneous breaking of a symmetry~\cite{Pobylitsa:1989uq}.
It differs slightly from the analytic coupling extracted from a naive random approximation, also used in our earlier analyses.
More specifically, (\ref{SQRTN}) gives
\be  G_{tHooft}=2\bigg ( \frac {\Delta M^2_q}{n} \bigg )\approx 35 \, GeV^{-2} \ee 
which is to be compared to $17\,{\rm GeV}^{-2}$ using the naive approximation (see Eq. 62 in~\cite{Shuryak:2021fsu}).

\section{Sea production, to the first order in 't Hooft vertex}~\label{THOOFTDGLAP}
The kinematics and the expression for the probability of the quark sea pair production has been given in (\ref{eqn_dN}). 
This relatively complicated expression need to be integrated over the 3-quark phase space, which is basically a 5-dimensional
integral over $x,z,p_\perp,k_\perp$ and the azimuthal angle between those two vectors. 

The expression as written diverges at large momenta. If we would use the NJL model, the
integrals should go to an arbitrary  UV cutoff $\Lambda$. Fortunately, the instanton-induced coupling is naturally cutoff
by the instanton size $\rho$. Its exact form is given by the Fourier transform of fermionic zero modes, which
at large momenta asymptotes simply to $e^{-p\rho}$. Since both momentum integrals are basically $\int e^{-p\rho} p^2(dp/p) $, their main contribution is at the scale $p_*\approx 2/\rho\sim 1.2\, GeV$. So, the scale of this process
is close to our assumed $^\prime$unification scale$^\prime$ of 1 $GeV$.

Let us consider separately the domains in which $p>k$ and $p<k$, and expand in the $p/k$ (or $k/p$) ratio. 
It turns out that the large square bracket in (\ref{eqn_dN}) simplifies to
$${ x z (1 - x - z)) \over (1 - x)^2 }+ {\cal O}\bigg(\frac{k_\perp^2}{p_\perp^2}\bigg)$$
as the first order term $O(k_\perp/p_\perp)$ vanishes by angle average. Therefore the $x,z$ integrals separate from 
the momentum integrals. The former one needs to be taken over the physical region, meaning that all three $x_i\in [0,1]$
and their divergencies are regulated by a small $\epsilon$.
The result for the $p>k$ region is the same as for the $k>p$,  and their total sum is 
\be dN={G_{Hooft}^2 \over 4} {1 \over  16\pi^4 \rho^4}\,{\rm log}({ e^2 \over 4\epsilon}) \ee
where the regulator $\epsilon$ corresponds to the minimal momentum fraction $\sim m_q/P$.

\section{Reduction of 't Hooft interaction on the light front }~\label{app_REDUCTION}
To analyze the 't Hooft determinantal interaction~(\ref{DET1}) on the light
front, we use the free particle $u(p,s)$ and antiparticle $v(p,s)$ spinors, solutions to the Dirac equation
\bea
\label{SPINORS}
u(p,s)&=&\frac 1{\sqrt{2p^+}}(\slashed{p}+m_Q)\gamma^+\chi(s)\nonumber\\
v(p,s)&=&\frac 1{\sqrt{2p^+}}(\slashed{p}-m_Q)\gamma^+\chi(-s)
\eea
with $\gamma^+=\gamma^0+\gamma^3$ in the chiral representation, and the particle spin up-down
4-spinors $\sigma_z\chi(\pm)=\pm  \chi(\pm)$. The anti-particle spinor is tied to the particle spinor by 
$v(p,s)=iu(-p,-s)$,  and its conjugate by $\bar v(p,s)=i\bar{u}(-p, -s)$.

To reduce (\ref{DET1}) into a 2-body interaction, we formally factorize the isospin content,
and specialize to the particle-anti-particle channel, with 
\bea
(1-\tau_1\cdot \tau_2)(\bar{u}_Ru_L(1)\bar{v}_Rv_L(2)+\bar{u}_Lu_R(1)\bar{v}_Lv_R(2))\nonumber\\
\eea
for the instanton plus anti-instanton contribution. Using the light front spinors (\ref{SPINORS}) 
in momentum space, we have  for the particle entry
\begin{widetext}
\bea
\label{ULUR}
\bar{u}_L(p_2,s_2)u_R(p_1,s_1)=
\begin{pmatrix}
m_Q\sqrt{\frac{p_1^+}{p_2^+}}&0\\
\sqrt{\frac{p_2^+}{p_1^+}}p_{1R}-\sqrt{\frac{p_1^+}{p_2^+}}p_{2R}& m_Q\sqrt{\frac{p_2^+}{p_1^+}}
\end{pmatrix}\rightarrow 
\begin{pmatrix}
m_Q&0\\
q_R& m_Q
\end{pmatrix}=m_Q{\bf 1}+\frac 12\sigma^-q_R
\eea
with the convention for the spin-entries 
$$[s_2s_1]=\begin{pmatrix} ++&+-\\-+& --\end{pmatrix}$$
 Similarly, for the
anti-particle entry, we have
\bea
\bar{v}_L(p_2,s_2)v_R(p_1,s_1)=i^2
\begin{pmatrix}
m_Q\sqrt{\frac{p_2^+}{p_1^+}} &-\sqrt{\frac{p_2^+}{p_1^+}}p_{1R}+\sqrt{\frac{p_1^+}{p_2^+}}p_{2R}\\
0& m_Q\sqrt{\frac{p_1^+}{p_2^+}}
\end{pmatrix}\rightarrow 
\begin{pmatrix}
-m_Q&q_R\\
0&- m_Q
\end{pmatrix}=-m_Q{\bf 1}+\frac 12\sigma^+q_R\nonumber\\
\eea
\end{widetext}
with $q_R=p_{1R}-p_{2R}$, for a meson with net $P_T=0$. The rightmost equation follows from the eikonalization of the
particle line with $p_2^+\approx  p_1^+$ but $p_{\perp 2}\neq p_{\perp 1}$, as per our use of the straight Wilsonian lines
in the general derivation of the potentials.

To move back to light front space, we use the inverse Fourier transform, which gives
the local light front Hamiltonian (\ref{HLFqq}) since
\bea
\label{DELTA1}
\int_0^1 \frac{dx}\pi \frac{dk_\perp}{(2\pi)^2} e^{-ixP^+x^--ik_\perp b_\perp} =\delta(P^+x^-)\delta(b_\perp)\nonumber\\
\eea
The ultra-local and boost-invariant (\ref{DELTA1}),  is to be compared to
 \bea
 \label{DELTA2}
 2M\delta (\xi_x)\equiv 2M\delta(((\gamma x^-)^2+b_\perp^2)^{\frac 12})
 \eea
 with the Lorentz factor $\gamma=P^+/M$, 
which is also ultra-local and boost-invariant. The contribution $\gamma x^-$ reflects on the
time dilatation effect on the light front. $\xi_x$ is the natural invariant distance,
when the light front Hamiltonian is extracted from the Wilson lines, using the analytical
construction discussed above and in our preceding studies~\cite{Shuryak:2021hng,Shuryak:2021mlh,Shuryak:2022thi}.

\section{PDFs and formfactors in holography}

The QCD gravity dual or holographic models are not used or discussed in this series of works. 
However, to put our results in perspective with those nonperturbative treatments of QCD in the
double limit of large $N_c$ and strong $^\prime$t Hooft coupling, we will briefly comment on some
of their aspects in relation to LFWs and formfactors.

Formally, the holographic models  provide a  description of
hadrons as modes of some bulk (5d) effective fields. The hadronic masses and wave functions $\psi_n(z)$ 
are defined from Schroedinger-like equations in the 5-th holographic coordinate $z$.  ``Holograms" obtained by certain 
prescriptions,  project  bulk wave functions  to the boundary $z=0$. They yield 
distributions of certain quantities like e.g. the stress tensor $T^{\mu\nu}(x)$. 
While such a procedure does define certain hadronic sizes and shapes, there is no access to 
their internal substructure in terms of quarks/gluons.
 
Holography and pQCD have  little in common, and  their predictions do not agree in general.
While the latter describes hard processes in the weak coupling regime, the former addresses 
QCD processes at strongly coupling  and large $N_c$,  in the semi-hard and soft regimes.
 
 In holography, a struck quark with large momentum does not create a jet 
but rather a very wide flow of energy
\cite{Lin:2007fa}. In its evolution  all the partons migrate 
to small $x$~\cite{Polchinski:2001tt,Hatta:2008tx}, where the physics is well captured by open
and closed  string excitations (Pomerons and Reggeons).

{\it Inelastic  DIS} or {\em hard elastic} scattering  are  well defined in holography~\cite{Polchinski:2001tt},
with structure functions and formfactors. There is certain confusion in the  literature about  
 the relationship of these results to pQCD and partonic physics.  

Holography and pQCD both have scattering rules for  hard {\it elastic} processes,
as originally noted in~\cite{Polchinski:2001tt}. The hard elastic turn requires the involvement of  the full hadron in the conformal limit,  features that are shared by both
holography and QCD.
 In pQCD, this follows from the fact that all the partons in a given hadron,
have  to undergo a   ``turn-around" under a hard scattering, which implies the probability $(1/Q^2)^{n-1}$ with $n$
being the number of partonic constituents in the hadron~\cite{Brodsky:1973kr}.
 In holography, a virtual photon has extension $1/Q$ in the holographic z-coordinate,  and  for a hard hit in bulk, 
 the   hadron has also to shrink to a size $1/Q^{\tau-1}$ with probability $P\sim (1/Q^2)^{\tau-1}$. 
Here $\tau$ is  the ``bulk" dimension of the field describing a hadron (of the corresponding double trace 5d operators ). 
For the elastic processes, weak  and strong coupling
scaling laws are both known, and  they are similar with $\tau$ identified as $n$~\cite{Polchinski:2001tt}.

In {\it inelatic} DIS processes,
a virtual photon of size $1/Q$ scatters on a  quark parton with probability 1 because it is pointlike. In contrast, 
a scattering off a hadron can only happen if  it  shrinks to the same size.   The
corresponding wave function is $(1/Q^2)^{\tau-1}$. While  
 $\tau$ is also sometimes called ``anomalous dimention", it
 depends on the hadron field behavior in 5 dimensions,  and has nothing to do with  the 
``anomalous dimensions" of perturbative operators. Neither $\alpha_s(Q^2)$ nor quarks or gluons
are  present in the bulk  fields or actions. 
In DIS $s=Q^2(1/x-1)$ is constant. So 
 the $F_2(x)$ structure function at large $Q^2$ 
and large $x\rightarrow 1$ are related. In fact one can write the structure function in the generic form
\bea
\label{F2XX}
F_2(x, Q^2)\sim Q^2\bigg|\bigg(\frac 1{Q^2}\bigg)^{\tau-1}\bigg|^2\,\big(s=Q^2(1-x)\big)^\alpha\nonumber\\
\eea
where $\alpha$ is $arbitrary$. 
To reproduce the hard scattering rule asymptotically,  we  $may$ set  it to some value, e.g. $\alpha=\tau-2$,
\bea
\label{PS}
F_2(x, Q^2)\rightarrow \bigg(\frac 1{Q^2}\bigg)^{\tau-1} \, (1-x)^{\tau-2}
\eea
to reproduce the holographic DIS result on a nucleon with a  ``reasonable" $\tau=3$,   or for a meson with $\tau=2$~\cite{Polchinski:2001tt}.
Yet there is still no Bjorken scaling at large $Q^2$!

Furthermore, the  large $x$ behavior  in (\ref{PS}) is different from that expected from the Drell-Yan-West scaling rule~\cite{Drell:1969km,West:1970av}. To reproduce it
one would require $Q^2$ independence of the structure function (\ref{F2XX}) as per Bjorken scaling, which fixes 
another power of $\alpha=2\tau-3$
in (\ref{PS}). Indeed, this rule follows from the existence of the LFWFs in terms of constituents, which
is absent in holography.

\bibliography{diquarks3,diquarks2,2bar,baryons,mesons_at_CM1,mesons_at_CM}
\end{document}